\documentclass[11pt,twoside]{article}
\usepackage{amsmath,amsfonts,amssymb,latexsym,bbm}
\usepackage[dvips]{graphicx} % standard LaTeX graphics tool to include Figure files
\usepackage{array}
\usepackage{tabularx}
\usepackage{url,subfigure}
\usepackage{epsfig}
\input{psfig.sty}

\newcommand{\REAL}{\ensuremath{\mathbb{R}}}
\newcommand{\INT}{\ensuremath{\mathbb{Z}}}

\newcommand{\EREAL}{\ensuremath{\overline{\REAL}}}

\newcommand{\beq}{\begin{equation}}
\newcommand{\eeq}{\end{equation}}
\newcommand{\bea}{\begin{eqnarray}}
\newcommand{\eea}{\end{eqnarray}}
\newcommand{\bean}{\begin{eqnarray*}}
\newcommand{\eean}{\end{eqnarray*}}
\newcommand{\bcen}{\begin{center}}
\newcommand{\ecen}{\end{center}}
\newcommand{\bitm}{\begin{itemize}}
\newcommand{\eitm}{\end{itemize}}
\newcommand{\dsty}{\displaystyle}
\newtheorem{Theorem}{\bf Theorem}

\newtheorem{Examples}{\bf Examples}
\newcommand{\Aa}{\ensuremath{\mathcal{A}}}
\newcommand{\Bb}{\ensuremath{\mathcal{B}}}
\newcommand{\Cc}{\ensuremath{\mathcal{C}}}

\newcommand{\Ee}{\ensuremath{\mathcal{E}}}

\newcommand{\Kk}{\ensuremath{\mathcal{K}}}
\newcommand{\Ll}{\ensuremath{\mathcal{L}}}
\newcommand{\Mm}{\ensuremath{\mathcal{M}}}

\newcommand{\Oo}{\ensuremath{\mathcal{O}}}
\newcommand{\Pp}{\ensuremath{\mathcal{P}}}

\newcommand{\Ss}{\ensuremath{\mathcal{S}}}

\newcommand{\Vv}{\ensuremath{\mathcal{V}}}

\newcommand{\Ww}{\ensuremath{\mathcal{W}}}

\newcommand{\sbs}{\ensuremath{\subseteq \,}}    % subset
\newcommand{\keno}{\ensuremath{\varnothing}}    % emptyset
\newcommand{\defineq}{\ensuremath{\, \triangleq \,}}    % define equation
\newcommand{\defineql}{\ensuremath{\, \triangleq \,}}    % define equation from left

\newcommand{\dilt}{\ensuremath{\oplus}}

\newcommand{\tran}[2]{\ensuremath{{#1}_{#2}}} % pointwise set translation
\newcommand{\refl}[1]{\ensuremath{{#1}^s}} % reflection

\newcommand{\flatdom}{\ensuremath{E}} % gen domain of function lattices
\newcommand{\gimdom}{\ensuremath{E}} % gen domain of images (may be on graphs of arbitr spaces)
\newcommand{\sscnv}{\ensuremath{\oplus}}     % Sup-sum convolution/Dilation
\newcommand{\iscnv}{\ensuremath{\oplus '}}   % Inf-sum convolution
\newcommand{\spcnv}{\ensuremath{\otimes}}    % Sup-product convolution/Dilation
\newcommand{\ipcnv}{\ensuremath{\otimes '}}      % Inf-product convolution
\def\sgcnv{\hbox{$\, \bigcirc \,$\kern-0.9em\hbox{\mgop}$\;\,$}} %sup-general convoln (Final in Arxiv)
\def\igcnv{\hbox{$\, \bigcirc \,$\kern-0.9em\hbox{\mgop}$\;'\,$}} %inf-general convoln (Final in Arxiv)
\def\supgeno{\hbox{$\, \bigcirc \,$\kern-1.0em\hbox{$\wedge$}$\,$}} %sup-generating map
\def\infgeno{\hbox{$\, \bigcirc \,$\kern-1.0em\hbox{$\vee$}$\,$}} %inf-generating map
\newcommand{\pord}{\ensuremath{\leq}} % gen partial order
\newcommand{\spord}{\ensuremath{<}} % gen partial order strict
\newcommand{\ipord}{\ensuremath{\geq}} % inverse gen partial order
\newcommand{\dpord}{\ensuremath{\pord '}} % dual gen partial order
\newcommand{\ltle}{\ensuremath{\bot}}  % lattice least elem.
\newcommand{\ltge}{\ensuremath{\top}}  % lattice greatest elem.
\newcommand{\flatle}{\ensuremath{O}}  % weighted function lattice least elem.
\newcommand{\flatge}{\ensuremath{I}}  % weighted function lattice greatest elem.
\newcommand{\ltneg}{\ensuremath{\neg}}   % gen lattice Negation (superscript)
\newcommand{\oconverg}{\ensuremath{\, \stackrel{\mathrm{ord}}{\rightarrow}\, }}  % Order convergence
\newcommand{\dimpls}{\ensuremath{q}} % Impulse (DTI systems)
\newcommand{\eimpls}{\ensuremath{q'}} % Impulse (ETI systems)
\newcommand{\dlop}{\mbox{\large $\delta$}} % gen lattice dilation operator
\newcommand{\erop}{\mbox{\large $\varepsilon$}} % gen lattice erosion operator
\newcommand{\fdlop}{\ensuremath{\Delta}}  % Function dilation operator
\newcommand{\ferop}{\ensuremath{\mathcal{E}}}  % Function erosion operator
\newcommand{\sbdl}{\ensuremath{\eta}}  % scalar binary dilation (infix: 2-arg notn)
\newcommand{\asbdl}{\ensuremath{\zeta}} % adjoint of scalar binary dilation (infix: 2-arg notn)
\newcommand{\sber}{\ensuremath{\zeta}} % scalar binary erosion (infix: 2-arg notn)
\newcommand{\asber}{\ensuremath{\eta}} % adjoint of scalar binary erosion (infix: 2-arg notn)

\newcommand{\opop}{\mbox{\large $\alpha$}} % gen lattice opening operator
\newcommand{\clop}{\mbox{\large $\beta$}} % gen lattice closing operator
\newcommand{\idop}{\ensuremath{\mathbf{id}}} % gen latt. Identity operator
\newcommand{\lcompl}{\ensuremath{c}} % (set-type) complement for B.lattices (superscript)
\newcommand{\trop}{\mbox{\large $\tau$}} % gen translation operator
\newcommand{\vset}{\ensuremath{\mathcal{K}}} % Value set
\newcommand{\vsetle}{\ensuremath{\bot}} % V-set least elem.
\newcommand{\vsetge}{\ensuremath{\top}} % V-set greatest elem.
\newcommand{\mgid}{\ensuremath{e}} % "multiplic" group identity elem.
\newcommand{\dmgid}{\ensuremath{e'}} % dual "multiplic" group ident elem.
\newcommand{\vsgroup}{\ensuremath{G}} % V-set Group
\newcommand{\mgop}{\ensuremath{\star}}  % "multiplic" group operation
\newcommand{\dmgop}{\ensuremath{\mgop'}} % dual "multiplic" group oper

\newcommand{\vct}[1]{\ensuremath{\boldsymbol{#1}}}  % vector (optional)
\newcommand{\mtr}[1]{\ensuremath{\boldsymbol{#1}}}  % matrix (optional)
\newcommand{\mxgmp}{\ensuremath{\ \frame{\mgop}\ }} % gen max-* matr prod
\newcommand{\mxgmpid}{\ensuremath{\boldsymbol{I}}} %Identity Matrix in max-* matr algebra
\newcommand{\mngmp}{\ensuremath{\ \frame{\mgop}\, '\ }} % gen min-* matr prod
\newcommand{\mxsmp}{\ensuremath{\boxplus}} % max-sum matr prod
\newcommand{\mnsmp}{\ensuremath{\mxsmp'}} % min-sum matr prod
\newcommand{\mxpmp}{\ensuremath{\boxtimes}} % max-pro matr prod
\newcommand{\mnpmp}{\ensuremath{\mxpmp'}} % min-pro matr prod
\newcommand{\mxfmp}{\ensuremath{\mxgmp}} %max-fuznorm mtrx product
\newcommand{\mnfmp}{\ensuremath{\mngmp}} %min-dualfuznorm mtrx prod
\newcommand{\mnasbdmp}{\ensuremath{\Box'_{\asbdl}}}  %min-adjscalarbindil mtrx product
\newcommand{\mxasbemp}{\ensuremath{\Box_{\asber}}} %max-adjscalarbinero mtrx product
\newcommand{\glconj}[1]{\ensuremath{{#1}^\ast}} % general lattice conjugate of elem/matrix
\newcommand{\gramtr}[1]{\mbox{${\rm Gr}({#1})$}}    % graph of matrix
\newcommand{\eresid}[1]{\ensuremath{{#1}^\sharp}} % erosion-type residual operator
\newcommand{\conjtranmtr}[1]{\ensuremath{{#1}^\ast}} % adjoint (=conjugate transpose) matrix
\newcommand{\metmtr}[1]{\ensuremath{\boldsymbol{\Gamma} ({#1})}} % metric matrix
\newcommand{\prnceval}[1]{\ensuremath{\lambda ({#1})}} % princ e-value of matrix
\newcommand{\pathl}{\ensuremath{\ell}} % graph path length
\newcommand{\pathw}{\ensuremath{w}} % graph path weight

\newcommand{\inorm}{\ensuremath{T}} % Fuzzy Intersection Norm = Fuzzy Conjunction
\newcommand{\dinorm}{\ensuremath{U}} % CoNorm Dual (general) to \inorm
\newcommand{\cinorm}{\ensuremath{\inorm^\ast}} % CoNorm Dual (Complement) to \inorm
\newcommand{\unorm}{\ensuremath{U}} % Fuzzy Union Norm
\newcommand{\fun}{\ensuremath{\mathrm{Fun}}}    % Function space
\newcommand{\spt}{\ensuremath{\mathrm{Spt}}}    % Support
\newcommand{\wsspan}{\ensuremath{\mathrm{span}_\vee}} % Weighted Sup Span
\newcommand{\wispan}{\ensuremath{\mathrm{span}_\wedge}} % Weighted Inf Span
\begin{document}

% --- For  ARXIV --------------
\pagestyle{myheadings}
\markboth{\small Submitted to arXiv.org}{\small Submitted to arXiv.org}
%\markboth{\small Submitted to arXiv.org, \today}{\small Submitted to arXiv.org, \today}
%
\begin{center}
{\Large\bf DYNAMICAL SYSTEMS ON WEIGHTED LATTICES: GENERAL THEORY}
\\[5mm]
{\large\sc Petros Maragos}
\\[3mm]
School of Electrical \& Computer Engineering, \\
National Technical University of Athens, \\
 Zografou campus,  15773 Athens, Greece.\\
Email: \ \texttt{maragos@cs.ntua.gr}
\\[1cm]
\end{center}

%-------------- TABLE Of CONTENTS  (for ARXIV) ----------------
{\small \tableofcontents }

\vspace*{10mm}
\noindent
{\bf Keywords:}
nonlinear dynamical systems, lattice theory, minimax algebra, control, signal processing.

\newpage

\begin{abstract}
In this work a theory is developed for unifying large classes of nonlinear
discrete-time dynamical systems obeying a superposition of a weighted maximum or minimum type.
The state vectors and input-output signals evolve on
nonlinear spaces which we call complete weighted lattices  and
include as special cases the nonlinear vector spaces of minimax algebra.
Their algebraic structure has a polygonal geometry.
Some of the special cases unified include max-plus, max-product,
and probabilistic dynamical systems.
We study problems of representation in state and input-output spaces using lattice monotone operators,
 state and output responses using nonlinear convolutions,
solving nonlinear matrix equations using lattice adjunctions, stability and controllability.
We outline applications in state-space modeling of nonlinear filtering;
dynamic programming (Viterbi algorithm) and shortest paths (distance maps); fuzzy Markov chains;
and  tracking audio-visual salient events
in multimodal information streams using generalized hidden Markov models with control inputs.

\end{abstract}

\section{Introduction}

Linear dynamical systems \cite{Broc70,Brog74,Kail80}
can be described in discrete-time by the state space equations
%$\vct{x}(t)=\mtr{A}\vct{x}(t-1)+\mtr{B}\vct{u}(t)$
%and $\vct{y}(t)=\mtr{C}\vct{x}(t)+\mtr{D}\vct{u}(t)$
\beq
\begin{array}{rcl}
\vct{x}(t) & = & \mtr{A}\vct{x}(t-1)+\mtr{B}\vct{u}(t)
\\
\vct{y}(t) & = & \mtr{C}\vct{x}(t)+\mtr{D}\vct{u}(t)
\end{array}
\label{ldtse}
\eeq
where $t \in \INT$ shall denote a discrete time index, % throughout our analysis,
$\vct{x}(t)$ is an evolving state vector, $\vct{u}(t)$ is the
input  signal (scalar or vector),
and $\vct{y}(t)$ is an output signal (scalar or vector).
$\mtr{A},\mtr{B},\mtr{C},\mtr{D}$ are appropriately sized matrices
and all the matrix-vector products are defined
in the standard linear way.
%In the deterministic case, the main tools for their analysis have been
%linear matrix algebra and vector spaces.
%
Linear systems have proven useful for a plethora
of problems in communications, control and signal processing.
%, such as
%frequency-selective or adaptive filtering, prediction, smoothing,
%estimation in Gaussian noise, tracking and stabilization.
The strongest motivation for using linear systems as models has been
the great familiarity of all sciences with linear mathematics,
e.g. linear algebra, linear vector spaces, and linear differential equations,
as well as the availability of computational tools and algorithms to solve problems with linear systems.

However, in the 1980s and 1990s,
  several broad classes of \emph{nonlinear} systems were developed
whose state-space dynamics can be described by equations
 whose structure resembles (\ref{ldtse}) but has nonlinear operations.
These were motivated by a broad spectrum of applications,
 such as scheduling and synchronization, operations research, dynamic programming, shortest paths on graphs, image processing, and  non-Gaussian estimation,
 for which nonlinear systems were more appropriate.
These nonlinearities involve two major elements:
1)~a nonlinear superposition of vectors/matrices via pointwise maximum $(\vee)$
or minimum ($\wedge$) which plays the role of a generalized `addition',
 and 2)~a  binary operation $\mgop$ among scalars that plays the role of
a generalized `multiplication'.
Thus, with the above generalized `addition' and `multiplication',
the set  of scalars has a conceptually similar arithmetic structure
as the field of reals with standard addition and multiplication
underlying the linear vector spaces over which linear systems act.
This alternative arithmetic  structure (with operations $\vee,\mgop$) is minimally an idempotent semiring.
Examples of `multiplication' include the sum and the product,
but $\mgop$ may also be only a semigroup operation.
The resulting algebras include 1)~the  \emph{max-plus algebra} $(\REAL\cup \{-\infty\},\max,+)$
used in scheduling and operations research  \cite{Cuni79},
 discrete event systems (DES)  \cite{BCOQ01,CaLa99,CDQV85,Ho92}, %\cite{CaHo90},
 automated manufacturing \cite{CDQV85,DoKa95,Kame93}, % \cite{Kame93,KaDo94,DoKa95}
 synchronization and transportation networks \cite{BCOQ01,Butk10,HOW06,BoSc12},
max-plus control \cite{Butk10,CDQV85,GaKa99,HOW06,BoSc12},
optimization \cite{BCOQ01,BeKa61,Butk10,CGQ04,GoMi08,McEn06}, geometry \cite{CGQ04,GaKa07},
morphological image analysis \cite{Heij94,MaSc90,Serr88,Ster86},
and neural nets with max-plus or max-min combinations of inputs \cite{RSD98,RiUr03,ChMa17,YaMa95};
2)~the min-plus algebra or else known as \emph{tropical semiring} $(\REAL\cup \{+\infty\},\min,+)$
used in shortest paths on networks \cite{Cuni79} and in speech recognition and natural language processing \cite{MPR02},\cite{HoNa13};
%\emph{probability semirings} used in speech recognition \cite{RaJu93}
%either of the sum-product type $([0,\infty],+,\times)$ or
%of the max-times type $([0,\infty],\vee,\times)$;
this is a logarithmic version of 3)~the underlying \emph{max-times} semiring $([0,+\infty),\vee,\times)$
used for  inference  with belief propagation in graphical models \cite{Pear88,Bish06};
4)~the \emph{fuzzy logic or probability semiring} $([0,1],\vee,T)$ with statistical $T$-norms used in
probabilistic automata and fuzzy neural nets \cite{KlYu95},\cite{KaPe00},
fuzzy image processing and dynamical systems \cite{BlMa95,Mara05a,MST00},
and fuzzy Markov chains \cite{AvSa02}.
Max-plus algebra is also a major part of \emph{idempotent mathematics} \cite{LMS01,Masl87},
a vibrant area with contributions to mathematical physics and optimization.
Further, in multimodal processing for cognition modeling,
%which has been a main motivation for this work,
several psychophysical and computational
experiments indicate that the superposition of sensory signals or cognitive
states seems to be better modeled using max or min rules, possibly weighted.
Such an example is the  work \cite{Eva+13} on attention-based multimodal
video summarization where a (possibly weighted) min/max fusion
 of  features from the audio and visual signal channels and
of salient events from various modalities seems to outperform linear fusion schemes.
Tracking of these salient events was  modeled in \cite{MaKo15} using
 a max-product dynamical system.
Finally, % heterogeneous / hierarchical
the problem of bridging the semantic gap in multimedia requires integration
of continuous sensory   modalities (like audio and/or vision)
with discrete language symbols and semantics extracted from text.
Similarly, in control and robotics there are efforts to develop hybrid systems
that can model interactions between heterogeneous information streams like continuous inputs
and symbolic strings, %  \cite{Broc94}.
 e.g. motion control with language-driven variables \cite{Broc94}.
In both of these applications we need models where the computations among modalities/states can handle
both real numbers and Boolean-like variables; this is possible using max/min rules.

%The above topics can be loosely classified into four areas (with possible overlaps):
% 1) Max-plus and minimax algebra with applications in DES, max-linear systems, control, and operations research;
%2)~Signal processing and machine learning with applications in nonlinear filtering, image processing
%and computer vision;
%3)~Idempotent analysis with applications in problems of mathematical physics and optimization;
%4)~Tropical geometry as a subdiscipline of algebraic geometry.
%Our work in this paper lies in the first and second areas.

Motivated by the above applications,
in this work we develop a theory and some tools to unify the representation and analysis
of nonlinear  systems whose dynamics evolve based on the following
 \textbf{max-$\boldsymbol{\mgop}$} model
\beq
\begin{array}{rcl}
\vct{x}(t) & = &
 \mtr{A}(t) \mxgmp \vct{x}(t-1) \  \vee \ \mtr{B}(t) \mxgmp \vct{u}(t)
\\
\vct{y}(t) & = &
 \mtr{C}(t) \mxgmp \vct{x}(t) \ \vee \ \mtr{D}(t) \mxgmp \vct{u}(t)
\end{array}
\label{mxgse}
\eeq
where $\vct{x}=[x_1,x_2,...,x_n]^T\in \vset^n$
is a $n$-dimensional state vector with elements from the scalars' set $\vset$,
% whose structure will be described in Sec.~\ref{sc-algss} and
which will generally be a subset of the extended reals
$\EREAL = \REAL \cup \{ -\infty,\infty\}$.
The linear matrix product in (\ref{ldtse}), which is based on a sum of products,
is replaced in (\ref{mxgse}) by a nonlinear matrix product ($\mxgmp$) based on a max of $\mgop$ operations,
where $\mgop$ shall denote our general scalar operation
discussed in Sec.~\ref{sc-algss}.
%whose structure is described in Sec.~\ref{sc-algss}.
%
%In general,
The max-$\mgop$  `multiplication' of
a matrix $\mtr{A}=[a_{ij}]\in \vset ^{m\times n}$
 with a vector $\vct{x}=[x_i]\in \vset ^n$ yields a vector $\vct b=[b_i]\in \vset ^m$
defined by:
\beq
\mtr{A}\mxgmp \vct{x}=\vct{b} \; \; \; , \; \; \;
b_i=\bigvee _{j=1}^n a_{ij}\mgop x_j
%\{ \mtr{A}\mxgmp \vct{x}\}_i =\bigvee _{j=1}^n a_{ij}\mgop x_j
\label{mxgeqaxb}
\eeq
Further, the pointwise `addition' of vectors (and possibly matrices) of same size in (\ref{ldtse})
is replaced in (\ref{mxgse}) by  their pointwise $\vee$:
\beq
\begin{array}{rcl}
\vct{x} \vee \vct{y} & = &  [x_1\vee y_1, \dots , x_n \vee y_n]^T
\\ \mtr{A} \vee \mtr{B} & = &  [a_{ij} \vee b_{ij}]
\end{array}
\eeq
A max-plus $2\times 2$ example of (\ref{mxgeqaxb}) is
\beq
%\left[ \begin{array}{cc} 4 &  -1 \\ 2  &  -\infty \end{array} \right]
%\mxgmp \left[ \begin{array}{c} x_1 \\ x_2 \end{array} \right] =
%\left[ \begin{array}{c} 3 \\ 1 \end{array} \right]
%, \;
%\Longleftrightarrow
%\begin{array}{r}
%\max (x_1+4, x_2-1) = 3 \\
%x_1+2 =  1
%\end{array}
\left[ \begin{array}{cc} 4 &  -1 \\ 2  &  -\infty \end{array} \right]
\mxgmp \left[ \begin{array}{c} x \\ y \end{array} \right] =
\left[ \begin{array}{c} 3 \\ 1 \end{array} \right]
, \;
%\Longleftrightarrow
\begin{array}{r}
\max (x+4, y-1) = 3 \\
x+2 =  1
\end{array}
\eeq
with solution $x=-1$ and $y\leq 4$.

By replacing maximum ($\vee$) with minimum ($\wedge$) and the $\mgop$ operation with a dual operation $\dmgop$
 we obtain a \emph{dual}  model
that describes the state-space dynamics of \textbf{min-$\boldsymbol{\dmgop}$} systems:
\beq
\begin{array}{rcl}
\vct{x}(t) & = &
 \mtr{A}(t) \mngmp \vct{x}(t-1) \  \wedge \ \mtr{B}(t) \mngmp \vct{u}(t)
\\
\vct{y}(t) & = &
 \mtr{C}(t) \mngmp \vct{x}(t) \ \wedge \ \mtr{D}(t) \mngmp \vct{u}(t)
\end{array}
\label{mngse}
\eeq
where the min-$\dmgop$  matrix-vector `multiplication'
%`multiplication' of a matrix $\mtr{A}$  with a vector $\vct{x}$
is defined by:
\beq
\mtr{A}\mngmp \vct{x}=\vct{b} \; \; \; , \; \; \;
b_i=\bigwedge _{j=1}^n a_{ij}\dmgop x_j
%\{ \mtr{A}\mxgmp \vct{x}\}_i =\bigvee _{j=1}^n a_{ij}\mgop x_j
\label{mngeqaxb}
\eeq
%In (\ref{mxgse}) and (\ref{mngse})  the `addition' of vectors in (\ref{{ldtse}})
%is replaced by their pointwise $\vee$ or $\wedge$, respectively:
%\beq
%\begin{array}{rcl}
%\vct{x} \vee \vct{y} & = &  [x_1\vee y_1, \dots , x_n \vee y_n]^T \\
%\mtr{A} \vee \mtr{B} & = &  [a_{ij} \vee b_{ij}]
%\end{array}
%\eeq
%
The state equations (\ref{mxgse}) and (\ref{mngse}) % for max and min control systems
have \emph{time-varying} coefficients.
For constant
matrices $\mtr{A},\mtr{B},\mtr{C},\mtr{D}$, we obtain
the \emph{constant-coefficient} case.

By specifying the scalar `multiplication' $\mgop$ and its dual $\dmgop$,
we obtain a large variety of classes of nonlinear dynamical systems
that are described by the above unified algebraic models of the max or min type.
%
%Special cases include systems formed by the following algebras:
% max-sum ($\mgop=+$), max-product ($\mgop=\times$),
% Boolean ($\mgop$ is Boolean product), and max-fuzzy intersection ($\mgop$ is a $T$ norm).
%
%Next we summarize three such choices,
%%and the respective types of systems,
%whose characteristic operations are
%
%In the above general models and related matrix algebra,
The most well-known special case is $\mgop=+$,
the principal interpretation of  minimax algebra,
which has been extensively studied  in scheduling,
 DES, max-plus control  and optimization
 \cite{BCOQ01,Butk10,CDQV85,CGQ04,Cuni79,GaKa99,HOW06,LMS01}.
In typical  applications of DES in automated manufacturing,
the states $x_i(t)$
represent starting times of the $t$-th cycle of machine $i$, % $i=1,...,n$,
the input $\vct{u}$ represents availability times of parts,
 $\vct{y}$  represents completion times,
and the elements of $\mtr{A},\mtr{B},\mtr{C},\mtr{D}$
represent  activity durations. %  service/delay times or
The homogeneous state dynamics of (\ref{mxgse}) are modeled by max-plus recursive equations:
\beq
%x_i(t+1)=(\max_{1\leq j\leq n} x_j(t)+a_{ij}) \vee (\max_{1\leq j\leq m} u_j(t)+b_{ij})
x_i(t)=\max_{1\leq j\leq n} a_{ij}+x_j(t-1), \quad i=1,...,n
\label{mxsrecurs}
\eeq
Another special case is $\mgop=\times$, which is however less frequently related to minimax algebra
 and rarely viewed as a dynamical system.
This is used in communications (e.g. Viterbi algorithm),
 in probabilistic networks \cite{Pear88,Bish06} as the max-product belief propagation,
and %(its log version)
in speech  recognition and language processing \cite{MPR02,HoNa13}.
%or weighted finite-state automata in language processing \cite{MPR02},
%where the modeling part involved only the homogeneous state equation %$\vct x(t)=\mtr A\mxpmp \vct x(t-1)$.
But  both its max-product algebra and its general dynamics
with control inputs have not been studied.
Other cases are much less studied or relatively unknown.

%In analogy to the linear spaces underlying linear systems,
 Our theoretical analysis is based on a relatively new type of nonlinear space we have developed
in recent work \cite{Mara13} and further refine herein,
which we call \emph{complete weighted lattice (CWL)}.
This combines two vector or signal generalized `additions' of the
supremum ($\vee$)  or the infimum ($\wedge$) type and two generalized scalar multiplications,
$\mgop$ and its dual $\dmgop$, which distribute over $\vee$ and $\wedge$ respectively.
%The scalar arithmetic is based on an algebraic structure we had introduced in \cite{Mara05a} called
% \emph{clodum} (complete lattice-ordered double monoid).
 The axioms of CWLs bear a remarkable similarity with those of linear spaces,
 the major difference being the lack of inverses for the sup/inf operations
 and sometimes for the $\mgop$ operation too.
The present work focuses on analyzing max/min dynamical systems
using CWLs, whose advantages  over the  minimax algebra \cite{Cuni79},
which has been so far the main algebraic framework for DES and max-plus control, include the following:
\\
%1)~CWLs can apply both to finite- and infinite-dimensional spaces whereas minimax algebra
% is mainly a matrix algebra on finite-dimensional vector spaces. \\
%2)~Their theoretical foundation contains a built-in duality which allows
% co-existence of the max-$\mgop$ and its dual min-$\dmgop$ algebra.
% \\
% 3)~We believe that lattice theory \cite{Birk67} offers a conceptually elegant and more compact way to express
% the combined rich algebraic structure instead of viewing it as two idempotent semirings of minimax algebra.
 1)~We believe that the  theory of lattices and lattice-ordered monoids \cite{Birk67}
 offers a conceptually elegant and compact way to express
 the combined rich algebraic structure instead of viewing it
 as a pair of two idempotent ordered semirings of minimax algebra. Although in several previous works
 the $(\max,+)$ and $(\min,+)$ algebras have been used at the same time,
 expressing and exploiting the coupling between the two becomes simpler by using the
 built-in duality of lattices which is at the core of their theoretical foundation.
 \\
 2)~Lattice monotone operators of the dilation ($\dlop$) or erosion ($\erop$)
 type can be defined, as done in morphological image analysis \cite{Heij94,Mara05a,Serr88},
 which play the role of `linear operators' on CWLs and
  can represent systems obeying a max-$\mgop$ or a min-$\dmgop$ superposition respectively.
Such operators can represent both state vector transformations by matrix-vector generalized products
of the max-$\mgop$ or min-$\dmgop$ type, as in (\ref{mxgeqaxb}) and (\ref{mngeqaxb}),
as well as input-output signal mappings
in the form of nonlinear convolutions of two signals $f$ and $g$:
%of the following sup-$\mgop$ type
sup-$\mgop$ convolution
\beq
(f\sgcnv g)(t) \defineq \bigvee _k f(k)\mgop g(t-k),
%\; \; \; (f\igcnv g)(t) \defineq  \bigwedge _k f(k)\dmgop g(t-k)
\eeq
or inf-$\dmgop$ convolution
\beq
(f\igcnv g)(t) \defineq  \bigwedge _k f(k)\dmgop g(t-k)
\eeq
The only well-known special case $\mgop=+$ is called supremal or infimal convolution in convex analysis and optimization
 \cite{BeKa63a,Luce10,Rock70} as well as weighted (Minkowski) signal dilation or erosion in
morphological image analysis and vision \cite{Heij94,MaSc90,Serr88,Ster86}. % \cite{Ster86,Serr82,Heij94,Mara05b}.
Other cases are much less studied or relatively unknown.
\\
3)~Modeling the information flow in these dynamical systems via the above lattice operators is
greatly enabled by the concept of \emph{adjunction},
which is a  pair
$(\erop,\dlop)$ of erosion and dilation operators forming a type of duality expressed by the following
\beq
\dlop (\vct x)\leq \vct y \Longleftrightarrow \vct x \leq \erop (\vct y)
\label{ivecadj}
\eeq
for any vectors or signals $\vct x,\vct y$.

\begin{figure}[h!]
\centerline{ \psfig{figure=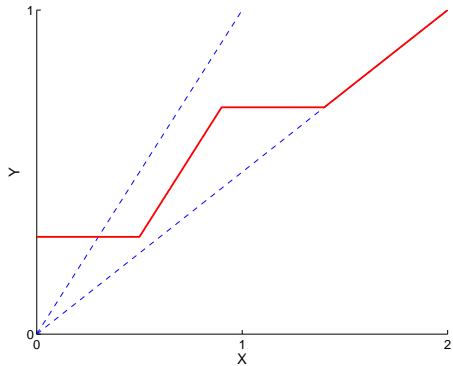,width=6cm}}
\caption{Geometry of basic superposition via a straight line (dashed) in linear spaces
versus the polygonal line (solid line) in complete weighted lattices - the \emph{polygonal spaces}.
The polygonal line is $y=\min [ \max(x-0.2,0.3), \max(x/2,0.7)]$.}
\label{fg-cwlgeometry}
\end{figure}

From a geometrical viewpoint, we may call the CWLs \emph{polygonal spaces} because of the
geometric shape of the corner-forming piecewise-straight lines $y=\max (a+x,b)$ or $y=\max (ax,b)$
and their duals (by replacing $\max$ with $\min$) which express the basic algebraic superpositions
in CWLs, in analogy to the geometry of the straight line $y=ax+b$ which expresses
in a simplified way the basic superposition in linear spaces.
See Fig.~\ref{fg-cwlgeometry} for an example.

\vspace*{3mm}
\textbf{Contributions of our work:}\\
(i)~Unify all types of max-$\mgop$ and min-$\dmgop$ systems under a
common theoretical framework of \emph{complete weighted lattices} (CWLs).
Further, while previous work focused mainly on the $(\max,+)$ or $(\min,+)$ formalism,
 we join both using CWLs and generalize them by replacing $+$ with any
  operation $\mgop$ that distributes over $\vee$ and a dual operation $\dmgop$ that distributes over $\wedge$.
  The corresponding generalized scalar arithmetic is governed by a rich algebraic structure, called \emph{clodum},
  which we developed in previous work \cite{Mara05a,Mara13} and further refine herein.
  This clodum serves as the `field of scalars' for the CWLs and binds together a pair of dual `additions'
  with a pair of dual `multiplications'; as opposed to max-plus,
  in some cases the `multiplications' do not have inverses.
   Two examples different from max-plus,
  which we analyze in some detail with applications, are the max-product and the max-min cases.
  %
%Special cases include systems formed by the following algebras:
% max-sum ($\mgop=+$), max-product ($\mgop=\times$),
% Boolean ($\mgop$ is Boolean product), and max-fuzzy intersection ($\mgop$ is a $T$ norm).

(ii)~Analyze the nonlinear system dynamics both in state space using a CWL matrix-vector algebra as well as
in the input-output signal space using sup/inf-$\mgop$ convolutions, represented
via lattice monotone  operators in adjunction pairs.
For the above, we have used the common formalism of CWLs to model both finite- and infinite-dimensional spaces.
% whereas minimax algebra is mainly a matrix algebra on finite-dimensional vector spaces.

(iii)~Enable and simplify the analysis and proofs of various results in system representation
using lattice adjunctions. Further, use the latter to generate lattice projections that provide
optimal solutions for max-$\mgop$ equations
$\mtr A \mxgmp \vct x = \vct b.$
Since the constituent operators of the lattice adjunctions are dilations and erosions which
have a geometrical interpretation and have found numerous applications in image analysis,
the above perspective to nonlinear system analysis also offers some geometrical insights.

%(iv)~Begin exploration of some interesting problems in these nonlinear spaces (CWLs) such as
%operator representation, bases,  solving max-$\mgop$ matrix equations, and spectral analysis.

(iv)~Study causality, stability, and controllability of max-$\mgop$ and min-$\dmgop$ systems
and link stability with spectral analysis in max-$\mgop$ algebra and controllability with lattice projections.

(v)~Advance the study of special cases employed in many application areas:
a)~Nonlinear systems represented by max/min-sum ($\mgop=+$) difference equations, as applied to geometric filtering
and shortest path computation. State equations and stability analysis of recursive nonlinear filters.
b)~Max-product systems ($\mgop=\times$) that extend the Viterbi algorithm of hidden Markov models
to cases with control inputs and can model cognitive processes related to audio-visual attention.
%that track the audio-visual saliency states in multimodal videos
%as compared with human attention.
c)~%Outline  in the Conclusions the possibility of our CWL framework to model the dynamics of
Probabilistic automata and fuzzy Markov chains governed by max/min rules and
with arithmetic based on triangular norms. % encountered in fuzzy logic.
% iii)~Boolean dynamical systems for tracking time-evolving binary decisions.

\textbf{Notation}: \ We think that the currently used notation in max-plus algebra of
$\oplus$ and $\otimes$ to denote the maximum $\vee$ (`addition') and the $\mgop$ (`multiplication') respectively
obscures the lattice operations; in contrast our proposed notation
is simpler and more realistic since it uses the well-established symbols  $\vee, \wedge$ for sup/inf operations
and does not bias the arbitrary scalar binary operation $\mgop$ with the symbol $\otimes$.
Further, the symbol $\oplus$ has been extensively used  in signal and image processing
for the max-plus signal convolution; herein, we continue this notation.
Table~\ref{tb-notation} summarizes the main symbols of our notation.
We use roman letters for functions, signals and their arguments and greek letters for operators.
Also, boldface roman letters for vectors (lowcase)  and matrices (capital).
If $\mtr{M}=[m_{ij}]$ is a matrix, its $(i,j)$th element is also denoted
as  $\{ \mtr{M} \} _{ij} = m_{ij}.$ Similarly,
$\vct{x}=[x_i]$ denotes a column vector, whose $i$-th element
is denoted as $\{ \vct{x} \} _i$ or simply $x_i$.

\begin{table}
\caption{Notation for the main algebraic operations}
\centerline{
\begin{tabular}{|c|c|}
\hline
Operation & Meaning \\ \hline
\hline
$\bigvee$ & Maximum/Supremum:  applies for scalars, vectors and matrices \\ \hline
$\bigwedge$ & Minimum/Infimum:  applies for scalars, vectors and matrices \\ \hline
$\mxgmp \ (\mngmp)$ & General max-$\mgop$ (min-$\dmgop$) matrix multiplication \\ \hline
$\mxsmp \ (\mnsmp)$ & Max-sum (min-sum) matrix multiplication \\ \hline
$\mxpmp \ (\mnpmp)$ & Max-product (min-product) matrix multiplication \\ \hline
$\sgcnv \ (\igcnv)$ & General max-$\mgop$ (min-$\dmgop$) signal convolution \\ \hline
$\sscnv \ (\iscnv)$ & Max-sum (min-sum) signal convolution \\ \hline
$\spcnv \ (\ipcnv)$ & Max-product (min-product) signal convolution \\ \hline
\end{tabular}
}
\label{tb-notation}
\end{table}

% #############################################################
\section{Lattices and Monotone Operators} \label{sc-latt}

Most of the background material in this section follows \cite{Birk67},
  \cite{Heij94}, \cite{HeRo90}, \cite{Mara13}, and \cite{Serr88}.

\subsection{Lattices} \label{sc-lattices}

A partially-ordered set, briefly \textit{poset} $(\Pp ,\pord)$,
 is a set $\Pp$ with a binary  relation $\pord$   that is
 a \textit{partial ordering}, i.e.
is reflexive, antisymmetric and transitive.
% satisfies  three properties for all $X,Y,Z\in \Pp$:
%reflexive ($X\pord X$), antisymmetric ($X\pord Y$ and $Y\pord X$ imply $X=Y$)
% and transitive  ($X\pord Y$ and $Y\pord Z$ imply $X\pord Z$).
If, in addition, % $\pord$ has the additional property  that,
for any two elements $X,Y\in \Pp$ we have either $X\pord Y$ or $Y\pord X$,
then $\Pp$ is called a % totally-ordered set or
\textit{chain}.
To every partial ordering $\pord$ there corresponds a
\emph{dual partial ordering} $\dpord$ defined by
``$X\dpord Y$ iff $X\ipord Y$''.
% If $(\Pp,\pord)$ is a poset, then
%$(\Pp,\dpord)$ is also a poset, called the \emph{dual poset}.
%
Let $\Ss$ be a subset of  $(\Pp, \pord)$;
an upper bound of  $\Ss$  is an element $B\in \Pp$
such that $X\pord B$ for all $X\in \Ss$.
%; if $B\in \Ss$, then it is the \emph{greatest element} of $\Ss$.
The least upper bound of $\Ss$ is called its \textbf{supremum}
and denoted by $\sup \Ss$ or $\bigvee \Ss$.
By duality,
we define the  greatest lower bound of $\Ss$,
 called its \textbf{infimum}
and denoted by $\inf \Ss$ or $\bigwedge \Ss$.
%The supremum and infimum are unique if they exist.
If the supremum (resp. infimum) of $\Ss$ belongs to $\Ss$, then it
is called the \emph{greatest element} or \emph{maximum}
(resp. \emph{least element} or \emph{minimum}) of $\Ss$.
An element $M$ of $\Ss$  is called \emph{maximal} (resp. \emph{minimal})
if there is no other element in $\Ss$ that is greater (resp. smaller) than $M$.
%with respect to $\pord$.

A \textbf{lattice} is a poset $(\Ll ,\pord)$ any two of whose elements
have a supremum, denoted by $X\vee Y$, and an infimum, denoted by $X\wedge Y$.
We often denote the lattice structure by $(\Ll ,\vee , \wedge)$.
A lattice  $\Ll$ is \emph{complete} if each of its subsets
(finite or infinite) has a supremum and
an infimum in $\Ll$. Any nonempty complete lattice is universally bounded because
it contains its supremum $\ltge=\bigvee \Ll$ and infimum $\ltle=\bigwedge \Ll$
which are its greatest (top) and least (bottom) elements, respectively.
%Sometimes $\ltle$ is called the `null element' and $\ltge$ is called the
%`all element'.
%If  $\Ll$ is not easily inferred from the context,  we may denote
%its least and greatest elements by $\ltle_\Ll$ and $\ltge_\Ll$ respectively.
%
%\emph{Duality in Lattices:}
%In any lattice $\Ll$, by replacing the partial ordering
%$\pord$ with its dual $\dpord$ and by interchanging the roles of the supremum and
%infimum, i.e., by considering the dual operations $\vee '=\wedge$ and $\wedge '=\vee$,
% we can form a new lattice  $(\Ll , \vee ', \wedge ')$,
%called the \emph{dual lattice} and often denoted just by $\Ll '$.
%The duality principle dictates that to every definition, property and statement
%that applies to the lattice $\Ll$ there corresponds a dual one that applies
%to the dual lattice $\Ll '$ by interchanging $\pord$ with $\dpord$ and
%$\vee$ with $\wedge$.
%
In any lattice $\Ll$, by replacing the partial ordering
 with its dual  and by interchanging the roles of the supremum and
infimum we obtain a \emph{dual lattice}  $\Ll '$.
\emph{Duality principle}: to every definition, property and statement
that applies to $\Ll$ there corresponds a dual one that applies
to  $\Ll '$ by interchanging $\pord$ with $\dpord$ and $\vee$ with $\wedge$.
A bijection between two lattices $\Ll$ and $\Mm$ is called an \emph{isomorphism}
(resp. dual isomorphism) if it preserves (resp. reverses) suprema and infima.
If $\Ll=\Mm$, a (dual-) isomorphism on $\Ll$ is called \emph{(dual-) automorphism}.
%A lattice is called \emph{self-dual} if there is a dual automorphism on it;
%it can be shown that a lattice $\Ll$ is self-dual iff
%it is isomorphic with its dual lattice $\Ll'$.

%--------------------------------------
%\subsubsection{Lattice Properties}

The lattice operations satisfy many properties, %the most fundamental of which are
as summarized in Table~\ref{tb-wlat}.
Conversely, a set $\Ll$ equipped with two binary operations $\vee$ and $\wedge$
that satisfy  properties (L1,L1$'$)--(L5,L5$'$)
 is a lattice whose
supremum is $\vee$, the infimum is $\wedge$, and partial ordering $\pord$ is
given by (L6).
%
% ----------- Semilattices -----------
%
% Note that, a lattice $(\Ll , \vee ,\wedge)$ contains two weaker substructures,
% two semilattices.
%Specifically, a sup-semilattice $(\Ll ,\vee)$ that
%satisfies properties (L$1-$L4) of Table~\ref{tb-wlat},
%an inf-semilattice $(\Ll,\wedge)$ that satisfies properties (L1$'-$L4$'$),
%and the two binary operations of supremum and infimum are related via properties
%(L5,L5$'$) that make them dual to each other.
A lattice $(\Ll , \vee ,\wedge)$ contains two weaker substructures:
 a sup-semilattice $(\Ll ,\vee)$
that satisfies properties (L$1-$L4) %of Table~\ref{tb-wlat},
and an inf-semilattice $(\Ll,\wedge)$ that satisfies properties (L1$'-$L4$'$).
%and the two binary operations of supremum and infimum are related via properties
%(L5,L5$'$) that make them dual to each other.
%Further, these binary operations are consistent with two dual corresponding ordering relations
%as revealed by properties (L6,L6$'$).

%
The additional properties (L7,L7$'$) and (L8,L8$'$) in Table~\ref{tb-wlat} hold only if
the lattice contains a least and a greatest element, respectively.
A lattice  $\Ll$ is called \textit{distributive} if
it satisfies properties (L9,L9$'$); % of Table~\ref{tb-wlat};
if these also hold over infinite set collections,
then the lattice is called \emph{infinitely distributive}.
%or equivalently
%\beq
%A\wedge \left( \bigvee_{i\in J} X_i\right) =
%\bigvee _{i\in J}(A\wedge X_i) \quad \mathrm{and} \quad
%A\vee \left( \bigwedge _{i\in J} X_i\right) = \bigwedge _{i\in J}(A\vee X_i)
%\label{latdistr} \eeq
%for any finite index set $J$ and for all $A,X_i\in \Ll$.
%If the above also holds for infinite index sets, then the lattice is called
%\emph{infinitely distributive}.
%
The rest of the properties of Table~\ref{tb-wlat}, labeled as ``WL\#'', refer to
a richer algebra defined as `weighted lattices' in Section~\ref{sc-cwl}.

%-------------------------------------------------------------
%\subsubsection{Sublattices}

%Let $(\Ll,\vee,\wedge)$ be a lattice with partial ordering $\pord$ and consider a
%nonempty subset $\Ss$  of $\Ll$. Under the same partial ordering, $(\Ss,\pord)$ is a
%poset. If $\Ss$ is also closed under finite suprema and infima that are induced by the
%partial ordering of $\Ll$, then $\Ss$ is a called an \textit{underlattice} of $\Ll$.
%If $\Ss$ is an underlattice whose  supremum and infimum induced by $\pord$ are
%the same operations ($\vee,\wedge$) as those of $\Ll$,
%then $\Ss$ is called a \textit{sublattice} of $\Ll$.
%A sublattice is always an underlattice, but the converse is not always true.
%Underlattices and sublattices are complete if they remain closed under
% infinite  suprema and infima.

%-----------------------------------------
%\subsubsection{Lattice Examples} \label{sc-latexamples}

%\emph{Lattice Examples}: \
\begin{Examples}\ \label{ex-lattices} {\rm
(a)~Any \emph{chain} is an infinitely distributive lattice.
% under the chain order,
%because the supremum and infimum of any pair of elements
%exist and equal their maximum and minimum, respectively.
Thus, the  chain $(\REAL, \leq)$ of real numbers equipped with the natural order $\leq$ is a lattice, but not complete.
 The set of extended real numbers
$
\EREAL = \REAL \cup \{ -\infty , +\infty\}
$ is a complete lattice.
%since all its subsets have a supremum and infimum.
%For example,  $\sup \REAL = +\infty$ and $\inf \REAL =-\infty$.
% $\EREAL$, as well as any other complete chain,
% is completely distributive,
% which is an even stronger property than infinite distributivity \cite{Birk67}.
%The set $\EINT =\INT\cup \{ -\infty , +\infty\}$ of extended integers
%is a complete sublattice of $\EREAL$.

(b)~The \emph{power set} $\Pp (E)=\{ X:X\sbs E\}$ of an arbitrary set $E$
equipped with the partial order of set inclusion  is an infinitely distributive lattice under
the supremum and infimum induced by set inclusion which are the
set union and intersection, respectively.
%Any collection of subsets of $E$ that is closed under finite unions and intersections
%is a sublattice of the power set $\Pp (E)$.
%%
%If $E$ is the Euclidean space $\REAL^m$, the collection of its closed subsets
% is  a complete underlattice
%but also an incomplete sublattice of $\Pp (\REAL^m)$.

%
%-----------------  Boolean Lattices  ------------
%
%(c)~\textit{Boolean Lattices}:
(c)~In a lattice $\Ll$ with universal bounds $\ltle$ and $\ltge$, an element
$X\in \Ll$ is said to have a \textit{complement} $X^\lcompl \in \Ll$ if
$X\vee X^\lcompl =\ltge$ and $X\wedge X^\lcompl =\ltle$.
%For example, if $\Ll$ is a power set $\Pp(E)$, then
% $X^\lcompl$ is the  set complement $E\setminus X$.
If all the elements of $\Ll$ % a general lattice $\Ll$
have complements, then $\Ll$ is called complemented.
%A \textit{Boolean lattice}  is defined as any
%lattice that is  complemented and distributive.
Any complemented and distributive lattice $\Bb$ is called
a \textit{Boolean lattice}.

%---------------  Function Lattices  ------------------------
%
(d)~Let $\Ll^\flatdom=\fun (\flatdom,\Ll)$ denote the set of all functions $f:\flatdom \rightarrow \Ll$.
If $\pord$ is the partial ordering of $\Ll$,
we can equip the function space $\Ll^\flatdom$  with the pointwise partial ordering
$f\pord g$, which means  $f(x)\pord g(x)\; \forall  x\in \flatdom$,
 the  pointwise supremum $(\bigvee _if_i)(x)= \bigvee _{i}f_i(x)$,
and pointwise infimum
$( \bigwedge _{i} f_i)(x)= \bigwedge _{i}f_i(x)$.
% $x\in \flatdom$.
Then, $(\Ll^\flatdom,\vee,\wedge)$ becomes a  \textit{function lattice},
which retains possible properties of $\Ll$ of being
complete, or (infinitely) distributive, or Boolean.
}  \end{Examples}

%%-------------------------------------------------------
%\subsubsection{Lattice-Ordered Monoids}

%++++++++++++++++++++++++++++++++++++++++++++++++++++++++++++++++++
\subsection{Operators on Lattices}

%       Operator Lattices
%
Let  $\Oo (\Ll )$ be the set of all  \textit{operators}
on a complete  lattice $\Ll$,   i.e., mappings from $\Ll$ to itself.
Given two such operators $\psi$ and $\phi$ we can define  a  partial
ordering  between them ($\phi \pord \psi$),
their  supremum ($\psi \vee \phi$) and
infimum ($\psi \wedge \phi$) in a pointwise way,
as done in Example~\ref{ex-lattices}(d). % Example~\ref{sc-latexamples}(d).
This makes $\Oo (\Ll )$  a complete function lattice
which inherits many of the possible properties of $\Ll$.
Further, we define the composition of two operators as an operator product:
$\psi \phi(X) \defineql \psi (\phi (X))$;
special cases are the operator powers $\psi ^n = \psi \psi^{n-1}$.
%
%   Elementary Defns of Operators (Id, Ext., Idemp, Invol)
%
Some useful types and properties of lattice operators $\psi$ include the following:
(i)~identity:     $\idop (X) = X$  $\; \forall X\in \Ll$.
%- constant operators:   $\infop (X) = \Ll_\ltle$,
%                $\supop (X) = \Ll_\ltge$ $\; \forall X\in \Ll$ \\
(ii)~extensive:  $\psi \ipord \idop$.
(iii)~antiextensive:  $\psi \pord \idop$.
(iv)~idempotent:  $\psi^2= \psi$.
(v)~involution: $\psi^2 = \idop$.

%+++++++++++++++++++++++++++++++++++++++++++++++++++++++++++++++++++++++++++++
\subsubsection{Monotone  Operators}

Of great interest are the monotone operators, whose collections form sublattices of $\Oo (\Ll)$.
They come in three basic kinds
according to which of the following lattice structures they preserve (or map to its dual):
(i)~partial ordering, (ii)~supremum, (iii)~infimum.

A lattice operator $\psi$ is called \textit{increasing} or \textit{isotone}
if it is order-preserving, i.e. % $\forall X,Y$
$X\pord Y \Longrightarrow \psi(X)\pord \psi (Y)$.
A lattice operator $\psi$ is called \textit{decreasing} or \textit{antitone}
if it is order-inverting, i.e.
$X\pord Y \Longrightarrow \psi(X)\ipord \psi (Y)$.

%-----------------------------------------------------------------------
%\subsubsection{Increasing Operators}

Examples of increasing operators are the lattice homomorphisms
which preserve suprema and infima over finite collections.
If a lattice homomorphism  is also a bijection,
then it becomes an automorphism.
%if it is also a bijection,
%then it is called a \textit{lattice automorphism}.
A bijection $\phi$ is an automorphism
if both $\phi$ and its inverse $\phi^{-1}$ are increasing.

Four  types of  increasing operators, fundamental for
 unifying  systems on lattices,  are the following:
\[
\begin{array}{rcl}
\mbox{\dlop\ is \textit{dilation}} & \mathrm{iff} &
\dlop (\bigvee _{i\in J} X_i ) =\bigvee _{i\in J} \dlop (X_i)
\\
\mbox{\erop\ is \textit{erosion}} & \mathrm{iff} &
\erop (\bigwedge _{i\in J} X_i ) =\bigwedge _{i\in J} \erop (X_i)
\\
\mbox{\opop\ is \textit{opening}} & \mathrm{iff} &
%\mbox{\opop\ is increasing, idempotent \& antiextensive}
\mbox{increasing, idempotent, antiextensive}
\\
\mbox{\clop\ is \textit{closing}} & \mathrm{iff} &
%\mbox{\clop\ is increasing, idempotent \& extensive}
\mbox{increasing, idempotent, extensive}
\end{array}
\]
Dilations and erosions require arbitrary (possibly infinite)
 collections $\{ X_i:i\in J\}$ of lattice elements; hence, they  need complete lattices.
The special case of an empty collection equips each dilation and erosion with
the following necessary properties: %$\dlop (\ltle)=\ltle$ and $\erop (\ltge)=\ltge$.
%Note: they can also be proved from the adjunction inequality [Heij94]
\beq
\dlop (\ltle)=\ltle, \quad \erop (\ltge)=\ltge
\label{dilboterotop}
\eeq
The four above types of lattice operators were originally
 defined in \cite{Serr88,Heij94} as generalizations of the
corresponding Minkowski-type morphological  operators
and have been applied in numerous image processing tasks.

%------------------------------------------------------------------
%\subsubsection{Decreasing Operators}

Examples of decreasing operators are the dual homomorphisms,
which interchange suprema with infima.
A lattice dual-automorphism is a bijection $\theta$
that interchanges suprema with infima, %.
or equivalently
%An operator $\eta$ is a lattice dual automorphism
iff it is a bijection
and both $\theta$ and its inverse $\theta^{-1}$ are decreasing.
A \textit{negation} $\nu$
is a dual automorphism that is also involutive;
we may write $X^\ltneg = \nu (X)$ for the negative of a lattice element.
Given an operator $\psi$ in a lattice equipped with a negation,
its corresponding \emph{negative}  (a.k.a. \emph{dual}) operator  is defined as
 $
 \psi^\ltneg (X)\defineql [\psi (X^\ltneg)]^\ltneg
 $.
For example, the most well-known negation on the set lattice $\Pp (\gimdom)$ is the
complementation $\nu (X)=X^c=\gimdom \setminus X$, whereas on
the function lattice $\fun(\gimdom, \EREAL)$ the most well-known negation is $\nu(f)=-f$.

The above definitions allow broad classes of  operators on vector or signal spaces  to be
grouped as parallel or sequential combinations of lattice monotone operators
%dilations, or erosions, or openings, or closings,
%possibly composed with negations,
and their common properties to be studied under the unifying lattice framework.
In this work we shall find them very useful for representing
the state and output responses or for approximating solutions of systems
obeying a supremal or infimal superposition.

%+++++++++++++++++++++++++++++++++++++++++++++++++++++++++
\subsubsection{Order Continuity}
\label{sc-ordercont}

Consider an arbitrary sequence $(X_n)$ of  elements in a complete lattice $\Ll$.
The following two limits can be defined using only sup/inf combinations:
%\begin{eqnarray}
%\limsup X_n & \defineql & \bigwedge _{n\geq 1}\bigvee_{k\geq n} X_k
%\label{ltlimsup}
%\\
%\liminf X_n & \defineql & \bigvee _{n\geq 1}\bigwedge_{k\geq n} X_k
%\label{ltliminf}
%\end{eqnarray}
\beq
\limsup X_n  \defineql  \bigwedge _{n\geq 1}\bigvee_{k\geq n} X_k
\; \; , \; \;
\liminf X_n  \defineql  \bigvee _{n\geq 1}\bigwedge_{k\geq n} X_k
\label{ltlimsupinf}
\eeq
In general, $\liminf X_n \leq \limsup X_n$.
A sequence $(X_n)$  is defined to \textit{order converge} to
a lattice element $X$, written as $X_n\oconverg X$, if $\liminf X_n = \limsup X_n=X$.

%Now we also define \emph{order continuity}:
%An operator $\psi: \Ll \rightarrow \Mm$ between two complete lattices
An operator $\psi$ on $\Ll$
 is called $\downarrow$-\emph{continuous} if
$(X_n)\oconverg X$ in $\Ll$ implies that $\limsup \psi (X_n) \leq \psi (X)$.
Dually,  $\psi$ is called $\uparrow$-\emph{continuous} if
$(X_n)\oconverg X$ implies that $\liminf \psi (X_n) \geq \psi (X)$.
Finally, $\psi$ is called % $\updownarrow$-\emph{continuous} or
\emph{order continuous} if it
is both $\downarrow$-continuous and $\uparrow$-continuous.
On a chain, e.g. $(\EREAL,\leq)$, the concepts of order convergence and order continuity
coincide with their topological counterparts.

There is a stronger form of  order convergence  applicable to
monotone sequences \cite{Heij94}.
We write $X_n\downarrow X$ to mean a \emph{monotonic convergence} where
$(X_n)$ is a decreasing sequence ($X_{n+1}\leq X_n$) and $X=\bigwedge_nX_n$.
Dually, we write $X_n\uparrow X$ to mean that
$(X_n)$ is an increasing sequence ($X_{n+1}\geq X_n$) and $X=\bigvee_nX_n$.
This monotonic convergence allows to easily examine the order continuity
of increasing operators.
Specifically,
an increasing operator $\psi$ on  a complete lattice $\Ll$
is $\downarrow$-continuous iff $X_n\downarrow X$ implies that
$\psi (X_n)\downarrow \psi(X)$ for any sequence $(X_n)$. Dually,
 $\psi$ is $\uparrow$-continuous iff
$X_n\uparrow X$ implies that $\psi (X_n)\uparrow \psi(X)$.
This result implies that,
since dilations (resp. erosions)
 distribute over arbitrary suprema (resp. infima),
dilations are $\uparrow$-continuous, whereas
erosions are $\downarrow$-continuous.
%
%The distributivity over sup or inf, viewed as some type of order continuity,
%is actually stronger than being $\uparrow$-continuous or  $\downarrow$-continuous.
% This makes dilations  l.s.c. and  erosions u.s.c. w.r.t. Scott-topology.
%Actually in \cite{BCOQ01} the distributivity of dilations over suprema
%and erosions over infima is called respectively l.s.c. and u.s.c.
%We may add "lattice l.s.c." and "lattice u.s.c." and use this instead of
% the defns of l.l.s.c. and l.u.s.c. in [Heij94] which are not used elsewhere.
%

%\begin{proposition} {\rm \cite{Heij94}.} \\
%(a)~If $(X_n)$ is a monotone sequence in a complete lattice with
%$X_n\downarrow X$ or $X_n\uparrow X$, then $X_n\oconverg X$.
%\\
%(b)~An increasing operator $\psi: \Ll \rightarrow \Mm$ between two complete lattices
%is $\downarrow$-continuous iff $X_n\downarrow X$ implies that
%$\psi (X_n)\downarrow \psi(X)$ for any sequence $(X_n)$. Dually,
%an increasing operator $\psi$ is $\uparrow$-continuous iff
%$X_n\uparrow X$ implies that $\psi (X_n)\uparrow \psi(X)$.
%\end{proposition}

%+++++++++++++++++++++++++++++++++++++++++++++++++++++++++
\subsubsection{Residuation and Adjunctions}
\label{sc-resid}

An increasing operator $\psi$ on a complete lattice $\Ll$ is called \textbf{residuated}
\cite{Blyt05,BlJa72} if there exists
an increasing operator $\eresid{\psi}$ such that
\beq
\psi \eresid{\psi} \pord \idop \pord \eresid{\psi} \psi
\label{resopcl}
\eeq
$\eresid{\psi}$ is called the \textbf{residual} of $\psi$ and is the closest to being an inverse of $\psi$.
%in the sense that $\hat{X}= \eresid{\psi} (Y)$ is the greatest solution of $\psi (X)\pord Y$ and
%the greatest solution of $\psi (X)= Y$ if a solution exists. Thus,
Specifically, the residuation pair ($\psi, \eresid{\psi})$ can solve inverse problems
of the type $\psi (X)= Y$ either exactly since $\hat{X}= \eresid{\psi} (Y)$ is
 the greatest solution of $\psi (X)= Y$ if a solution exists,
 or approximately since $\hat{X}$ is the greatest \emph{subsolution} in the sense that
$\hat{X}=\bigvee \{ X:\psi (X)\pord Y \}$.
On complete lattices an increasing operator $\psi$ is residuated (resp. a residual $\eresid{\psi}$)
if and only if it is a dilation (resp. erosion).
The residuation theory has been used for solving inverse problems
 in  matrix algebra \cite{Cuni79,BCOQ01,CGQ04} over the max-plus or other idempotent semirings.

Dilations and erosions come in pairs as the following concept reveals.
The pair $(\dlop,\erop)$ of operators
 on a complete lattice $\Ll$ is called an \textbf{adjunction}\footnote{As explained in \cite{Heij94,HeRo90},
the  adjunction is  related to a concept in poset and
lattice theory called `\emph{Galois connection}'.
%It is also related to \emph{residuation theory} \cite{BlJa72}.
In \cite{Heij94,Serr88} an adjunction pair is denoted as $(\erop,\dlop)$,
but  in this paper we prefer to reverse the positions of its two operators,
 so that it agrees with the structure of a residuation
  pair $(\psi, \eresid{\psi})$.
}
on $\Ll$ if
\beq
\dlop (X)\leq Y \Longleftrightarrow X \leq \erop (Y)
\quad
\forall X, Y\in \Ll
\label{adjunction}
\eeq
In any adjunction, (\ref{adjunction}) implies that $\dlop$ is a dilation and $\erop$ is an erosion.
It can be shown that this double inequality is equivalent to the inequality (\ref{resopcl})
satisfied by a residuation pair of increasing operators
if we identify the residuated map $\psi$ with $\dlop$ and its residual $\eresid{\psi}$ with $\erop$.
To view $(\dlop,\erop)$ as an adjunction instead of a residuation pair has
the advantage of the additional  geometrical intuition and visualization afforded
by the dilation and erosion operators,
which are well-known in image analysis and can be interpreted as augmentation and shrinkage
respectively of input sets or of hypographs of functions.
%
% -- To Prove ---------------------------------
% The adjunction inequality is stronger condition bcs it does not need the assumption
% that the two operators are increasing as required by the residuation inequality.
%

In any adjunction $(\dlop,\erop)$, $\erop$ is called the
\emph{adjoint erosion}  of $\dlop$, whereas $\dlop$ is
the \emph{adjoint dilation}  of $\erop$.
There is a one-to-one correspondence between the two operators
of an adjunction pair, since,
given a dilation $\dlop$, there is a unique erosion
%(a.k.a. \emph{right adjoint})
\beq
\erop (Y)
%\ltadj{\dlop} (Y)
= \bigvee \{ X\in \Ll : \dlop (X)\pord Y\}
\label{adjero}
\eeq
such that $(\dlop,\erop)$ is adjunction.
%such that $(\ltadj{\dlop},\dlop)$ is adjunction.
Conversely, given an erosion $\erop$, there is a unique dilation
%(a.k.a. \emph{left adjoint})
\beq
\dlop (X)
%\ltadj{\erop} (X)
= \bigwedge \{ Y\in \Ll: \erop (Y) \ipord X \}
\label{adjdil}
\eeq
such that $(\dlop,\erop)$ is adjunction.
Adjunctions  create operator duality pairs  that are different than negation
in the sense that one operator is the closest to being the inverse of the other,
either from below or above.
%%
%% ---- Sub-Inverse, Super-Inverse Opers
%%
%\begin{Lemma} \label{lm-adjinv}
%Given an adjunction $(\dlop,\erop)$, the  adjoint of
%%an operator
%the operator $\dlop$ (resp. $\erop$)
%is its \emph{greatest sub-inverse}\footnote{Given an increasing operator $\psi$,
%we define an operator $\phi$ to be a \emph{sub-inverse} (resp. \emph{super-inverse}) of $\psi$
%if $\psi \phi \pord \idop$ (resp. $\psi \phi \ipord \idop$).}
%(resp.  \emph{least super-inverse}); as a corollary,
%\beq
%%\dlop \ltadj{\dlop} \pord \idop \pord \erop \ltadj{\erop}
%\dlop \erop \pord \idop \pord \erop \dlop
%\label{adjopcl}
%\eeq
%\end{Lemma}
%\emph{Proof}: It can be deduced from (\ref{adjero}) and (\ref{adjdil}).
%\hfill Q.E.D.

%+++++++++++++++++++++++++++++++++++++++++++++++++++++++++
\subsubsection{Projections on Lattices}

%---- \subsubsection{Openings, Closings, Order Projections} ------------

A large variety of useful lattice operators share two properties:
 \emph{increasing and idempotent}.
Such operators were called \emph{morphological filters} in \cite{Heij94,Serr88}.
We shall call them
\textit{lattice projections} of the order type, since they preserve the lattice ordering and are idempotent
 in analogy with the linear projections that preserve the algebraic structure
of linear spaces and are idempotent.
Two well-studied special cases of lattice projections are the
 openings and closings, each of which has an additional property.
Specifically, openings are lattice projections that are anti-extensive,
whereas closings are extensive  projections.
Combinations of such generalized filters have proven to be very
useful for signal denoising, image enhancement, simplification,
segmentation,  and object detection.
From the composition
of the erosion and dilation of any adjunction $(\dlop,\erop)$ we can generate a projection
$\opop=\dlop\erop$  that is also an opening since $\opop(X)\leq X$ and $\opop^2=\opop$.
To prove this note that, by (\ref{adjunction}),
% $\dlop \erop \pord \idop \pord \erop \dlop$
\beq
\dlop \erop \pord \idop \pord \erop \dlop
\label{adjopcl}
\eeq
 which implies that $\dlop\erop\dlop\erop=\dlop\erop$.
Dually, any adjunction can also generate a closing projection
$\clop = \erop\dlop$, which always satisfies $\clop (X)\geq X$ and $\clop^2=\clop$.
%The opening and closing generated by an adjunction satisfy  (\ref{adjopcl}).
There are also other types of lattice projections that are studied in \cite{CGQ04}.

%+++++++++++++++++++++++++++++++++++++++++++++++++++++++++++
\subsection{Lattice-Ordered Monoids and Clodum}
\label{sc-lomclodum}

%A lattice $\Mm$ is often endowed with additional structure
%of the group type \cite{Birk67}. Namely, it may have an additional binary operation,
%called symbolically the `multiplication' $\mgop$, under which
%$(\Mm,\mgop)$ is  a  semigroup or monoid
%% (if it is a semigroup with an identity element)
%or group.

A lattice $(\Mm,\vee,\wedge)$ is often endowed with a third binary operation,
called symbolically the `multiplication' $\mgop$, under which
$(\Mm,\mgop)$ is  a group or  monoid or just semigroup \cite{Birk67}.

Consider now an algebra $(\Mm,\vee,\wedge,\mgop,\dmgop)$ with four binary operations,
which we call a \emph{lattice-ordered double monoid},
where $(\Mm,\vee,\wedge)$ is a lattice,
$(\Mm,\mgop)$ is a monoid whose `multiplication' $\mgop$ distributes over $\vee$,
and $(\Mm,\dmgop)$ is a  monoid whose `multiplication' $\dmgop$ distributes over $\wedge$.
%\beq
%\begin{array}{ccc} A\mgop (X\vee Y) &=&  (A\mgop X)\vee (A\mgop Y) \\
%A\dmgop (X\wedge Y) &=& (A\dmgop X )\wedge (A\dmgop Y) \end{array}
%\label{lodmdistr}
%\eeq
%We call the above structure  a \emph{lattice-ordered double monoid}.
These distributivities   imply that both $\mgop$ and $\dmgop$ are increasing.
To the above definitions we add the word \emph{complete}
if $\Mm$  is a complete lattice and the distributivities involved are infinite.
%Assuming also that the monoids are commutative,
We call the resulting algebra % $(\Mm ,\vee,\wedge,\mgop,\dmgop)$
a  \emph{complete lattice-ordered double monoid}, in short \textit{clodum} \cite{Mara05a,Mara13}.

Previous works on minimax or max-plus algebra and their applications
have  used alternative names\footnote{Minimax algebra  \cite{Cuni79} % (Cuninghame-Green, 1979)
has been based on bands (idempotent semigroups) and belts (idempotent pre-semirings),
whereas max-plus algebra and its application to DES \cite{BCOQ01,Butk10,CMQV89,GoMi08}
is based on dioids (canonically ordered semirings).
In \cite{Cuni79},
a semilattice is called a \textit{commutative band} and a lattice is called \emph{band with duality}.
Further,  a \emph{belt} is a semilattice-ordered semigroup, and
a \emph{belt with duality} \cite{Cuni79} is a pair of two idempotent pre-dioids \cite{GoMi08}
whose `additions' are dual and form a lattice.
%
% A \emph{belt with duality} is a lattice-ordered double semigroup. Note (PM):
% nobody except from PM has called them as such bcs everybody
%who wrote about "lattice-ordered" structures talked about one only structure,
%e.g. lattice-ordered group or monoid, etc.
%
%Adding to a belt $(\Bb , \vee , \mgop )$ an identity element
%for the operation $\mgop$ and an element  $\epsilon$ that is both the
% least element  and also an absorbing null, i.e.
%  $a\vee \epsilon=a$ and  $a\mgop \epsilon =\epsilon$, $\forall a\in \Bb$,
%creates a \emph{dioid}  \cite{BCOQ01,CMQV89,GoMi08} which is an idempotent semiring.
%More general (including non-idempotent) dioids are studied in \cite{GoMi08}.
Adding to a belt $(\Bb , \vee , \mgop )$ identity elements for $\mgop$ and $\vee$,
the latter of which is also an absorbing null for $\mgop$,
creates an idempotent \emph{dioid}  \cite{BCOQ01,CMQV89,GoMi08}. % which is an idempotent semiring.
More general (including non-idempotent) dioids are studied in \cite{GoMi08}.
%Finally, division belts \cite{Cuni79} are lattice-ordered groups.
Finally,  belts that are groups under the `multiplication' $\mgop$  and
as lattices have global bounds are called \emph{blogs }(bounded lattice-ordered groups) in \cite{Cuni79}.
}
for  algebraic structures similar to the above definitions which emphasize semigroups and semirings
instead of lattices.
If $\mgop=\dmgop$, we have a self-dual `multiplication'.
 This always happens
if $(\Mm ,\mgop)$ is a group, i.e. a monoid where each element has an inverse;
in this case we obtain a \emph{lattice-ordered group},
%the inequalities (\ref{lodmineq}) become equalities,
and the group `multiplication' $X\mapsto A\mgop X$ is a lattice automorphism.

%+++++++++++++++++++++++++++++++++++++++++++++++++++++++++++
%\subsection{Clodum}

We give a precise definition of a general clodum and some examples since this will be one of the
fundamental algebraic structures to build the nonlinear spaces in our work.
An algebraic structure $(\Kk, \vee ,\wedge, \mgop,\dmgop )$ is called a \textbf{clodum} if: \\
(C1)~$(\Kk,\vee,\wedge)$ is a complete distributive lattice. \\
(C2)~$(\Kk,\mgop)$ is a monoid whose operation $\mgop$ is a dilation. \\
%distributes over any $\vee$ and has as null the bottom element of $\Kk$. \\
(C3)~$(\Kk,\dmgop)$ is a monoid whose operation $\dmgop$ is an erosion.
%distributes over any $\wedge$ and has as null the top element of $\Kk$.

Remarks:
(i)~As a lattice, $\Kk$ is not necessarily infinitely distributive, although in this paper
all our examples will be such.  \\
(ii)~The clodum `multiplications' $\mgop$ and $\dmgop$ do not have to be commutative. \\
(iii)~The least (greatest) element $\ltle$ ($\ltge$) of $\Kk$ is both the identity element
for $\vee$ ($\wedge$) and an absorbing null for $\mgop$ ($\dmgop$) due to  (\ref{dilboterotop}).

%If the `multiplications' $\mgop$ and $\dmgop$ are the same operation
%over $\vsgroup=K\setminus \{ \latle,\latge\}$ and $(\vsgroup,\mgop)$ is a group,
If  $\mgop=\dmgop$
over $G=\Kk\setminus \{ \ltle,\ltge\}$ where $(G,\mgop)$ is a group and $(G,\vee,\wedge)$
a conditionally complete lattice,
then the clodum $\Kk$ becomes a richer structure
which we call a \emph{complete lattice-ordered group}, in short \textbf{clog}.
 By extending properties of lattice-ordered groups \cite{Birk67} to clogs,
 we can show that in any clog the distributivity between $\vee$ and $\wedge$ is of
the infinite type and the `multiplications' $\mgop$ and $\dmgop$ are commutative.
Thus, a clog
has a richer structure than a \emph{blog} (bounded lattice-ordered group) as defined in \cite{Cuni79},
because a clog is a complete and commutative blog.

\begin{Examples}\ \label{ex-cloda} {\rm
%\emph{Clodum Examples:}
\ (a)~Our scalar arithmetic in this paper will use a numeric commutative clodum. Two such examples follow:\\
(a1)~Max-plus clog\footnote{In every clodum and clog we have
a pair of dual `additions' and a pair of dual `multiplications'.
However, for brevity, we assign them shorter names that contain only one `addition' (max)
and one `multiplication'; e.g. `max-plus clog'.}: \ $(\EREAL,\vee,\wedge,+,+')$,
where $\vee/\wedge$ denote the standard sup/inf on $\EREAL$,
$+$ is the standard addition on the set $\EREAL$ of extended reals playing the role of a `multiplication' $\mgop$
with $+'$ being the `dual multiplication' $\dmgop$;
 the operations $+$ and $+'$ are identical for finite reals, but $a+(-\infty)=-\infty$
and $a+'(+\infty)=+\infty$ for all $a\in \EREAL$. \\
%(a2)~Max/min-product clog: $([0,+\infty],\vee,\wedge,\times,\times')$. \\
(a2)~Max-min clodum: \ $([0,1],\vee,\wedge,\min,\max)$, where $\mgop=\min$ and $\dmgop=\max$.
%(a3)~Max/min-fuzzy norm clodum: \ $([0,1],\vee,\wedge,\inorm,\dinorm)$ where $\inorm$ and $\dinorm$
%denote intersection and union fuzzy norms respectively, discussed in Sec.~\ref{sc-fuzzy}.
\\
(b)~Matrix max-sum clodum: $(\EREAL^{n\times n},\vee,\wedge,\mxsmp,\mnsmp)$
where $\EREAL^{n\times n}$ is the set of $n\times n$ matrices with entries from $\EREAL$,
$\vee$ and $\wedge$ denote here element-wise matrix supremum and infimum,
and $\mxsmp,\mnsmp$ denote max-sum and min-sum matrix `multiplications':
\bea
\mtr{C} = \mtr{A} \mxsmp \mtr{B}=[c_{ij}]  & , &
 \; c_{ij} = {\dsty \bigvee _{k=1}^n} a_{ik}+b_{kj}
\label{mxsmpr} \\
\mtr{C} = \mtr{A} \mnsmp \mtr{B}=[c_{ij}]  & , &
 \; c_{ij} = {\dsty \bigwedge _{k=1}^n} a_{ik}+'b_{kj}
\label{mnsmpr}
\eea
%where $+$ and $+'$ are identical for finite reals, but $a+(-\infty)=-\infty$
%and $a+'(+\infty)=+\infty$ for all $a\in \EREAL$.
This is a clodum with non-commutative  `multiplications'.
} \end{Examples}

%################################################################
\section{Representations of Vector and Signal Operations on Weighted Lattices}
\label{sc-repsvsowl}

\subsection{Algebraic Structures on the Scalars}
\label{sc-algss}

%In this and the following subsection we outline the
%algebraic structures necessary to exist in the set of scalars
%as well as the induced generalized vector and matrix operations.

We assume that all elements of the vectors, matrices, or signals
involved in the description of the  systems examined herein
take their values from a set $\vset$ of \emph{scalars}, which in general
will be a subset of  $\EREAL = \REAL \cup \{ -\infty ,\infty\}$
with the natural ordering $\leq$ of extended real numbers.
%will be a subset of the set $\EREAL = \REAL \cup \{ -\infty ,\infty\}$ of extended real numbers.
% $\vset$ is  a chain under the standard real number ordering $\geq$.
We assume that the chain $(\vset, \leq)$ is universally bounded, i.e.,
contains its least $\vsetle \defineq \bigwedge \vset$
  and greatest element $\vsetge \defineq \bigvee \vset$. %:
%$
%\vsetle \leq a\leq \vsetge \; , \quad \forall a\in \vset.
%$
For the weighted lattice model we need to equip  $\vset$
with  four  binary operations:

(A).\ The standard maximum or supremum $\vee$ on $\EREAL$,
which plays the role of a generalized \textit{`addition'}.
% and makes $\vset$ a complete sup-semilattice.

(A$'$).\
The standard minimum or infimum $\wedge$ on $\EREAL$,
which plays the role of a generalized \textit{`dual addition'}.
%which makes $\vset$ a complete  inf-semilattice.
%It plays the role of a generalized \textit{`dual addition'}
% and is related to the `addition' via the
%absorption law L5 of Table~\ref{tb-gscalar}.

(M).\ A commutative generalized \textit{`multiplication'} $\mgop$
under which: (i) \vset\ is a monoid with (`unit') identity element $\mgid$ and
 (`zero') null element $\vsetle$, i.e.,
\beq
a\mgop \mgid = a, \quad
a\mgop \vsetle = \vsetle, \quad  \forall a\in \vset ,
\label{multidnl}
\eeq
and (ii) $\mgop$
is a scalar dilation, i.e., distributes over any supremum:
\beq
a\mgop ( \bigvee _{i}x_i ) = \bigvee _{i} a\mgop x_i
%a\mgop \left( \bigvee _{i\in J}x_i\right) = \bigvee _{i\in J} (a\mgop x_i)
\eeq
%for any (possibly infinite) index set $J$.

(M$'$).\ A commutative \textit{`dual\,\footnote{\rm It is simply a matter a convention that we selected to call
$\wedge$ and $\dmgop$ as `\emph{dual} addition and multiplication' (instead of $\vee$ and $\mgop$).}
multiplication'} $\dmgop$ under which:
(i) $\vset$ is a monoid  with identity $\dmgid$ and null element $\vsetge$, i.e.,
\beq
a\dmgop \dmgid = a, \quad
a\dmgop \vsetge = \vsetge , \quad \forall a\in \vset ,
\label{dmultidnl}
\eeq
and (ii) $\dmgop$
is a scalar erosion, i.e.,  distributes over any infimum:
\beq
a\dmgop ( \bigwedge _{i}x_i ) = \bigwedge _{i} a\dmgop x_i
\label{dmultdistr}
\eeq

%We group the above requirements into the
%following sets of conditions:\\
%(C1).\ $(\vset ,\vee,\wedge)$
%is a complete infinitely-distributive lattice.
%\\
%(C2).\ $(\vset,\mgop)$ is a commutative  monoid,
%and $\mgop$ is a dilation.
%\\
%(C3).\ $(\vset,\dmgop)$ is a commutative  monoid,
%and $\dmgop$ is an erosion.

Under the above assumptions
 $(\vset , \vee ,\wedge, \mgop,\dmgop )$ becomes a  \textbf{scalar clodum}.
 Note that, in addition to the minimal requirements of a general clodum in Sec.~\ref{sc-lomclodum},
 we assume \emph{commutative}  operations $\mgop,\dmgop$. Further, the rich structure of $\EREAL$
 endows the set  $\vset$ to be infinitely distributive as a lattice.
% \footnote{In this paper
% we assume \emph{commutative}  operations $\mgop,\dmgop$ and that $\vset$ is infinitely distributive as a lattice,
% but a general clodum does not have to be infinitely distributive and its operations $\mgop,\dmgop$
%  do not have to be commutative.}.
%commutative \emph{complete lattice-ordered double  monoid (CLODUM)}.
This will be the most general and minimally required
%\footnote{If we  deal only with max  systems we can weaken
%this requirement by dropping the dual `multiplication',
% in which case $(\vset , \vee ,\wedge, \mgop)$
%needs only to be a complete lattice-ordered monoid. But to have both
%max- and min-type  systems we need the structure of a CLODUM.}
algebraic structure we consider for the set of scalars.
%We avoid degenerate cases by henceforth assuming that
%each `addition' is different from its corresponding `multiplication',
%i.e., $\vee \not = \mgop$ and $\wedge \not = \dmgop$.
We avoid degenerate cases by  assuming that
 $\vee \not = \mgop$ and $\wedge \not = \dmgop$.
However,  $\mgop$ may be the same as $\dmgop$, in which case
we have a self-dual `multiplication'.

A clodum $\vset$ is called \emph{self-conjugate}
if  it has a lattice negation (i.e. involutive dual automorphism)
% $a\mapsto \glconj{a}$
that  maps each element $a$ to its  \emph{conjugate}  $\glconj{a}$ such that
%\beq
%\begin{array}{rcl}
%\glconj{\bigvee_i a_i} & = & \bigwedge_i \glconj{a_i} \\
%\glconj{\bigwedge_i b_i} & = & \bigvee_i \glconj{b_i} \\
%\glconj{a\mgop b} & = & \glconj{a}\dmgop \glconj{b}
%\end{array}
%\label{clodconj}
%\eeq
\beq
%\glconj{\left( \bigvee_i a_i \right)}  =  \bigwedge_i \glconj{a_i} \; \; , \; \;
%\glconj{\left( \bigwedge_i b_i\right)} =  \bigvee_i \glconj{b_i}\; \; , \; \;
\glconj{( \bigvee_i a_i )}  =  \bigwedge_i \glconj{a_i} \; \; , \; \;
\glconj{( \bigwedge_i b_i )} =  \bigvee_i \glconj{b_i}\; \; , \; \;
\glconj{(a\mgop b)}  =  \glconj{a}\dmgop \glconj{b}
\label{clodconj}
\eeq
%The first two above properties are generalization of De Morgan's laws in Boolean algebras.
% As for the `multiplication' operations, we assume that the negation also
We assume that the suprema and infima in (\ref{clodconj}) may be over any (possibly infinite) collections.

The  set of scalars can be partitioned as
 $\vset = \vsgroup \cup \{ \vsetle , \vsetge \}$;
the members of $\vsgroup$ are called the \emph{finite scalars},
borrowing terminology from the case when $\vset =\EREAL$.
This is useful for cases where $(\vsgroup, \mgop)$ is a commutative \emph{group}.
Then, for each $a\in \vsgroup$ there exists its `multiplicative inverse'
$a^{-1}$ such that $a\mgop a^{-1}=\mgid $.
%
%In some cases the scalars set may be partitioned as
% $\vset = \vsgroup \cup \{ \vsetle , \vsetge \}$
%where $(\vsgroup, \mgop)$ is a commutative group.
%Then, for each $a\in \vsgroup$ there exists its `multiplicative inverse'
%$a^{-1}$ such that $a\mgop a^{-1}=\mgid $.
%The members of $\vsgroup$ are called the \emph{finite scalars}.
%Further,  $(\vsgroup , \vee ,\wedge, \mgop,\mgop)$
%is a lattice-ordered group with self-dual `multiplication'.
Further, the `multiplication' $\mgop$ and its self-dual
$\dmgop$ (which coincide over $\vsgroup$) can be extended over the whole
$\vset$ by adding the rules in (\ref{multidnl}) and (\ref{dmultidnl})
involving the null elements.
As defined in Sec.~\ref{sc-lomclodum},
 the resulting richer structure  is a
 %\emph{complete lattice-ordered group}, in short
 \textbf{clog}.
 %\footnote{A clog
%has a richer structure than a blog (bounded lattice-ordered group) defined in \cite{Cuni79},
%because a clog is a complete and commutative blog.}
%
%(In this paper a scalar clodum or a clog will always be a chain.)
Whenever $\vset$ is a clog,
it becomes self-conjugate  by setting
\beq
\glconj{a} =\left\{ \begin{array}{lll}
a^{-1} & \mathrm{if} & \vsetle \spord a \spord \vsetge \\
\vsetge & \mathrm{if} & a=\vsetle \\
\vsetle & \mathrm{if} & a=\vsetge
\end{array} \right.
\eeq
%Table~\ref{tb-blog} summarizes the results of all types of scalar arithmetic operations in a blog.
%%(In this table we have dropped the subscript `$\vset$' from the extreme elements.)
%We see that
%
%In a clog the $\mgop$ and $\dmgop$ coincide in all cases with only one exception: the combination
%of the least and greatest elements. Thus, henceforth when a clodum $\vset$ is a clog we can
%denote the algebra as $(\vset, \vee,\wedge,\mgop)$ using only one `multiplication' operation
%and the case $\vsetle \mgop \vsetge$ will have value $\vsetle$ ($\vsetge$)
%if it is combined with other terms via  a supremum (infimum).

Next we further elaborate on three main examples used in this paper for a scalar clodum.

\begin{Examples}\ \label{ex-numcloda} {\rm
%\emph{Clodum Examples:}
(a)~\emph{Max-plus} clog \ $(\EREAL,\vee,\wedge,+,+')$: This is the archetypal example of a clog.
The identities are $\mgid=\dmgid=0$, the nulls are $\vsetle=-\infty$ and $\vsetge=+\infty$, and the
conjugation mapping is $\glconj{a}=-a$.
% $+$ and $+'$ are identical for finite reals, but $a+(-\infty)=-\infty$
%and $a+'(+\infty)=+\infty$ for all $a\in \EREAL$.
\\
(b)~\emph{Max-times} clog $([0,+\infty],\vee,\wedge,\times,\times')$:
The identities are $\mgid=\dmgid=1$, the nulls are $\vsetle=0$ and $\vsetge=+\infty$, and the
conjugation mapping is $\glconj{a}=1/a$.
%The scalar multiplications $\times$ and $\times'$ coincide over $(0,+\infty)$,
%but $a\times 0=0$ and $a\times '(+\infty) =+\infty$ for all $a\in [0,+\infty]$.
%Henceforth, we shall use the same symbol $\times$ for both scalar operations.
\\
(c)~\emph{Max-min} clodum \ $([0,1],\vee,\wedge,\min,\max)$:
As `multiplications' we have $\mgop=\min$ and $\dmgop=\max$.
The identities and nulls are $\dmgid=\vsetle=0$, $\mgid=\vsetge=1$. A possible
conjugation mapping is $\glconj{a}=1-a$.
Additional cloda that are not clogs are discussed in Sec.~\ref{sc-fuzzy} using
more general fuzzy intersections and unions.
} \end{Examples}

\begin{table}[!h]
\caption{Scalar Arithmetic in a CLOG}
\centerline{
\begin{tabular}{|c|c|c|c|c|c|c|}
\hline
$a\in \vset$ & $b\in \vset$ & $\vee $ & $\wedge $ & $\mgop $ & $\dmgop $  \\
\hline \hline
$\vsetle$ & $\vsetle$ & $\vsetle$ & $\vsetle$ & $\vsetle$ & $\vsetle$ \\ \hline
$\vsetle$ & $y\in \vsgroup$ & $y$ & $\vsetle$ & $\vsetle$ & $\vsetle$ \\ \hline
$\vsetle$ & $\vsetge$ & $\vsetge$ & $\vsetle$ & $\vsetle$ & $\vsetge$ \\ \hline
% $a\in \vsgroup$ & $\vsetle$ & $a$ & $\vsetle$ & $\vsetle$ & $\vsetle$ \\ \hline
$x\in \vsgroup$ & $y\in \vsgroup$ & $x\vee y$ & $x\wedge y$ & $x\mgop y$ & $x\dmgop y$ \\ \hline
% $a\in \vsgroup$ & $\vsetge$ & $\vsetge$ & $a$ & $\vsetge$ & $\vsetge$ \\ \hline
$\vsetge$ & $\vsetle$ & $\vsetge$ & $\vsetle$ & $\vsetle$ & $\vsetge$ \\ \hline
$\vsetge$ & $y\in \vsgroup$ & $\vsetge$ & $y$ & $\vsetge$ & $\vsetge$ \\ \hline
$\vsetge$ & $\vsetge$ & $\vsetge$ & $\vsetge$ & $\vsetge$ & $\vsetge$ \\ \hline
\end{tabular}
}
\label{tb-clog}
\end{table}

Table~\ref{tb-clog} summarizes the results of all scalar binary operations in a clog.
We see that
%Note: In a clog
in a clog the $\mgop$ and $\dmgop$ coincide in all cases with only one exception: the combination
of the least and greatest elements. Henceforth when a clodum $\vset$ is a clog we can
denote the algebra as $(\vset, \vee,\wedge,\mgop)$ using only one `multiplication' operation
and the case $\vsetle \mgop \vsetge$ will have value $\vsetle$ (resp. $\vsetge$)
if it is combined with other terms via  a supremum (resp. infimum).

\subsection{Complete Weighted Lattices}
\label{sc-cwl}

Consider a nonempty collection $\Ww$ of mathematical objects, which will be our
space; examples of such objects include the vectors in $\EREAL^n$ or signals in
$\fun(E,\EREAL)$. Thus, we shall symbolically refer to the space elements as
`vectors/signals', although they may be arbitrary objects. Also, consider a
clodum $(\vset, \vee, \wedge, \mgop, \dmgop )$ of `scalars'.\footnote{In this paper
as `scalars' we use numbers from $\EREAL$ or its subsets,
 but the general definition of a weighted lattice allows for
an arbitrary clodum as the set of `scalars'.} We define two internal
operations among vectors/signals $X,Y$ in $\Ww$:
their supremum $X\vee Y:\Ww^2\rightarrow \Ww$ and infimum $X\wedge Y:\Ww^2\rightarrow \Ww$,
which we denote using the same supremum symbol ($\vee$) and infimum symbol ($\wedge$) as
in the clodum, hoping that the differences will be clear to the reader from the
context. Further, we define two external operations among any vector/signal $X$ in $\Ww$
and any scalar $c$ in $\vset$:
a `scalar multiplication' $c\mgop X:(\vset,\Ww)\rightarrow \Ww$ and
a `scalar dual multiplication' $c\dmgop X:(\vset,\Ww)\rightarrow \Ww$,
again by using the same symbols as in the clodum.
Now, we define $\Ww$ to be a \textbf{weighted lattice} space
 over the clodum $\vset$ if for all $X,Y,Z\in \Ww$ and  $a,b\in \vset$ all the axioms of
Table~\ref{tb-wlat} hold. Note\footnote{If in our definition of a weighted lattice,
one focuses only on one vector `addition', say the vector supremum, and its corresponding scalar `multiplication',
then the weaker algebraic structure becomes an idempotent semimodule over an idempotent semiring.
This has been studied in \cite{CGQ04,GoMi08,LMS01}.}
 that: (a)~Under axioms L1-L9 and their duals L1$'$-L9$'$,
$\Ww$ is a distributive lattice with a least element ($\flatle$) and a greatest element ($\flatge$).
(b)~These  axioms
bear a striking similarity with those of a linear space.
One difference is that the vector/signal addition ($+$) of linear spaces
is now replaced by two dual superpositions,
the lattice supremum ($\vee$) and infimum ($\wedge$);
further, the scalar multiplication ($\times$) of linear spaces is now replaced
by two  operations $\mgop$ and $\dmgop$ that are dual to each other.
Only one major property of the linear
spaces is missing from the  weighted lattices: the existence of `additive inverses';
i.e., the supremum and infimum operations do not have inverses.
%(c)~Not all axioms of Table~\ref{tb-wlat} are logically independent.
%For example,  the properties L6 and L8 are
%not independent axioms because they result from others.
%
%\noindent
%(WL1)~$(\Ww, \vee, \wedge)$ is
%a distributive lattice with a least element ($O$) and a greatest element ($I$). \\
%(WL2)~The two multiplications $\mgop$ and $\dmgop$ between scalars and vectors are associative.
%\\
%(WL3)~The two multiplications $\mgop$ and $\dmgop$ between scalars and vectors distribute over
%(scalar and vector) suprema and infima.
%

\begin{table*}[!h]
\caption{Axioms of Weighted Lattices}
%\resizebox{0.88\textwidth}{!}{\begin{minipage}{\textwidth}
%\begin{tabular}{|l|l|c|} \hline
\begin{tabularx}{\textwidth}{|X|X|X|} \hline
Sup-Semilattice & Inf-Semilattice & Description \\ \hline \hline
L1. $\; X\vee Y\in \Ww$ & L1$'. \; X\wedge Y\in \Ww$ & Closure of $\vee, \wedge$
 \\ \hline
L2. $\; X\vee X=X$ & L2$'. \; X\wedge X=X$ & Idempotence of $\vee, \wedge$ \\ \hline
L3. $\; X\vee Y=Y\vee X$ & L3$'. \;  X\wedge Y=Y\wedge X$ & Commutativity of $\vee, \wedge$\\ \hline
%L4. $\;  X\vee (Y\vee Z)=(X\vee Y)\vee Z$ & L4$'. \;  X\wedge (Y\wedge Z)
%     =(X\wedge Y)\wedge Z$ & Associativity  of $\vee, \wedge$ \\ \hline
L4. $\;  X\vee (Y\vee Z)=$ & L4$'. \;  X\wedge (Y\wedge Z)=$ & Associativity  of $\vee, \wedge$ \\
    \hspace*{5mm} $(X\vee Y)\vee Z$ & \hspace*{5mm} $(X\wedge Y)\wedge Z$ & \\ \hline
L5. $\; X\vee (X\wedge Y)=X$ & L5$'.\; X\wedge (X\vee Y)=X$ & Absorption between $\vee,\wedge$\\ \hline \hline
%L6. $\;  X\pord Y \Longleftrightarrow Y=X\vee Y$ & L6$'. \;  X\dpord Y
%   \Longleftrightarrow Y=X\wedge Y$ &  Consistency of $\vee,\wedge$ with $\pord$ \\ \hline
L6. $\;  X\pord Y \Longleftrightarrow$  & L6$'. \; X\dpord Y \Longleftrightarrow$ & Consistency of $\vee,\wedge$
\\ %with $\pord$ \\
   \hspace*{5mm} $Y=X\vee Y$ & \hspace*{5mm} $Y=X\wedge Y$  & with partial order $\pord$
   \\ \hline
L7. $\; \flatle \vee X=X$ & L7$'. \; \flatge \wedge X=X$ & Identities of $\vee, \wedge$ \\ \hline \hline
L8. $\; \flatge \vee X=\flatge$ & L8$'. \; \flatle \wedge X=\flatle$ &
      Absorbing Nulls of $\vee, \wedge$ \\ \hline
%L9. $\; X\vee (Y\wedge Z)=(X\vee Y)\wedge (X\vee Z)$ &
%  L9$'. \; X\wedge (Y\vee Z)=(X\wedge Y)\vee (X\wedge Z)$ & Distributivity of $\vee, \wedge$
%\\ \hline \hline
L9. $\; X\vee (Y\wedge Z)=$  & L9$'. \; X\wedge (Y\vee Z)=$ & Distributivity of $\vee, \wedge$ \\
\hspace*{5mm} $(X\vee Y)\wedge (X\vee Z)$ & \hspace*{5mm} $(X\wedge Y)\vee (X\wedge Z)$ & \\ \hline \hline
WL10. $\; a\mgop X\in \Ww$ & WL10$'. \; a\dmgop X\in \Ww$ & Closure of $\mgop, \dmgop$
 \\ \hline
%WL11. $\; a\mgop (b\mgop X)=(a\mgop b)\mgop X$ &
%   WL11$'. \; a\dmgop (b\dmgop X)=(a\dmgop b)\dmgop X$ & Associativity of $\mgop, \dmgop$
% \\ \hline
 WL11. $\; a\mgop (b\mgop X)=$ & WL11$'. \; a\dmgop (b\dmgop X)=$ & Associativity of $\mgop, \dmgop$ \\
   \hspace*{10mm} $(a\mgop b)\mgop X$ & \hspace*{10mm} $(a\dmgop b)\dmgop X$ & \\ \hline
%WL12 . $\; a \mgop (X\vee Y)=a \mgop X\vee a \mgop Y$ &
%  WL14$'. \; a \dmgop (X\wedge Y)=a \dmgop X\wedge a \dmgop Y$ & Distributivity scalar-vector
% \\ \hline
 WL12. $\; a \mgop (X\vee Y)=$ & WL12$'. \; a \dmgop (X\wedge Y)=$ & Distributive  scalar-vector   \\
 \hspace*{10mm} $a \mgop X\vee a \mgop Y$ & \hspace*{10mm} $a \dmgop X\wedge a \dmgop Y$ & mult over vector sup/inf
 \\ \hline
%WL13. $\; (a\vee b) \mgop X=a \mgop X\vee b \mgop X$ &
%  WL15$'. \; (a\wedge b) \dmgop X=a \dmgop X\wedge b \dmgop X$ & Distributivity vector-scalar
% \\ \hline
 WL13. $\; (a\vee b) \mgop X=$ & WL13$'. \; (a\wedge b) \dmgop X=$ & Distributive scalar-vector  \\
 \hspace*{10mm} $a \mgop X\vee b \mgop X$ & \hspace*{10mm} $a \dmgop X\wedge b \dmgop X$ & mult over scalar sup/inf\\ \hline
 WL14. $\; \mgid \mgop X=X$ & WL14$'. \; \dmgid \dmgop X=X$ & Scalar Identities
 \\ \hline
WL15. $\; \vsetle \mgop X=\flatle$ & WL15$'. \; \vsetge \dmgop X=\flatge$ & Scalar Nulls
 \\ \hline
 WL16. $\; a \mgop \flatle=\flatle$ & WL16$'. \; a \dmgop \flatge=\flatge$ & Vector Nulls
 \\ \hline
%\end{tabular}
\end{tabularx}
\label{tb-wlat}
%\end{minipage}}
 \end{table*}

We define the weighted lattice $\Ww$ % over the clodum $\vset$
to be a \textbf{complete weighted lattice (CWL)} space
 if all the following hold:\\
 (i)~\ $\Ww$ is closed under any, possibly infinite, suprema and infima.\\
%(ii)~Its distributivity laws between supremum and infimum are of the infinite type.\\
(ii)~The distributivity laws between the scalar operations  $\mgop$ ($\dmgop$)
and the supremum (infimum)  are of the infinite type.
\\
Note that, a  clodum is by itself a  complete weighted lattice over itself.

%-----------------------------------------------------------------
%\subsubsection{Sup/Inf Span, Independence, Basis, Dimension}

% --- CWL Span -----------------

Consider a  subset $\Aa$  of a complete weighted lattice  $\Ww$ over a clodum $\vset$.
 A space element $F$
is called a \textbf{sup-$\mgop$ combination} of points in $\Aa$ if there exists
an indexed set of space elements $\{ F_i\}$ in $\Aa$ and a corresponding set of  scalars
$\{ a_i\}$ in $\vset$ such that
\beq
F= \bigvee_{i} a_i\mgop F_i,
\eeq
Dually, we can form an \textbf{inf-$\dmgop$ combination} $G= \bigwedge_{i} b_i\dmgop G_i$
 of points  $G_i$ in $\Aa$  with scalars $b_i$.
The \textit{sup-$\mgop$ span} of $\Aa$, denoted by $\wsspan (\Aa)$, is
the set of all sup-$\mgop$ combinations of elements in $\Aa$.
If $\Aa=\keno$, then $\wsspan (\Aa)=\{ \flatle \}$.
Dually, the set of all inf-$\dmgop$ combinations of elements in $\Aa$
is called its \textit{inf-$\dmgop$ span}, denoted by  $\wispan (\Aa)$.
If $\Aa=\keno$, then $\wispan (\Aa)=\{ \flatge \}$.

% --- Independence is FINITE -----------------

If the above sup-$\mgop$ and inf-$\dmgop$ combinations are based on a
\emph{finite} set of space elements, we shall call them
\emph{max-$\mgop$} and \emph{min-$\dmgop$ combination}, respectively.
A set $\Ss$ in a complete weighted lattice
is called \textbf{max-$\mgop$ independent}
%(resp. \textbf{min-$\dmgop$ independent})
if each point $F\in \Ss$ is not a max-$\mgop$ %(resp. min-$\dmgop$)
combination of points in $\Ss\setminus \{ F\}$;
otherwise, the set is called \emph{max-$\mgop$ dependent}.
Dually for the min-$\dmgop$ (in)dependence.

% --- CWL Basis -----------------

%In linear spaces, a Hamel basis is a subset of the space that is linearly independent
%and its linear span makes up all the space.
%%
%In the nonlinear spaces under discussion,
A max-$\mgop$ independent subset $\Bb$ of a CWL $\Ww$ is called an \emph{upper basis}
 for the space if each space element $F$  can be represented as a
sup-$\mgop$ combination of  basis elements:
\beq
F= \bigvee_i c_i\mgop B_i, \; \; B_i\in \Bb
\label{cwlubasrep}
\eeq
Dually, a min-$\dmgop$ independent subset $\Bb'$ of $\Ww$ is called a \emph{lower basis}
   if $\Ww=\wispan (\Bb')$.
%i.e. each element of the space can be represented as an inf-$\dmgop$ combination of  basis elements.
Examples of upper and lower bases are given later for CWLs of functions.
%
%\beq
%F= \bigwedge_i d_i\dmgop B'_i, \; \; B'_i\in \Bb'
%\label{cwlubasrepd}
%\eeq
%
%If the space $\Ww$ is self-conjugate, then (\ref{cwlubasrep}) implies that
%$
%\glconj{F}= \bigwedge_i \glconj{c_i}\dmgop \glconj{B_i}.
%%\label{cwlubasconj}
%$
%Thus, if the space possesses an upper basis, it will possess a lower basis too.
%We conjecture that the upper and lower bases of a self-conjugate
%complete weighted lattice
%have the same cardinality, which is called the \emph{dimension} of $\Ww$.
%%This cardinality is called the \emph{dimension} of $\Ww$.
%If this is finite, the space is called finite-dimensional; otherwise, it is called infinite-dimensional.
%%Examples of an upper and a lower basis are mentioned in Section~\ref{sc-wmmsigspace}
%%and Section~\ref{sc-wmmvecspace} for signal and vector spaces respectively;
%%in the first case the basis is infinite-dimensional, whereas  the second case
%%is finite-dimensional.

%-----------------------------------------------------------------
%\subsubsection{Complete Weighted Lattices of Functions}

%Herein we are  working on complete weighted lattice
% (CWL) \emph{spaces of signals and vectors}.
%Thus, the underlying set of our CWL space is a function space $\Ww=\fun(\gimdom,\vset)$
In this paper we shall focus on CWLs
 whose underlying set  is a \emph{function space }$\Ww=\fun(\gimdom,\vset)$
where $\gimdom$ is an arbitrary nonempty set serving as the domain of our functions
and the values of these functions are from a clodum
$(\vset ,\vee,\wedge,\mgop,\dmgop)$ of scalars as described in Sec.~\ref{sc-algss}.
Then, we extend \emph{pointwise} the  supremum, infimum and scalar multiplications
of $\vset$ to the functions: for $F,G\in \Ww$,  $a\in \vset$ and $x\in \flatdom$
\beq
\begin{array}{ccl}
(F\vee G) (x) & \defineq & F(x)\vee G(x) %, \quad x\in \flatdom
\\
(F\wedge G) (x) & \defineq & F(x)\wedge G(x) %, \quad x\in \flatdom
\\
(a\mgop F) (x) & \defineq & a\mgop F(x) %, \quad x\in \flatdom
\\
(a\dmgop F) (x) & \defineq & a\dmgop F(x) %, \quad x\in \flatdom
\end{array}
\label{flatsism}
\eeq

Under the first two operations $\Ww$ becomes a complete infinitely distributive
lattice that inherits many properties from the lattice structure of $\vset$.
The least ($\flatle$) and greatest ($\flatge$) elements of $\Ww$ are the
constant functions $\flatle (x) = \vsetle$ and $\flatge (x)=\vsetge$, $x\in \flatdom$.
%\beq
%\ltle (x) = \vsetle, \quad \ltge (x)=\vsetge, \; \; \forall x\in \flatdom .
%\eeq
Further, the scalar operations $\mgop$ and $\dmgop$, extended pointwise to functions,
distribute over any suprema and infima, respectively.
Thus, the function space $\fun(\flatdom,\vset)$ is by construction
a \emph{complete weighted lattice of functions} over the clodum $\vset$.
The collection of all its properties creates a rich algebraic structure.

If the clodum $\vset$ is  self-conjugate,
% i.e. has a negation $\glconj{(\cdot)}$ satisfying (\ref{clodconj}),
then we can extend the conjugation $\glconj{(\cdot)}$ to functions $F$  pointwise:
$\glconj{F}(x) \defineq \glconj{(F(x))}$.
%\beq
%\glconj{F}(x) \defineq \glconj{F(x)}, \quad x\in \flatdom
%\eeq
This obeys the same rules as the scalar conjugation on the clodum.
Namely,
\beq
\glconj{\left( \bigvee_i F_i \right)}  =  \bigwedge_i \glconj{F_i} \; \; , \; \;
\glconj{\left( \bigwedge_i G_i\right)} =  \bigvee_i \glconj{G_i}\; \; , \; \;
\glconj{(a\mgop F)}  =  \glconj{a}\dmgop \glconj{F}
\label{fcwlconj}
\eeq
In such a case we have a \emph{self-conjugate} complete weighted lattice.

%\subsection{Function Lattices (from old LCGT)}

The space of vectors and the space of signals with values from  $\vset$
are special cases of function lattices.
%The underlying set of these lattices is the set  $\Ww =\fun(\vset, \flatdom)$
%of all functions mapping an arbitrary nonempty set $\flatdom$ into $\vset$.
In particular, if $\flatdom =\{1,2,...,n\}$, then
$\Ww$ becomes the set of all $n$-dimensional
vectors  with elements from $\vset$.
If $\flatdom =\INT$, then  $\Ww$ becomes the set of all
discrete-time
%\footnote{Our representation results in
%Section~\ref{sc-repsvsowl} do not change if we use \emph{multidimensional} signals,
%with discrete ($\flatdom=\INT^d$) or continuous domain ($\flatdom=\REAL^d$).
%Henceforth, in this work we focus on one-dimensional signals.}
signals with values from  $\vset$.

Elementary increasing operators on
$\Ww$ are those that act as \textbf{vertical translations}
(in short V-translations)  of functions.
Specifically, pointwise `multiplications' of functions $F\in \Ww$
by scalars $a\in \vset$ yield the \emph{V-translations} $\trop _a$
and \emph{dual V-translations} $\trop' _a$,  defined by
\beq
[\trop_a (F)](x) \defineq  a\mgop F(x), \; \;
[\trop'_a (F)](x) \defineq  a\dmgop F(x) %, \quad x\in \flatdom
\eeq
%Actually, the set $\tgroup =\{ \trop _a: a\in \vset  \}$ of
%translations forms a commutative
%monoid under composition
%\beq
%\trop _a \trop _b =\trop _a (\trop _b )=
%\trop _{a\mgop b}
%\eeq
A function operator $\psi$ on $\Ww$ is called \textbf{V-translation invariant}
if it commutes with any V-translation $\trop$, i.e.,
$
\psi \trop  = \trop \psi .
$
Similarly, $\psi$ is called \emph{dual V-translation invariant}
if $\psi \trop'  = \trop' \psi$ for any dual V-translation $\trop'$.

The above CWL $\Ww$ of functions contains an upper basis $\Bb$ and a lower basis $\Bb'$
which consist of the \emph{impulse functions} $\dimpls$
and the \emph{dual impulses} $\eimpls$, respectively:
\beq
\dimpls_{y} (x)\defineq \left\{  \begin{array}{ll}
 \mgid , & x=y \\ \vsetle , & x\not = y
\end{array} \right.
, \; \; \;
\eimpls_{y} (x) \defineq \left\{  \begin{array}{ll}
 \dmgid , & x=y \\ \vsetge , & x\not = y
\end{array} \right.
\label{tvimpsig}
\eeq
Then, every function $F(x)$ admits a representation as a supremum
of V-translated impulses placed at all points
or as infimum of dual V-translated impulses:
\beq
F(x) = \bigvee _{y\in \flatdom} F(y)\mgop  \dimpls_y(x)
= \bigwedge _{y\in \flatdom} F(y)\dmgop  \eimpls_y(x)
\label{sigimprep}
\eeq
By using the % above defined
V-translations and the basis representation of functions with impulses,
we can build more complex increasing operators,
% such as translation-invariant dilations and erosions of vectors or signals,
 as explained next.

In general, if the space $\Ww$ is self-conjugate and has an upper basis $\Bb$,
then it will also possess a lower basis since (\ref{cwlubasrep}) implies that
$
\glconj{F}= \bigwedge_i \glconj{c_i}\dmgop \glconj{B_i}.
%\label{cwlubasconj}
$
Thus,
in the case of function CWLs that are self-conjugate, the upper and lower bases
have the same cardinality, which is called the \emph{dimension}\footnote{A dimension theory for semimodules
 has been developed in \cite{Wagn91}. Further, the concept of an upper basis has been used in \cite{Butk10}
 to define the dimension of finite-dimensional subspaces of max-plus matrix algebra.} of $\Ww$.
%This cardinality is called the \emph{dimension} of $\Ww$.
If this is finite, the space is called finite-dimensional; otherwise, it is called infinite-dimensional.
Specific examples of finite- and infinite-dimensional upper and lower basis are mentioned
in Sec.~\ref{sc-cwlvec} and Sec.~\ref{sc-cwlsig}
 for vector and signal spaces respectively.

Consider systems that are V-translation invariant dilations or erosions
over  $\Ww$.
This invariance implies that they obey an interesting
\emph{nonlinear superposition principle} which has direct conceptual analogies
with the well-known linear superposition.
Specifically, we define  $\dlop$ to be  a
\textbf{dilation V-translation invariant  (DVI)} system iff
for any $c_i\in \vset,\; F_i \in \Ww$
\beq
\dlop (\bigvee _{i\in J}c_i\mgop F_i) = \bigvee _{i\in J} c_i \mgop \dlop ( F_i),
%\quad c_i\in \vset,\; F_i \in \Ww
\label{vtidilfunop}
\eeq
for any (finite or infinite) index set $J$. We also define  $\erop$ to be  an
\textbf{erosion V-translation invariant  (EVI)} system  iff
\beq
\erop (\bigwedge _{i\in J}c_i\dmgop F_i) = \bigwedge _{i\in J} c_i \dmgop \erop ( F_i)
%, \quad c_i\in \vset,\; F_i \in \Ww
\label{vtierofunop}
\eeq

Compare the two above nonlinear superpositions with the \emph{linear} superposition
obeyed by a linear system $\Gamma$:
\beq
\Gamma (\sum_{i\in J}a_i\cdot F_i) = \sum_{i\in J}a_i\cdot \Gamma (F_i)
\label{linsprpos}
\eeq
where $J$ is a finite index set, $a_i$ are constants from a field
(of real or complex numbers)
 and $F_i$ are field-valued signals from a linear space.

The structure of a DVI or EVI system's output is
simplified if we express it via the system's impulse
responses, defined next. Given a dilation system $\dlop$, its \textbf{impulse response map}
is the map $H:\flatdom \rightarrow \fun (\flatdom,\vset)$ defined at each $y\in \flatdom$
as the output function $H(x,y)$ from $\dlop$ when the input is the impulse $\dimpls_y(x)$.
Dually, for an erosion operator $\erop$ we  define its
 \textit{dual impulse response map} $H'$ via its outputs when excited by dual impulses:
 for $x,y\in \flatdom$
\beq
H(x,y)  \defineq  \dlop (\dimpls_y)(x),
\quad H'(x,y)  \defineq  \erop (\eimpls_y)(x) %, \quad x,y\in \flatdom
\label{impresp}
\eeq
Applying a DVI operator $\dlop$ or an EVI operator $\erop$
to (\ref{sigimprep}) and using
the definitions in (\ref{impresp}) proves the following unified representation
 for all V-translation invariant dilation or erosion systems.
\begin{Theorem} \label{th-devtirep}
%Let $\Ww =\fun(\flatdom,\vset)$ be a complete weighted lattice
%of functions over a
%clodum $(\vset, \vee,\wedge,\mgop,\dmgop)$. Then: \\
(a)~A system $\dlop$ on $\Ww$ is DVI, i.e.
obeys the sup-$\mgop$ superposition of (\ref{vtidilfunop}),
  if and only if its output can be expressed as
\beq
\dlop (F)(x) = \bigvee _{y\in \flatdom} H(x,y)\mgop F(y)
\label{tvdil}
\eeq
 where $H$ is its impulse response map in (\ref{impresp}). % \\
(b)~A system $\erop$ on $\Ww$ is EVI, i.e.
obeys the inf-$\dmgop$ superposition of (\ref{vtierofunop}),
if and only if its output can be expressed as
\beq
\erop (F)(x) = \bigwedge _{y\in \flatdom} H'(x,y)\dmgop F(y)
\label{tvero}
\eeq
 where $H'$ is its dual impulse response map in (\ref{impresp}).
\end{Theorem}

The result (\ref{tvdil}) for the max-plus dioid is analyzed in \cite{BCOQ01}.
In the case of a signal space where $\flatdom=\INT$,
 the operations in (\ref{tvdil}) and (\ref{tvero}) are like
\emph{time-varying nonlinear convolutions} where a dilation (resp. erosion) system's output
is obtained as supremum (resp. infimum) of various impulse response signals produced by
exciting the system with impulses at all points and weighted by the input signal values
via a  $\mgop$-`multiplication'.

%++++++++++++++++++++++++++++++++++++++++++
\subsection{Weighted Lattice of Vectors}
\label{sc-cwlvec}

 Consider now the nonlinear vector space $\Ww =\vset ^n$, equipped
with the pointwise partial ordering $\vct{x} \leq \vct{y}$,
%which  means $x_i \leq y _i$ $\forall i$,
supremum $\vct{x}\vee \vct{y}=[x_i\vee y_i]$ and infimum
$\vct{x}\wedge \vct{y}=[x_i\wedge y_i]$ between any vectors
$\vct{x},\vct{y}\in \Ww$. Then, $(\Ww , \vee , \wedge, \mgop, \dmgop )$ is
 a complete weighted lattice.
%These are essentially complete minimax\footnote{By `minimax vector spaces'
%we mean the finite-dimensional nonlinear vector spaces of minimax algebra that are equipped
%with max-plus arithmetic or its dual and corresponding nonlinear matrix operations
%\cite{Cuni79}.} vector spaces.
Elementary increasing operators are the \emph{vector V-translations}
%$\trop$ and their duals $\trop '$
%\[
%\trop _a (\vct{x}) = a\mgop \vct{x}=[a\mgop x_i],
%\quad
%\trop' _a (\vct{x}) = a\dmgop \vct{x}=[a\dmgop x_i]
%\]
$\trop _a (\vct{x}) = a\mgop \vct{x}=[a\mgop x_i]$
and their duals $\trop' _a (\vct{x}) = a\dmgop \vct{x}$,
which `multiply' a scalar $a$ with a vector $\vct{x}$  componentwise.
A vector transformation on $\Ww$ is called (dual) V-translation invariant
%called simply a $(\tgroup')\tgroup$-operator,
if it commutes with any vector (dual) V-translation.

By defining as `impulse functions'  the impulse vectors
%$\vct{e}$ and their duals $\vct{e}'$
%\[
%\vct{e}_i \defineq [\vsetle , ..., \vsetle ,\mgid , \vsetle , ... , \vsetle ]^T,
%\quad
%\vct{e}'_i \defineq [\vsetge , ..., \vsetge ,\dmgid , \vsetge , ... , \vsetge ]^T,
%\]
$\vct{q}_j=[\dimpls_j(i)]$ and their duals $\vct{q}'_j=[\eimpls_j(i)]$,
where the index $j$ signifies the position of the identity,
each vector $\vct{x} = [x_1,...,x_n]^T$
has a basis representation as a max  of V-translated impulse
vectors or as a min of V-translated dual impulse vectors:
\beq
\vct{x} = \bigvee _{j=1}^n x_j  \mgop \vct{q}_j
%=\bigvee _{i=1}^n \trop _{x_i}(\vct{q}_i)
= \bigwedge _{j=1}^n x_j \dmgop  \vct{q}'_j
%=\bigwedge _{i=1}^n \trop' _{x_i}(\vct{q}'_i)
\eeq

More complex examples of increasing operators on this vector space
%are the max-$\mgop$ `product' $\mtr{M} \mxgmp \vct{x}$ and the
%min-$\dmgop$ `product' $\mtr{M} \mngmp \vct{x}$ of a matrix $\mtr{M}$
%with an input vector $\vct x$,
are the max-$\mgop$   and the
min-$\dmgop$ `multiplications'  of a matrix $\mtr{M}$
with an input vector $\vct x$,
\beq
\dlop _{\mtr M} (\vct{x}) \defineq \mtr{M} \mxgmp \vct{x},
\; \; \;
\erop _{\mtr M} (\vct{x}) \defineq \mtr{M} \mngmp \vct{x}
\label{vecdilero}
\eeq
which are, respectively, a vector dilation and a vector erosion.
These two nonlinear matrix-vector `products'
 are the prototypes of any vector transformation
that obeys a sup-$\mgop$ or an inf-$\dmgop$ superposition, as proven next.
%
%The following theorem
% establishes a two-way correspondence between vector transformations that are
% the above max-\mgop\  and min-\dmgop\ matrix-vector `products'
%and V-translation invariant dilations and erosions of vectors.
%
\begin{Theorem} \label{th-devtirepvec}
(a)~Any vector transformation on the complete weighted lattice
$\Ww=\vset ^n$ is DVI, i.e.
obeys the sup-$\mgop$ superposition of (\ref{vtidilfunop}),
iff it can be represented as a matrix-vector max-$\mgop$ product
$\dlop _{\mtr M}(\vct x)=\mtr M\mxgmp \vct x$ where
$\mtr{M}=[m_{ij}]$ with $m_{ij}=\{ \dlop (\vct{q}_j)\}_i$.
\\
(b)~Any vector transformation  on $\vset ^n$ is EVI, i.e.
obeys the inf-$\dmgop$ superposition of (\ref{vtierofunop}),
iff it can be represented as
a matrix-vector min-$\dmgop$ product  $\erop _{\mtr M}(\vct x)=\mtr M \mngmp \vct x$ where
$\mtr{M}=[m_{ij}]$ with $m_{ij}=\{ \erop (\vct{q}'_j)\} _i$.
%(b)~Any dual V-translation invariant erosion $\erop$ on $\Ww$ can be represented as
%a matrix-vector min-\dmgop\ `product'  $\erop _{M'}=\mtr M \mngmp \vct x$ where
%$\mtr{M}'=[m'_{ij}]$ with $m'_{ij}=\{ \erop (\vct{q}'_j)\} _i$,
%and vice-versa.
\end{Theorem}
{\em Proof:\/} This is a special case of Theorem~\ref{th-devtirep}
where  the domain points $x,y\in \flatdom$
become indices $i,j\in \{1,...,n\}$ and the impulse response
values $H(x,y)$ become matrix elements $m_{ij}$.
Thus, the operations (\ref{tvdil}) and (\ref{tvero})
%, which remind time-varying convolutions,
become the max-$\mgop$ and min-$\dmgop$ products (\ref{vecdilero}) of input vectors
with the matrix $\mtr M=[m_{ij}]$.
%vector dilation and erosion of (\ref{vecdilero}).
\hfill Q.E.D.

Given a vector dilation
$\dlop(\vct{x})=\mtr{M}\mxgmp \vct{x}$ with matrix $\mtr{M}=[m_{ij}]$,
there corresponds a unique  adjoint vector erosion $\erop$
so that $(\dlop,\erop)$ is a \emph{vector adjunction} on $\Ww$, i.e.
\beq
\dlop (\vct x)\leq \vct y \Longleftrightarrow \vct x \leq \erop (\vct y)
\label{vecadj}
\eeq
(We seek adjunctions because they can easily generate projections.)
 % such as openings.)
 We can find the adjoint vector erosion by decomposing both vector operators
 based on scalar operators $(\sbdl,\asbdl)$ that form a \emph{scalar adjunction} on $\vset$:
 \beq
\sbdl (a,v)\leq w \Longleftrightarrow v \leq \asbdl (a,w)
\label{scalaradj}
\eeq
If we use as scalar `multiplication'  a commutative binary operation
$\sbdl (a,v)=a\mgop v$ %$\sbdl: \vset^2 \rightarrow \vset$
that is a dilation on $\vset$,  its  scalar adjoint erosion becomes
\beq
\asbdl (a,w)= \sup \{ v\in \vset : a\mgop v\leq w\}
\label{sadjerop}
\eeq
which is a (possibly non-commutative) binary operation on $\vset$.
% $\asbdl :\vset^2 \rightarrow \vset$
%such that $(\sbdl,\asbdl)$ is an adjunction:
Then,
%according to  (\ref{fleddcmps}) and (\ref{vdlopfadjdec})
the original vector dilation $\dlop (\vct x)=\mtr M \mxgmp \vct x$ is decomposed as
\beq
\{ \dlop (\vct x)\} _i =\bigvee_j \sbdl (m_{ij},x_j)=m_{ij}\mgop x_j, \quad i=1,...,n
\label{vdlopij}
\eeq
whereas its adjoint vector erosion % that satisfies (\ref{vecadj})
is decomposed as
\beq
\{ \erop (\vct{y})\}_j=\bigwedge _{i} \asbdl (m_{ji},y_i), \quad j=1,...,n
\label{vadjeropij}
\eeq
The latter can be written as a min-$\asbdl$ matrix-vector multiplication
\beq
\erop (\vct y) = \mtr M^T \mnasbdmp \vct y
\label{vadjerop}
\eeq
where the symbol $\mnasbdmp$ denotes the following nonlinear  product
of a matrix  $\mtr{A}=[a_{ij}]$ with a matrix $\mtr{B}=[b_{ij}]$:
\[
\{ \mtr{A} \mnasbdmp\mtr{B}\}_{ij} \defineq \bigwedge _{k} \asbdl (a_{ik},b_{kj})
\]
Further, if $\vset$ is a \emph{clog}, % then $\asbdl (b;c)=b\dmgop \glconj{c}$.
it can be shown that  $\asbdl (a,w)=\glconj{a} \dmgop w$ and hence
\beq
\erop (\vct{y}) = \conjtranmtr{\mtr M} \mngmp \vct y,
\label{vadjeropclog}
\eeq
where $\conjtranmtr{\mtr M}$ is the \emph{adjoint} (i.e. conjugate transpose)\footnote{Despite
its notation \cite{Butk10,Cuni79},  $\mtr M^\ast$ is not the element-wise conjugate
 of the matrix $\mtr M$ but is obtained via transposition and element-wise conjugation of $\mtr M$.
 To avoid the above ambiguity, we prefer the terminology `adjoint'
 which is  based on some conceptual similarities with
 the adjoint of a linear operator in Hilbert spaces \cite{Cuni79}.}
%\footnote{In minimax algebra
%the conjugate transpose $\conjtranmtr{\mtr M}$ is called the adjoint matrix and
% denoted by $\mtr M^*$.}
 of $\mtr M=[m_{ij}]$:
\beq
\conjtranmtr{\mtr M} \defineq % \glconj{(\mtr M^T)}=
[\glconj{m_{ji}}]
\label{mtradj}
\eeq

\begin{Examples}\ \label{ex-mtrmuladj} {\rm
(a)~In the max-plus clog $(\EREAL,\vee,\wedge,+)$,
consider the max-sum product (\ref{mxsmpr}) of a matrix $\mtr M$ and a vector $\vct x$:
\beq
\mtr M=\left[ \begin{array}{ccc} 1 & 0.4 & 0 \\ 0.3 & 1 & 0.5 \\ 0.7 & 0.2 & 1 \end{array} \right],
\; \; \vct x =\left[ \begin{array}{c} -0.2 \\ -0.6 \\ -0.3  \end{array} \right]
\Longrightarrow \dlop_s (\vct x)=\mtr M \mxsmp \vct x
=\left[ \begin{array}{c} 0.8 \\ 0.4 \\ 0.7  \end{array} \right]
=\vct y
\label{exmxsmpr}
\eeq
Let us apply to the result $\vct y$ the adjoint erosion. By (\ref{vadjeropclog}) and (\ref{mnsmpr}),
\beq
\conjtranmtr{\mtr M} =
\left[ \begin{array}{ccc} -1 & -0.3 & -0.7 \\ -0.4 & -1 & -0.2 \\ 0 & -0.5 & -1 \end{array} \right]
\Longrightarrow \erop_s (\vct y)= \conjtranmtr{\mtr M}\mnsmp \vct y
=\vct x
\label{exadjmnsmpr}
\eeq
Thus, in this example we have $\erop_s \dlop_s =\idop$. \\
(b)~In the clodum $([0,1],\vee,\wedge,\min,\max)$, let us use
a vector dilation $\dlop_f$ as in (\ref{vdlopij})
with max-min arithmetic (common in fuzzy systems), i.e. with $\sbdl(a,v)=a\mgop v=\min (a,v)$,
to multiply the same matrix $\mtr M=[m_{ij}]$ as above
with a different vector $\vct z$ so as to reach the same result $\vct y$:
\beq
\vct z =\left[ \begin{array}{c} 0.8 \\ 0.4 \\ 0.4  \end{array} \right]
\Longrightarrow
[ \dlop_f (\vct z) _i] =[\bigvee_j \min(m_{ij}, z_{j})] = \vct y
\label{exmxfmpr}
\eeq
To apply now the adjoint vector erosion (\ref{vadjeropij}), we need first to find
the adjoint scalar erosion:
\beq
\asbdl (a,w)=\sup \{ v\in [0,1]:\min(a,v)\leq w \}
 =\left\{ \begin{array}{ll} w, & w < a \\ 1, & w \geq a \end{array} \right.
\label{asbdlmin}
\eeq
Then, by (\ref{vadjeropij}) we can construct the adjoint vector erosion $\erop _f$,
from which we obtain $\erop_f (\vct y)=\vct z$; i.e., again the adjoint vector erosion
happened to be the inverse of the vector dilation.
} \end{Examples}

Dually, given a vector erosion $\erop'(\vct{y})=\mtr{M}\mngmp \vct{y}$
% with matrix $\mtr{M}=[m_{ij}]$,
%we can find  its \emph{adjoint dilation} $\dlop'$
%so that $(\erop' , \dlop')$ is an adjunction.
%Starting from the `dual multiplication' $\sber(a,w)=a\dmgop w$ as a scalar erosion,
%its right adjoint is the scalar dilation $\asber (a,v)=\inf \{ w: a\dmgop w\geq v\}$,
%such that $(\sber,\asber)$ is a scalar adjunction.
we can obtain  its adjoint vector  dilation $\dlop'$
%so that $(\erop' , \dlop')$ is an adjunction.
by starting from the `dual multiplication' $\sber(a,w)=a\dmgop w$
as a scalar erosion and finding its  adjoint scalar dilation
\beq
\asber (a,v)=\inf \{ w: a\dmgop w\geq v\}
\label{sadjdlop}
\eeq
 Then the min-$\sber$ matrix-vector multiplication
 $\erop' (\vct y)=\mtr M \mngmp \vct y$ with
\beq
\{ \erop' (\vct y)\} _i =\bigwedge_j \sber (m_{ij},y_j)=m_{ij}\dmgop y_j, \quad i=1,...,n
\eeq
has as adjoint a max-$\asber$ matrix-vector  multiplication
$\dlop'(\vct x)$ with
\beq
\{ \dlop' (\vct{x})\}_j \defineq \bigvee _{i} \asber (m_{ji},x_i), \quad j=1,...,n
\eeq
We can write this as a max-$\asber$ matrix-vector multiplication
\beq
\dlop' (\vct x) = \mtr M^T \mxasbemp \vct x
\eeq
where the symbol $\mxasbemp$  denotes the following nonlinear  product
of a matrix  $\mtr{A}=[a_{ij}]$ with a matrix $\mtr{B}=[b_{ij}]$:
\[
\{ \mtr{A} \mxasbemp\mtr{B}\}_{ij} \defineq \bigvee _{k} \asber (a_{ik},b_{kj})
\]
Further, if $\vset$ is a  \emph{clog},
it can be shown that  $\asber (a,v)=\glconj{a} \mgop v$ and hence
\beq
\dlop' (\vct{x}) = \conjtranmtr{\mtr M} \mxgmp \vct x
\label{vecadjdlop}
\eeq

%+++++++++++++++++++++++++++++++++++++++++++++++++++
\subsection{Weighted Lattice of Signals}
\label{sc-cwlsig}

Consider the set $\Ww=\fun (\INT , \vset )$ of all discrete-time
signals $f:\INT \rightarrow \vset$ with values from $\vset$.
Equipped with pointwise supremum $\vee$ and infimum $\wedge$,
and two pointwise scalar multiplications ($\mgop$ and $\dmgop$), this becomes
a complete weighted lattice $\Ww$ with partial order the pointwise
signal relation $\leq$. The signal translations are the operators
$\trop _{k,v}(f)(t)=f(t-k)\mgop v$, where $(k,v)\in \INT \times \vset$
and $f(t)$ is an arbitrary input signal.
Similarly, we define dual signal translations
$\trop' _{k,v}(f)(t)=f(t-k)\dmgop v$.
A signal operator on $\Ww$
is called \emph{(dual) translation invariant\/} iff it commutes with any
such (dual) translation.
Note that, the above translation-invariance contains both a vertical translation
and a horizontal translation; the horizontal part is the well-known \emph{time-invariance}.
Consider  two
elementary signals, called the \emph{impulse} $\dimpls$
and the \emph{dual impulse} $\eimpls$:
\[
\dimpls (t)\defineq \left\{  \begin{array}{ll}
 \mgid , & t=0 \\ \vsetle , & t\not = 0
\end{array} \right.
, \; \; \;
\eimpls (t)\defineq \left\{  \begin{array}{ll}
 \dmgid , & t=0 \\ \vsetge , & t\not = 0
\end{array} \right.
\]
Then every signal $f$ has a basis representation
as a supremum of translated impulses
or as infimum of dual translated  impulses:
\beq
f(t) = \bigvee _{k} f(k)\mgop  \dimpls (t-k)
= \bigwedge _{k} f(k)\dmgop \eimpls (t-k)
\label{sigimprepti}
\eeq

Consider now operators $\fdlop$ on $\Ww$ that are dilations and translation-invariant.
Then $\fdlop$ is both DVI in the sense of (\ref{vtidilfunop}) and time-invariant. We call such operators
\textbf{dilation translation-invariant (DTI)} systems. Applying $\fdlop$ to an input signal $f$
decomposed as in (\ref{sigimprepti}) yields its output as the sup-$\mgop$ convolution $\sgcnv$
of the input with the system's impulse response $h=\fdlop (\dimpls )$:
\beq
\fdlop (f)(t)=(f\sgcnv h)(t)=\bigvee _{k\in \INT} f(k)\mgop h(t-k)
%=\bigvee _{k\in \INT} \sbdl [h(t-k),f(k)]
\eeq
Conversely, every sup-$\mgop$ convolution is a DTI system.
%Then, it follows  from (\ref{scalaradj}) that its adjoint signal erosion is
As done for the vector operators, we can also build  signal operator pairs $(\fdlop,\ferop)$ that form adjunctions:
\beq
\fdlop (f)\leq g \Longleftrightarrow f \leq \ferop (g)
\label{sigadj}
\eeq
Given $\fdlop$ we can find its adjoint $\ferop$  from scalar adjunctions $(\sbdl,\asbdl)$.
Thus, by (\ref{scalaradj}) and (\ref{sadjerop}), if $\sbdl (h,f)=h\mgop f$, the adjoint signal erosion becomes
\beq
%\ltadj{\fdlop}(g)(t)
\ferop (g)(t)=\bigwedge _{\ell \in \INT} \asbdl[h(\ell-t),g(\ell )]
\eeq
%such that $(\ltadj{\fdlop}, \fdlop)$ is an adjunction.
Further, if $\vset$ is a clog, then
\beq
%\ltadj{\fdlop} (g)(t)
\ferop (g)(t)=\bigwedge _{\ell \in \INT} g(\ell ) \dmgop \glconj{h}(\ell-t)
\eeq
%where $\glconj{h}(\cdot)=\glconj{h(\cdot)}$.

Dually, if we start from an operator $\ferop$ on $\Ww$ that is erosion and translation-invariant,
then $\ferop$ is both EVI in the sense of (\ref{vtierofunop}) and time-invariant. We call such operators
\textbf{erosion translation-invariant (ETI)} systems. Applying $\ferop$ to an input signal $g$
decomposed as in (\ref{sigimprepti}) yields the output as the inf-$\dmgop$  convolution $\igcnv$
of the input with the system's  dual impulse response $h'=\ferop (\eimpls )$:
\beq
\ferop (g)(t)=(g\igcnv h')(t)=\bigwedge _{k\in \INT} g(k)\dmgop h'(t-k)
\eeq
Setting $\sber (h',g)=h'\dmgop g$ and using (\ref{scalaradj}),(\ref{sadjdlop}) yields the adjoint signal dilation
\beq
\fdlop (f)(t)=\bigvee _{\ell \in \INT} \asber [h'(\ell-t),f(\ell)]
\eeq
which, if $\vset$ is a clog, becomes
\beq
\fdlop (f)(t)=\bigvee _{\ell \in \INT} f(\ell)\mgop \glconj{h'}(\ell-t)
\eeq

An outcome of the previous discussion is:
\begin{Theorem} \label{th-reptdesig}  % \cite{Mara05a}
(a)~An operator $\Delta$ on a CWL $\Ww$ of signals
%the CWL $\Ww=\fun(\INT,\vset)$ of signals
is a dilation translation invariant (DTI) system  iff it can be
represented as
the sup-$\mgop$ convolution of the input signal with
 the system's  impulse response $h=\Delta (\dimpls )$.
(b)~An operator $\Ee$ on $\Ww$
is an erosion dual-translation invariant (ETI) system iff it can be
represented as
the inf-$\dmgop$ convolution of the input signal with
 the system's dual impulse response $h'=\Ee (\eimpls )$.
\end{Theorem}
The above result for the max-plus clog was obtained in \cite{Mara94a}.

%#############################################################

\section{State and Output  Responses}
\label{sc-soresp} %\label{sc-resp}

Based on the
 state-space model of a max-$\mgop$ dynamical  system (\ref{mxgse}),
% which involves lattice superpositions and  max-$\mgop$ matrix operations,
we can compute its state response and output response if we know its
\emph{transition matrix}:
\beq
\mtr{\Phi}(t_2,t_1) \defineql \left\{ \begin{array}{lll}
\mtr{A}(t_2)\mxgmp \cdots \mxgmp \mtr{A}(t_1+1) & \mathrm{if} & t_2>t_1 \\
\mxgmpid _n & \mathrm{if} & t_2=t_1 \end{array} \right.
\label{tranmtr}
\eeq
for $t_2\ge t_1$, where $\mxgmpid _n$ is the $n\times n$ identity matrix in
max-$\mgop$ matrix algebra that has values equal to the identity element $\mgid$ on its diagonal and
least element (null) $\vsetle$ off-diagonally.
The importance of $\mtr \Phi$ is obvious by noticing that for a null input,
% i.e. $u_i(t)=\vsetle \; \; \forall i$, % $\{ \vct{u}(t)\}_i=\vsetle$,
the solution of the homogeneous state equation
\beq
\vct{x}(t)  = \mtr{A}(t)\mxgmp \vct{x}(t-1)
\eeq
equals
\beq
\vct x(t) = \mtr \Phi (t,0)\mxgmp \vct x (0)
\eeq
The transition matrix obeys a \emph{semigroup property}:
\beq
 \mtr \Phi (t_2,t_1) \mxgmp  \mtr \Phi (t_1,t_0)=\mtr \Phi (t_2,t_0),
\quad t_0\leq t_1\leq t_2
\eeq

\subsection{Time-Varying Systems}
\label{sc-soresptv}

By using induction on  (\ref{mxgse}) we can find the state and output responses
 of the general time-varying causal system; % nonhomogeneous system;
 for $t=0,1,2,...$,
\beq
\begin{array}{l}
\vct x(t)  =  \mtr \Phi (t,0) \mxgmp \vct x (0) \vee
\left( {\dsty \bigvee _{k=0}^{t} } \mtr \Phi (t,k) \mxgmp \mtr B(k)\mxgmp \vct u(k) \right)
\\
\vct y (t)  =  \underbrace{ \mtr C (t)\mxgmp \mtr \Phi (t,0) \mxgmp \vct x (0) }
_{\mbox{\rm $\vct{y}_{ni}(t)$\defineq`null'-input resp.}} \ \vee \
 \\
\underbrace{ \left(  \bigvee _{k=0}^{t} \mtr C (t)\mxgmp \mtr \Phi (t,k)\mxgmp \mtr B(k)\mxgmp \vct u(k) \right)
\vee \mtr D (t)\mxgmp \vct u(t)}_{\mbox{\rm $\vct{y}_{ns}(t)$\defineq `null'-state resp.}}
\end{array}
\label{sormxgsemtr}
\eeq
 where the  supremum $\bigvee_{k=0}^{t}$ is null if $t<0$.
 Henceforth, without loss of generality in (\ref{sormxgsemtr}),
  we shall assume that in practice $\vct u(0)$ is null (i.e. the input starts being active from $t\geq 1$)
  and use $\vct x(0)$ as the system's effective initial condition.
  (Otherwise, we use $\vct x(-1)$ as  initial condition.)
Thus, the output response is found to consist of two parts:
(i)~the `null'-input response which is due only to the initial conditions
$\vct{x}(0)$ and assumes a null input, i.e. equal to $\vct u(t)=\vsetle$,
and (ii)~the `null'-state response which is due only to the input
$\vct{u}(t)$ and assumes null initial conditions, i.e. $\vct{x}(0)=\vsetle$.

%We observe that the `null'-state response is essentially a time-varying sup-$\mgop$
%matrix convolution  of the input with a weight matrix
%$\mtr H(t,k)\defineq \mtr C (t)\mxgmp \mtr \Phi (t,k)\mxgmp \mtr B(k)\vee \dimpls(t-k)\mgop \mtr D(k)$:
%\beq
%\vct{y}_{ns}(t) = \bigvee _{k=0}^t \mtr H(t,k)\mxgmp \vct u (k)
%%,\quad \mtr H_k(t)\defineq \mtr C (t)\mxgmp \mtr \Phi (t,k)\mxgmp \mtr B(k)
%\label{irmmxgsemtr}
%\eeq
We observe that the `null'-state response is essentially a time-varying sup-$\mgop$
matrix convolution
\beq
\vct{y}_{ns}(t) = \bigvee _{k=0}^t \mtr H(t,k)\mxgmp \vct u (k)
%,\quad \mtr H_k(t)\defineq \mtr C (t)\mxgmp \mtr \Phi (t,k)\mxgmp \mtr B(k)
\label{irmmxgsemtr}
\eeq
of the input with a weight matrix
\[ \mtr H(t,k)\defineq \mtr C (t)\mxgmp \mtr \Phi (t,k)\mxgmp \mtr B(k)\vee \dimpls(t-k)\mgop \mtr D(k) \]
The response (\ref{irmmxgsemtr}) is a matrix version of the  scalar  time-varying sup-$\mgop$
 convolution  in (\ref{tvdil}).

The representation of the responses of time-varying  max-$\mgop$ systems over idempotent dioids
via the transition matrix has been developed in \cite{LBH04}.

\subsection{Time-Invariant Systems}
\label{sc-sorespti}

Most of the results in this section are well-known for time-invariant max-$\mgop$ systems over idempotent dioids,
especially  in the max-plus case \cite{BCOQ01}.
We present them using monotone operators over weighted lattices.

Let the  matrices $\mtr A, \mtr B, \mtr C, \mtr D$  be constant.
Then, the  max-$\mgop$ state equations become:
\beq
\begin{array}{rcl}
\vct{x}(t) & = &
\mtr A \mxgmp \vct x(t-1)\  \vee \ \mtr B \mxgmp \vct u(t)
\\
\vct y(t) & = &
 \mtr C \mxgmp \vct x(t) \ \vee \ \mtr D \mxgmp \vct u(t)
\end{array}
\label{mxgseti}
\eeq
Since the transition matrix simplifies to % $\Phi (t_2,t_1) = \mtr A ^{(t_2-t_1)}$
\beq
\mtr \Phi (t_2,t_1) = \mtr A ^{(t_2-t_1)}
\eeq
where
$A ^{(t)}$ denotes the $t$-fold max-$\mgop$ matrix product of $\mtr A$ with itself
for $t\geq 1$ and $\mtr{A}^{(0)}=\mxgmpid _n$,
the solutions of the constant-matrix state equations become
\begin{eqnarray}
\vct x(t)  = & \mtr A ^{(t)} \mxgmp \vct x (0) \vee
\left( \bigvee _{k=0}^{t} \mtr A ^{(t-k)} \mxgmp \mtr B \mxgmp \vct u(k) \right)
\nonumber \\ % \label{srmxgsemtrti}
\vct y (t)=  &  \underbrace{\mtr C \mxgmp \mtr A ^{(t)} \mxgmp \vct x (0)}_{\mbox{\rm $\vct{y}_{ni}(t)$=`null'-input resp.}}
\ \vee \ % \mtr D \mxgmp \vct u(t)
\label{sormxgsemtrti} \\
 & \underbrace{
  \mtr C \mxgmp \left( \bigvee_{k=0}^{t} \mtr A ^{(t-k)}\mxgmp \mtr B\mxgmp \vct u(k) \right) \vee \mtr D \mxgmp \vct u(t)
 }_{\mbox{\rm $\vct{y}_{ns}(t)$=`null'-state resp.}}
\nonumber % \label{ormxgsemtrti}
 \end{eqnarray}

By representing the matrix-vector $\mgop$-product as a dilation operator
$\vct x \mapsto \dlop _{\mtr A} (\vct x)=\mtr A \mxgmp \vct x$,
we can express the  state equations (\ref{mxgseti}) with vector operators:
\beq
\begin{array}{rcl}
\vct{x}(t) & = &
\dlop _{\mtr A} [\vct{x}(t-1)] \  \vee \ \dlop _{\mtr B} [\vct{u}(t)]
\\
\vct{y}(t) & = &
 \dlop _{\mtr C} [\vct{x}(t)] \ \vee \ \dlop _{\mtr D} [\vct{u}(t)]
\end{array}
\label{mxsedlop}
\eeq
and the state and output  responses (\ref{sormxgsemtrti}) in operator form:
\beq
\begin{array}{l}
\vct{x}(t) =  \dlop _{\mtr A}^{t}[\vct{x}(0)] \ \vee \
\left( {\dsty \bigvee } _{k=0}^{t}
\dlop _{\mtr A}^{t-k}\dlop _{\mtr B}[\vct{u}(k)] \right)  \\
\vct{y}(t) = \dlop _{\mtr C} \dlop _{\mtr A}^t[\vct{x}(0)]
\vee \dlop _{\mtr C} \left( {\dsty  \bigvee  _{k=0}^{t} }
 \dlop _{\mtr A}^{t-k}\dlop _{\mtr B}[\vct{u}(k)] \right)
\vee   \dlop _{\mtr D}[\vct{u}(t)]
\end{array}
\label{sormxsedlop}
\eeq

For {\em single-input single-output (SISO)\/} systems the mapping
$u(t)\mapsto y_{ns}(t)$ can be viewed as a causal translation-invariant
dilation system $\Delta$. Hence, the `null'-state response
can be found as the sup-$\mgop$ convolution of the input with the
system's impulse response $h=\Delta (\dimpls )$:
\beq
y_{ns} (t)= \Delta (u)(t) =(u\sgcnv h)(t)= \bigvee _{k=0}^t h(t-k)\mgop u(k)
\eeq
The impulse response can be found from (\ref{sormxgsemtrti})
% the general output (\ref{sormxgsemtrti})
by setting initial conditions  equal to null % $\vct x(0)=\vsetle$ %
and the input $u(t)=\dimpls(t)$:
\beq
 h(t) = \left\{
\begin{array}{ll} \vsetle , & t <0 \\
(\mtr{C}\mxgmp  \mtr{B}) \vee D , & t = 0 \\
\mtr{C}\mxgmp \mtr{A}^{(t)}\mxgmp \mtr{B} , & t>0
\end{array} \right.
\label{irsiso}
\eeq
where in this case $D$ is a scalar, $\mtr C$ is a row vector and $\mtr B$ a column vector.
The last two results can be easily extended to multi-input multi-output (MIMO)
systems by using an impulse response matrix as in (\ref{irmmxgsemtr}).

%######################################################
\section{Solving Max-$\boldsymbol{\mgop}$ Equations}
\label{sc-solmaxeqn}

Consider a scalar clodum  $(\vset,\vee,\wedge, \mgop, \dmgop)$,
a matrix  $\mtr{A}\in \vset ^{m\times n}$ and a vector $\vct{b}\in \vset ^m$.
The set of solutions of the max-$\mgop$ equation
\beq
\mtr{A} \mxgmp \vct{x} = \vct{b}
\label{mxgeq}
\eeq
over $\vset$ is either empty or forms a sup-semillatice. % semigroup under vector $\vee$.
In \cite{Cuni79} necessary and sufficient conditions  are given
for the existence and properties of such solutions in the max-plus case.
%for the existence and properties of such solutions, especially of the form
%\beq
%\vct x  = \conjtranmtr{\mtr A} \mngmp \vct b
%\label{mxgminizsolclog}
%\eeq
%where $\conjtranmtr{\mtr{A}}$ is  the conjugate transpose matrix, i.e.
%$\{ \conjtranmtr{\mtr{A}} \} _{ij} = \glconj{\{ \mtr{A} \} _{ji}}$.
%
%{\bf An Optimization Problem in Max-Plus Algebra:}\
A related problem in applications of max-plus algebra to scheduling is
when a vector $\vct x$  represents start times,
 a vector $\vct b$ represents finish times and the matrix $\mtr{A}$ represents processing
 delays. Then, if $\mtr A \mxgmp \vct x  = \vct b$ does not have an exact solution,
 it is possible to find the optimum $\vct x$ such that we
 minimize a norm of the earliness subject to zero lateness.
 We generalize this problem from max-plus to max-$\mgop$ algebra. The optimum
 will be the solution of the following constrained minimization problem:
\beq
\mathrm{Minimize} \; \| \mtr A \mxgmp \vct x -\vct b \| \; \;  \mathrm{s.t.} \; \;
\mtr{A} \mxgmp \vct x \leq  \vct b
%\begin{array}{c}
%{\rm Minimize} \; \; || \mtr A \mxgmp \vct x -\vct b || \\
%\mbox{\rm subject to} \; \;
%\mtr{A} \mxgmp \vct x \leq  \vct b
%\end{array}
\label{mxgminiz}
\eeq
where the norm $||\cdot ||$ is either the $\ell _\infty$ or the $\ell _1$ norm.
While the two above problems have been solved in \cite{Cuni79}
by using minimax algebra over the max-plus  $(\EREAL,\vee,\wedge,+)$ or other clogs,
%Instead of proving the above  over a clog as in \cite{Cuni79},
we provide next an alternative and shorter proof of both results using adjunctions and for the
general case when $\vset$ may not be a clog.

\begin{Theorem}  \label{th-mxgeqminiz} \ Consider a vector dilation $\dlop (\vct x)=\mtr A \mxgmp \vct x$
over a scalar clodum $\vset$ and let $\erop$ be its adjoint vector erosion.
(a)~If Eq.~(\ref{mxgeq}) has a solution, then
\beq
\hat{\vct x}  = \mtr A^T \mnasbdmp \vct b = [ \bigwedge _{i} \asbdl (a_{ji},b_i)]
\label{mxgminizsolclod}
\eeq
 is its greatest solution, where $\asbdl$ is the scalar adjoint erosion (\ref{sadjerop}) of $\mgop$.
 \\
(b)~If $\vset$ is a clog, the  solution (\ref{mxgminizsolclod}) becomes
\beq
\hat{\vct x}  = \conjtranmtr{\mtr A} \mngmp \vct b
\label{mxgminizsolclog}
\eeq
(c)~The solution to the minimization problem (\ref{mxgminiz}) is generally (\ref{mxgminizsolclod}),
or  (\ref{mxgminizsolclog}) in the special case of a clog.
\end{Theorem}
\emph{Proof:} \
%Instead of proving the above using minimax algebra over a clog as in \cite{Cuni79},
%we provide next an alternative and shorter proof of both results using adjunctions and for the
%general case when $\vset$ may not be a clog.
(a),(c):~We showed in (\ref{vadjeropij}),(\ref{vadjerop}) that the adjoint vector erosion of
 $\dlop (\vct x)=\mtr A \mxgmp \vct x$ is generally equal to
$\erop (\vct y)= \mtr{A}^T \mnasbdmp \vct y$.
Thus, the solution (\ref{mxgminizsolclod}) has the form of an erosion,
% and Lemma~\ref{lm-adjinv} completes the proof.
%Namely, by (\ref{adjero}), this  has the property
 which by (\ref{adjero})  has the property
\[
\erop (\vct b)=\bigvee \{ \vct x: \dlop (\vct x)\leq \vct b \}
\]
This implies that
\[
\dlop (\erop (\vct b))=\bigvee \{ \dlop (\vct x): \dlop (\vct x)\leq \vct b \}
\]
The above immediately suggests that if $\erop (\vct b)$ is a solution, then it is the greatest solution.
If not, then the difference $\vct b - \dlop(\erop (\vct b))$ is nonnegative and has the smallest $\ell_\infty$
or $\ell_1$ norm.
(b)~For a clog,  the scalar adjoint erosion of $\mgop$ is $\asbdl(a,w)=\glconj{a}\dmgop w$
which gives (\ref{mxgminizsolclod}) the simpler expression (\ref{mxgminizsolclog}). \hfill Q.E.D.

A main idea behind the method for solving (\ref{mxgminiz}) is to consider vectors $\vct x$
that are \emph{subsolutions} in the sense that $A \mxgmp \vct x \leq \vct b$ and find the greatest such
subsolution. The set of subsolutions forms a sup-semilattice % semigroup under vector $\vee$
whose supremum equals $\hat{\vct x}$,
which  yields either the greatest exact solution of (\ref{mxgeq}) or an optimum approximate solution
in the sense of (\ref{mxgminiz}).
Another attractive aspect of the adjunction-based solution is that it creates
a lattice projection onto the max-$\mgop$ span of the columns of $\mtr A$
via the opening $\dlop (\erop (\vct b))\leq \vct b$ that best approximates $\vct b$
from below.

\begin{Examples}\ \label{ex-solvmaxeq} {\rm
(a)~Consider solving $\dlop_s(\vct x)=\mtr A \mxgmp \vct x=\vct b$ in the max-plus clog $(\EREAL,\vee,\wedge,+)$ with
\beq
\mtr A=\left[ \begin{array}{ccc} 1 & 0.4 & 0 \\ 0.3 & 1 & 0.5 \\ 0.7 & 0.2 & 1 \end{array} \right],
\; \; \vct b =\left[ \begin{array}{c} 0.8 \\ 0.4 \\ 0.9  \end{array} \right]
\label{exmxseq}
\eeq
The algorithm (\ref{mxgminizsolclog}) yields the greatest solution
\beq
\hat{\vct x}_s =\erop_s(\vct b)=\conjtranmtr{\mtr A} \mngmp \vct b = [-0.2,-0.6,-0.1]^T
\eeq
 among all exact solutions,  which have the form
$\vct x = [-0.2,c,-0.1]^T$ with $c\leq -0.6$.
Note that in Example~\ref{ex-mtrmuladj}(a) we had the same matrix but a different
$\vct b=[0.8,0.4,0.7]^T$ which gave a unique solution. \\
(b)~Let us now try to solve $\dlop_f(\vct x)=\mtr A \mxgmp \vct x=\vct b$
in the max-min clodum $([0,1],\vee,\wedge,\min,\max)$
with the same $\mtr A,\vct b$ as above. Then, by working as in Example~\ref{ex-mtrmuladj}(b)
to construct the adjoint vector erosion, (\ref{mxgminizsolclod}) yields
\beq
\hat{\vct x}_f  = \erop_f(\vct b)=\mtr A^T \mnasbdmp \vct b = [0.8,0.4,0.4]^T
\eeq
where the specific $\asbdl$, i.e. the scalar adjoint erosion of $a\mgop v =\min(a,v)$, is given by  (\ref{asbdlmin}).
In this case, the algorithm gave an approximate solution since
$\mtr A \mxgmp \hat{\vct x}_f=[0.8,0.4,0.7]^T\leq \vct b$.
However, it is the greatest subsolution. Note that the same matrix but with a different $\vct b$ gave
an exact solution in Example~\ref{ex-mtrmuladj}(b).
} \end{Examples}

Further, by using adjunctions and duality,
 the CWL framework  allows us to easily formulate and solve a \emph{dual problem}
 of solving the min-$\dmgop$ equation
\beq
\mtr{A} \mngmp \vct{y} = \vct{b}
\label{mngeq}
\eeq
%whose set of solutions forms a semigroup under vector $wedge$,
either exactly if it has a solution,
or by finding \emph{supersolutions} $\vct y$ in the sense that $A \mngmp \vct y \geq \vct b$
and picking the smallest such supersolution.
Approximate solutions of (\ref{mngeq}) can always be found by solving the following problem
\beq
\mathrm{Minimize} \;  \| \mtr A \mngmp \vct y -\vct b \| \; \; \mathrm{s.t.} \; \;
\mtr{A} \mngmp \vct y \geq  \vct b
%\begin{array}{c}
%\mathrm{Minimize} \; \; || \mtr A \mngmp \vct y -\vct b || \\
%\mbox{\rm subject to} \; \;
%\mtr{A} \mngmp \vct y \geq  \vct b
%\end{array}
\label{mngminiz}
\eeq
where the norm $||\cdot ||$ is either the $\ell _\infty$ or the $\ell _1$ norm.
The set of supersolutions forms a semigroup under vector $\wedge$
whose infimum yields either the smallest exact solution of (\ref{mngeq}) if it exists
or an optimum approximate solution in the sense of (\ref{mngminiz}); this infimum is
\beq
\hat{\vct y}  = \mtr A^T \mxasbemp \vct b
\label{mngminizsolclod}
\eeq
where $\asber$ is the scalar adjoint dilation (\ref{sadjdlop}) of $\dmgop$.
For a clog this becomes
\beq
\hat{\vct y } = \conjtranmtr{\mtr A} \mxgmp \vct b
\label{mngminizsolclog}
\eeq
By viewing $\erop (\vct y)=A \mngmp \vct y$ as a vector erosion,
the operation in (\ref{mngminizsolclod}) or (\ref{mngminizsolclog}) is its corresponding
adjoint vector dilation $\dlop$. This adjunction yields as best approximation
 the  closing $\erop (\dlop (\vct b))\geq \vct b$  which is a lattice projection
that comes optimally close to $\vct b$ from above.

Solving the one-sided equation (\ref{mxgeq}) has direct applications in providing the
system reachability and observability problems with
exact or approximate solutions, as shown in Sec.~\ref{sc-reachobs}. There are also double-sided max-$\mgop$ equations
of the type
\beq
\mtr{A} \mxgmp \vct{x} = \mtr{B} \mxgmp \vct{y}
\label{mxgeq2s}
\eeq
which model synchronization problems and can be solved by iterating the method (\ref{mxgminizsolclog})
between left and right side,  as shown in \cite{CuBu03}. This has been extended in \cite{HCLC09,LiSo01}
to one- and two-sided equations
whose matrix elements are intervals representing numerical uncertainties.

%######################################################
\section{Spectral Analysis in Max-$\mgop$ Algebra}

There has been significant progress on eigenvalue analysis for the max-plus semiring
$(\REAL \cup \{ -\infty\},\vee,+)$; see  \cite{Cuni79,Butk10} and the references therein.
Herein, we extend some of the main results to any scalar clodum\footnote{Although
the main results \cite{Butk10} of max-plus eigenvalue analysis in the max-plus semiring
assume all scalars $< +\infty$, in the more general max-$\mgop$ eigenvalue analysis over a clodum
we allow scalars to equal $\vsetge$; this has direct applications for cloda $\vset=[0,1]$ in fuzzy systems,
like the max-min clodum, where $1=\mgid=\vsetge$.}
%For the max-$\mgop$ eigenvalue analysis we shall use the max-$\mgop$ subalgebra
%$(\vset\setminus \{ \vsetge\},\vee,\mgop)$;
%% , where $\vset_\ell=\vset\setminus \{ \vsetge\}$.
%practically,  we assume that all scalars and matrix elements are less than $\vsetge$.
 $\vset$ even in cases
 where the `multiplications'  do not have inverses. The only constraint on the clodum $\vset$ is to
be  \emph{radicable} w.r.t. operations $\mgop,\dmgop$: namely,
% --- I am not sure that finiteness and uniqueness are needed ----
%for each finite $a\in \vsgroup=\vset\setminus \{\vsetle,\vsetge\}$ and integer $p\geq 1$
%there is a unique $x\in \vsgroup$ such that
%its $p$-fold $\mgop$-multiplication with itself equals $a$,
%i.e. $x^{\mgop p}\defineq x\mgop x \mgop \cdots \mgop x=a$.
for each $a\in \vset$ and integer $p\geq 2$
there is some $x\in \vset$ such that
its $p$-fold $\mgop$-multiplication with itself equals $a$,
i.e. $x^{\mgop p}\defineq x\mgop x \mgop \cdots \mgop x=a$.
Note that both the max-plus clog and the max-min clodum are radicable.

% ---- {\bf Graph of a Matrix:} -----
%For the max-$\mgop$ eigenvalue analysis we shall use the max-$\mgop$ subalgebra
%$(\vset\setminus \{ \vsetge\},\vee,\mgop)$;
%% , where $\vset_\ell=\vset\setminus \{ \vsetge\}$.
%practically,  we assume that all scalars and matrix elements are less than $\vsetge$.
Consider a $n\times n$  matrix  $\mtr{A}=[a_{ij}]$, $n>1$. % $\mtr{A}=[a_{ij}]\in \vset_\vee ^{n\times n}$, $n>1$.
This can be represented
by a {\em directed weighted graph\/} $\gramtr{\mtr{A}}$ that has $n$ nodes
and arcs connecting pairs of nodes $(i,j)$ if the corresponding weights $a_{ij}>\vsetle$.
If $\gramtr{\mtr{A}}$ is strongly connected,
i.e. if there is a path from every node to every other node, %between each pair of nodes in each direction,
then $\mtr{A}$ is called \emph{irreducible}.
Consider a {\em path\/} on the graph, i.e. a sequence of nodes $\boldsymbol{\pi}=(i_0,i_1,...,i_t)$
with length $\pathl (\boldsymbol{\pi})=t$; its weight
is defined by $\pathw (\boldsymbol{\pi}) \defineq a_{i_0i_1}\mgop ...\mgop a_{i_{t-1}i_t}$.
%\[
%\pathl (P)\defineq \mbox{\# arcs on $P$}=t, \; \; \; \;
%\pathw (P) \defineq a_{i_0i_1}+...+a_{i_{t-1}i_t}
%\]
A path $\boldsymbol{\sigma}$ is called a \emph{cycle} if $i_0=i_t$;
the cycle is \emph{elementary} if the nodes $i_0,...,i_{t-1}$ are  distinct.
For any cycle $\boldsymbol{\sigma}$ we  define its \emph{cycle mean}\footnote{For
the max-plus clog $(\EREAL,\vee,\wedge,+)$ the mean of a cycle $\boldsymbol{\sigma}$ is given by
$\pathw (\boldsymbol{\sigma}) / \pathl (\boldsymbol{\sigma})$,
 for the max-product clog $([0,\infty],\vee,\wedge,\times)$
  it is given by $\pathw (\boldsymbol{\sigma})^{1 / \pathl (\boldsymbol{\sigma})}$,
  whereas for the max-min clodum $([0,1],\vee,\wedge,\min,\max)$
  the cycle mean is simply $\pathw (\boldsymbol{\sigma})$.}
 by $\pathw (\boldsymbol{\sigma})^{\mgop (1/\pathl (\boldsymbol{\sigma}))}$.
Let
\beq
\lambda (\mtr{A}) \defineq \bigvee_{\mbox{\rm all cycles $\boldsymbol{\sigma}$ of \mtr{A}}}
 \pathw (\boldsymbol{\sigma})^{\mgop (1/\pathl (\boldsymbol{\sigma}))}
\label{prnceval}
\eeq
be the {\em maximum cycle mean\/} in $\gramtr{\mtr{A}}$.
Since $\gramtr{\mtr{A}}$ has $n$ nodes, only elementary cycles
with length $\leq n$ need be considered in (\ref{prnceval}).
There is also at least one cycle whose average weight
coincides with the maximum value
(\ref{prnceval}); such a cycle is called {\em critical}.
The existence of $\lambda (\mtr{A})$ is guaranteed if $(\vset,\mgop)$ is radicable.

The max-$\mgop$  eigenproblem for the  matrix $\mtr A$  consists of finding
its \emph{eigenvalues} $\lambda$ % $\lambda \in \vset_\vee$
and \emph{eigenvectors} $\vct v \neq \vsetle$ % $\vct v \in \vset_\vee^n$
%(but $\vct v$ not identically equal to $\vsetle$)
such that
\beq
\mtr{A} \mxgmp \vct{v} = \lambda \mgop \vct{v}
\label{eigvvp}
\eeq
%If there exist finite $\lambda$ and $\vct{v}$ satisfying (\ref{eigvvp}),
%then  the eigenproblem is said to be {\em finitely soluble\/} for $\mtr{A}$.
The maximum cycle mean $\lambda (\mtr{A})$ plays a fundamental role in this eigenproblem
for many reasons \cite{Cuni79,Butk10}:
It is the largest eigenvalue of $\mtr A$ and
the only eigenvalue whose corresponding eigenvectors may be finite.
Thus, $\lambda (\mtr{A})$ is called the \emph{principal eigenvalue} of $\mtr A$.
Some further properties include the following.
Define the \emph{metric matrix} generated by $\mtr{A}$ as the  series
\beq
\metmtr{\mtr{A}} \defineq \bigvee _{k=1}^\infty \mtr{A}^{(k)}
\label{metmtrdef}
\eeq
If it converges, it conveys very useful information since its elements equal
the weights of the heaviest paths of any length for all pairs of nodes
(like a  graph of longest distances),
and  its columns can provide eigenvectors \cite{Cuni79,Butk10}.
However, its existence is controlled by $\lambda (\mtr{A})$ as explained by
%
%\begin{Lemma} \label{lm-metmtr}
%
\begin{Theorem} \label{th-metmtr}
Assume a $n\times n$ matrix $\mtr A=[a_{ij}]$ with elements from a radicable clodum $\vset$. % $\vset \setminus \{ \vsetge \}$.
(a)~The infinite series (\ref{metmtrdef}) converges in finite time to a matrix $\metmtr{\mtr{A}}=[\gamma_{ij}]$
if  $\lambda (\mtr{A})\leq \mgid$,
in which case for all $t\geq 1$
\beq
\mtr{A}^{(t)} \leq \metmtr{\mtr{A}} = \mtr{A} \vee \mtr{A}^{(2)} \vee \cdots \vee \mtr{A}^{(n)}
\label{metmtrfin}
\eeq
(b)~If all $a_{ij}<\vsetge$, then both (\ref{metmtrfin}) holds and all $\gamma_{ij}<\vsetge$
if and only if $\lambda (\mtr{A})\leq \mgid$. \\
(c)~If $\lambda (\mtr{A})\leq \mgid$ and $\mtr A$ is irreducible, then all $\gamma_{ij}>\vsetle$.
\end{Theorem}
\emph{Proof:} We extend the results of \cite{Cuni79,Butk10} to a general radicable clodum.
%(a)~Assume first cloda with $\mgid<\vsetge$.
(a)~If $\lambda (\mtr{A})\leq \mgid$, then a path $\boldsymbol{\pi}$ between any nodes $i,j$
of length  $>n$ contains cycles, all whose weights $\leq \mgid$.
By deleting these cycles we can create only heavier subpaths $\boldsymbol{\pi}'$
of length $\leq n$, i.e. $\pathw (\boldsymbol{\pi}')\geq \pathw (\boldsymbol{\pi})$.
Given the finite only number of paths (without cycles) between any nodes $i,j$,
if a path exists, then a heaviest such path also exists with length $\leq n$
and weight $\gamma_{ij}$;
if no path exists, then $\gamma_{ij}=\vsetle$.
(b)~If $\lambda (\mtr{A})\leq \mgid$, then in part (a) we proved convergence in finite time.
Further, since all elements of $\mtr A$ are $<\vsetge$, the finite-length heaviest path
between any nodes $i,j$ will have weight $\gamma_{ij}<\vsetge$.
 If $\lambda (\mtr{A})> \mgid$, then there exists a cycle with weight $>\mgid$ which will drive
at least one element in $\mtr{A}^{(t)}$ unbounded (i.e. $\vsetge$) as $t\rightarrow \infty$
and hence there is no finite convergence.
Thus, (\ref{metmtrfin}) holds iff $\lambda (\mtr{A})\leq \mgid$.
(c)~If $\mtr A$ is also irreducible, i.e. $\gramtr{\mtr{A}}$ is strongly connected,
then a path exists between any nodes $i,j$ and hence each $\gamma_{ij}>\vsetle$.
The above results also cover  the case of  cloda with $\mgid=\vsetge$ (like the max-min clodum)
 because then the condition $\lambda (\mtr{A})\leq \mgid$ always holds.
\hfill Q.E.D.

%
% --------- DUAl EIGENPROBLEM -------------
%
By using duality between the max-$\mgop$ and min-$\dmgop$ matrix subalgebras over a radicable scalar clodum
we can also solve the \emph{dual eigenproblem}
\beq
\mtr{A} \mngmp \vct{v}' = \lambda' \dmgop \vct{v}'
\label{deigvvp}
\eeq
%where all elements of the matrix  $\mtr A$ and the dual eigenvectors as
%well as the dual eigenvalues $\lambda'$  assume values only from
%$\vset \setminus \{\vsetle\}$.
The \emph{dual principal eigenvalue}, denoted by $\lambda'(\mtr{A})$,  is
the smallest of all dual eigenvalues and can be found as
the \emph{minimum cycle mean} of $\mtr{A}$.

%#########################################################
\section{Causality, Stability}

%Assume for brievity single-input single-output  systems.
Assume for brievity SISO  systems. (Our results can be easily extended for MIMO systems.)
Assume also that systems' matrices are constant.
A useful bound  for signals $f(t)$ processed by max-$\mgop$ systems
is their supremal  value $\bigvee _{t} f(t)$.
%Next we define these two signal bounds:
%\[
%B_\vee (f) \defineq \bigvee _{t} f(t),
%\quad B_\wedge (f) \defineq \bigwedge _{t} f(t)
%\]
We call max-$\mgop$ systems % bounded-input bounded-output (BIBO)
\emph{upper-stable} if an upper bounded input and initial condition yields an upper bounded output,
i.e. if
\beq
\vct x(0) < \vsetge \; \mathrm{and}\; \bigvee _t u(t) <\vsetge \Longrightarrow \bigvee_t y(t) <\vsetge
\label{uppstable}
\eeq
%where $\vsetge$ is the greatest scalar.
%
If initial conditions are null and (\ref{uppstable}) is satisfied,
we call the system bounded-input bounded-output (BIBO) upper-stable.
Dually, min-$\dmgop$ systems are called
\emph{lower-stable} if a lower bounded input and initial condition yields a lower bounded output,
i.e. if
\beq
\vct x(0) > \vsetle \; \mathrm{and}\; \bigwedge _t u(t) >\vsetle \Longrightarrow \bigwedge_t y(t) >\vsetle
\label{lowstable}
\eeq
A max-$\mgop$ (min-$\dmgop$) dynamical system with null initial conditions can be viewed
as a DTI (ETI) system  mapping
the input  $u$ to the output which is the sup-$\mgop$ (inf-$\dmgop$) convolution $y=u\sgcnv h$
($y=u\igcnv h'$)
of the input  with the (dual) impulse response $h$ ($h'$).
The following theorem provides us with simple algebraic criteria
for checking the causality and stability of DTI and ETI  systems
based on their impulse response.

\begin{Theorem} \label{th-causal-stabil}
%Consider systems on a CWL over  a scalar clodum $\vset$.
(a1)~Consider a DTI system $\fdlop$ and let
$h=\fdlop (\dimpls )$ be its impulse response. Then:
(a1)~The system is causal iff $h(t)=\vsetle$ for all $t<0$.
(a2)~The system is BIBO upper-stable iff $\bigvee_t h(t) < \vsetge$.
\\
(b)~Consider an ETI system $\ferop$ and let
$h'=\ferop (\eimpls)$ be its dual impulse response. Then:
(b1)~The system is causal iff $h'(t)=\vsetge$ for all $t<0$.
(b2)~The system is BIBO lower-stable iff $\bigwedge_t h'(t) >\vsetle$.
\end{Theorem}
{\em Proof:} Part~(a): (a1)~follows from the definition of causality
since the output can be written as
$\fdlop (u) (t)=\bigvee _k u(t-k)\mgop h(k)$.
(a2)~Sufficiency: If $u$ and $h$ have suprema  $<\vsetge$,
then their dilation $y=u\sgcnv h$ also has a supremum $<\vsetge$ because
\[
y(t)\leq \bigvee _{k}u(k) \mgop \bigvee _{k}h(k), \quad \forall t
\]
Necessity: Assume now that $\fdlop$ is upper-stable.
Then $\bigvee_th(t)$ must be $<\vsetge$, because otherwise we can find a bounded input
yielding an unbounded output. For example, the input
$u(t)=\dimpls (t)$ yields as output $y(t)=h(t)$.
Obviously, this $u$ is bounded,  but if $\bigvee_th(t)=\vsetge$ we get
an unbounded output. Part~(b) follows by duality.
\hfill Q.E.D.

The stability of a linear dynamical system can  be
expressed via the eigenvalues of its state transition matrix $\mtr{A}$.
For max-$\mgop$ (min-$\dmgop$) systems we derive below a conceptually similar result
that links the upper (lower)  stability of the
system with the (dual) principal eigenvalue of $\mtr{A}$.

\begin{Theorem}
(a)~Consider a max-$\mgop$  system  whose matrices
 do not contain any $\vsetge$ elements.
If $\lambda (\mtr{A})\leq \mgid$,
 the system is   upper-stable.
 (b)~If a min-$\dmgop$  system  has matrices
 without any $\vsetle$ elements and $\lambda' (\mtr{A})\geq \dmgid$,
 then the system is lower-stable.
\end{Theorem}
{\em Proof:\/}
% Assume a SISO system. (The proof can be easily extended for MIMO systems.)
(a)~By (\ref{irsiso}), if $\mtr C=[c_i]^T$ and $\mtr B=[b_j]$,
\beq
h(t)=\max_i\max_j c_i \mgop a_{ij}^{(t)}\mgop b_j
\label{irse}
\eeq
where $a_{ij}^{(t)}$ is the $(i,j)$ element of matrix $\mtr A^{(t)}$.
By Theorem~\ref{th-metmtr}, we have $\mtr A^{(t)} \leq \metmtr{\mtr A}=[\gamma_{ij}]$,
and equivalently $a_{ij}^{(t)}\leq \gamma_{ij}<\vsetge$
 for all $i,j,t\geq 1$. Thus,
\beq
\bigvee _t h(t)\leq \max_{i,j}\gamma_{ij}\mgop \max_ic_i \mgop \max_j b_j < \vsetge
\eeq
Hence, by Theorem~\ref{th-causal-stabil} the system is BIBO upper-stable.
%This takes care of
This upper bounds the null-state response $y_{ns}(t)$ of the output.
Now if $\vct x(0)\neq \vsetle$, the null-input response $y_{ni}(t)=\mtr C\mxgmp \mtr A^{(t)}\mxgmp \vct x(0)$
%of the output
will also be upper bounded via a similar proof as above.
Thus, the system is upper-stable.
Part~(b) follows by duality.
 \hfill Q.E.D.

From another viewpoint,
the useful information in a signal $f$ analyzed by a DTI system exists only
at times where $f(t)$ is not null. Thus, its
\emph{support} (or effective domain)
is defined by $\spt_\vee (f)\defineq  \{ t  : f(t) >\vsetle \}$.
An alternative useful bound  for signals $f(t)$ processed by such systems
is their supremal `absolute value' over their support:
\beq
M_f \defineq \bigvee _{t\in \spt_\vee (f)} \mu (f(t))
\eeq
where  $\mu (a)\defineq a\vee \glconj{a}$ is called the \emph{absolute value seminorm}
in \cite{Cuni79} and is `sublinear' over a self-conjugate clodum
 in the sense that $\mu (a\vee b) \leq \mu (a) \vee \mu (b)$.
We call max-$\mgop$ systems  BIBO
\emph{absolutely stable} iff a bounded input yields a bounded output
in the following sense:
\beq
M_u <\vsetge \Longrightarrow M_y<\vsetge
\eeq
This is controlled by the system's impulse response as shown next.

\begin{Theorem}
Consider a DTI system $\fdlop$  over a  self-conjugate clodum
whose matrices do not contain any $\vsetge$ elements.
Let $h=\fdlop (\dimpls )$ be its impulse response. Then,
the system is BIBO absolutely stable iff $M_h <\vsetge$.
\end{Theorem}
{\em Proof:} Sufficiency: If $u$ and $h$ have finite bounds $M_u$ and $M_h$
within their supports $U$ and $H$ respectively,
then their sup-$\mgop$ convolution $y=u\sgcnv h$ is also absolutely bounded because
\[
\mu (y(t))\leq \bigvee _{k\in U\cap \tran{(\refl{H})}{+t}} \mu[u(k)\mgop h(t-k)] \leq
 M_u\mgop M_h
\]
for all $t$ in
the Minkowski set addition $U\dilt H=\{k+\ell:k\in U, \ell \in H \}$ of the two supports,
where $\tran{(\refl{H})}{+t}=\{t-k:k\in H\}$ denotes the reflected $H$ translated by $t$.
Necessity: Assume  that $\fdlop$ is stable.
Then $M_h$ must be finite, because otherwise we can find a bounded input
yielding an unbounded output. For example, the bounded input
$u(t)=\dimpls (t)$ yields the output $y(t)=h(t)$
which is unbounded  if $M_h=\vsetge$.
\hfill Q.E.D.

The next theorem links absolute stability with the principal eigenvalue of the system.

\begin{Theorem} \label{th-stabil-peval}
Consider a max-$\mgop$  system over a clog whose matrices
%$\mtr{A},\mtr{B},\mtr{C},\mtr{D}$
do not contain any $\vsetge$ elements.
For matrix $\mtr{A}=[a_{ij}]$ assume that it is irreducible, $a_{ii}>\vsetle$ for some $i$,
and there is a unique critical cycle of length $d$
corresponding to  its finite principal eigenvalue $\lambda (\mtr{A})$.
Then:
(a) If $\lambda (\mtr{A})=\mgid$,
the impulse response of the system is eventually periodic with period  $d$.
(b) The system is BIBO absolutely stable iff  $\lambda (\mtr{A})=\mgid$.
\end{Theorem}
{\em Proof:\/} (a)~As shown for the max-plus case in \cite{CDQV85}
under the above hypotheses for $\mtr A$,
if $\lambda (\mtr{A})=0$,  then
$\mtr{A}$ is order-$d$-periodical, i.e.  there is an integer $k_0$ such that
$\mtr{A}^{(k+d)}=\mtr{A}^{(k)}$ $\forall k\geq k_0$.
The proof of the above in \cite{CDQV85} can  be extended to  general clogs.
Hence, by (\ref{irsiso}), there exists
$k_0$ such that $h(k+d)=h(k)$ for all $k\geq k_0$.
(b)~Let $\lambda=\lambda (\mtr{A})$. Then $\glconj{\lambda}\mgop \mtr A$ is order-$d$-periodical
and hence
 $\mtr{A}^{(k+d)}=\lambda^{\mgop d} \mgop \mtr{A}^{(k)}$ for all $k\geq k_0$. Hence,
\beq
h(k+d)=\lambda^{\mgop d} \mgop h(k), \; \; \; \forall k\geq k_0
\label{irprd}
\eeq
Further,
the absence of $\vsetge$ values in the system's matrices guarantees
that $h(k)$ does not have any such values. Now, if $\lambda =\mgid$,
then $h(k+d)=h(k)$ $\forall k\geq k_0$ and hence $M_h<\vsetge$.
In contrast, if $\lambda \not = \mgid$, then
(\ref{irprd})  will drive asymptotically (as $k\rightarrow \infty$)
the values of $\mu(h(k))$ unbounded, and hence $M_h=\vsetge$.
 \hfill Q.E.D.

%#############################################################
\section{Reachability, Observability}
\label{sc-reachobs}

Assume single-input single-output systems with constant matrices described by (\ref{sormxgsemtrti}),
acting on a CWL over a clodum $\vset$.
A max-$\mgop$ system is called {\em reachable\/} in $k$-steps if the following
system of nonlinear equations can be solved and provide the  control input sequence
$\vct{u}_k=[u(1),...,u(k)]^T$
required to drive the system from the initial state
$\vct{x}(0)$ to any desired state $\vct{x}(k)$ in $k$ steps:
\beq
\vct{x}(k) = \mtr{A}^{(k)}\mxgmp \vct{x}(0) \ \vee \Cc_k\mxgmp \vct{u}_k
\label{cntrlmxseq}
\eeq
 where $\Cc_k=[\mtr{A}^{(k-1)}\mxgmp \mtr{B},\cdots ,\mtr{A}\mxgmp \mtr{B}, \mtr{B}]$
 is called the \emph{controllability matrix}.
This  system of max-$\mgop$ equations
can be solved using the methods of Sec.~\ref{sc-solmaxeqn}.
However, we can simplify it first by assuming that the input is dominating the initial conditions
 (e.g. by assuming inputs with sufficiently large values);
 then, the second term is greater than the first term of the right hand side,
and we can rewrite (\ref{cntrlmxseq}) as % $\Cc_k \mxgmp \vct{u}_k = \vct{x}(k)$,
% or in a simpler notation
\beq
\Cc_k \mxgmp \vct{u}_k = \vct{x}(k)
\label{wcntrlmxseq}
\eeq
If there is an exact solution to (\ref{wcntrlmxseq}), the system is called \emph{weakly-reachable} \cite{GaKa99}.
Because of some dimensional anomalies in minimax algebra \cite{Cuni79},
there is no guarantee of exact solution even when $\Cc_k$ has adequate column rank\footnote{The column (row) rank
of a matrix over a clodum can be defined as the largest number of max-$\mgop$ independent columns (rows).
In \cite{Butk10,Cuni79,GoMi08} there are also weaker concepts of vector independence in minimax algebra.}
(i.e. $n$ max-$\mgop$ independent columns)
because the max-$\mgop$ span of its columns may be only a subset of $\vset ^n$,
unlike the linear system case where full rank of $\Cc_k$ makes the system reachable in at most $k= n$ steps.
Another difference with linear systems is that the max-$\mgop$ column rank may not be the same with the row rank.
Thus, by using $k>n$ one may obtain a matrix $\Cc_k$ that will give an exact solution.
By  Theorem~\ref{th-mxgeqminiz}, if there exists an exact solution,
the greatest solution is the lattice erosion
\beq
\hat{\vct{u}}_k=\erop (\vct{x}(k)) =\Cc_k ^T \mnasbdmp \vct{x}(k)
\label{cntrloptsln}
\eeq
where $\erop$ is the adjoint erosion of the dilation
$\dlop (\vct{y})=\Cc_k \mxgmp \vct{y}$. (See Sec.\ref{sc-cwlvec}.)
%
%Its optimality can be proven
%simply by noting that  $(\erop , \dlop )$ forms a lattice
%adjunction, and hence
%$\dlop \erop$ is an {\em opening\/} operator.
%Thus, $\dlop (\erop (\vct{x})) \leq \vct{x}$ and $\vct{u}=\erop (\vct{x})$
%is the largest solution with $\dlop (\vct{u}) \leq \vct{x}$.
%
If $\vset$ is a clog, the solution (\ref{cntrloptsln}) becomes
\beq
\hat{\vct{u}}_k = \conjtranmtr{\Cc}_k \mngmp \vct{x}(k)
\label{cntrloptsln2}
\eeq
%A necessary condition to find
%a  solution of (\ref{cntrlmxseq}), i.e., a unique input vector
%to drive the system to the state $\vct{x}$ in $n$ steps is the rank
%of $\Cc$ to be equal to $n$, i.e., the columns of $\Cc$ must be SMI.
However, in certain applications Eq.~(\ref{wcntrlmxseq}) may be too strong
of a condition and it may be sufficient to solve an approximate
reachability problem that has some optimality aspects.
Specifically, consider  finding an optimal control input sequence
$\vct{u}_k$ as solution to the following constrained optimization problem:
%\beq
%\begin{array}{c}
%{\rm Minimize} \; \; || \Cc \mxgmp \vct{u}-\vct{x} || \\
%\mbox{\rm subject to} \; \;
%\Cc \mxgmp \vct{u} \leq  \vct{x}
%\label{cntrloptprb}
%\end{array}
%\eeq
\beq
{\rm Min} \; \; \| \Cc_k \mxgmp \vct{u}_k-\vct{x}(k) \| \; \;
\mbox{\rm s.t.} \; \; \Cc_k \mxgmp \vct{u}_k \leq  \vct{x}(k)
\label{cntrloptprb}
\eeq
where the norm $\|\cdot \|$ is either the $\ell _\infty$ or the $\ell _1$ norm.
Then the optimal  solution is  (\ref{cntrloptsln}) or (\ref{cntrloptsln2}).

\begin{Examples}\ \label{ex-reachabil} {\rm Consider a max-sum system over the max-plus clog  with
\beq
%A=\matr{{rrr}    -4 &-1&-3\\
%	-4&-3 & 0\\
%	1 &-2 &-1}
\mtr A=\left[ \begin{array}{rrr} -4 &-1&-3\\
	-4&-3 & 0\\
	1 &-2 &-1 \end{array}\right],
\quad
%B=\matr{{r}-1\\2\\-1}
\mtr B=\left[ \begin{array}{r}-1\\2\\-1 \end{array}\right]
\eeq
The controllability matrix for $k=5$ steps (shown below)
 has full column rank ($5$ and larger than the row rank):
\beq
\Cc_5
%=\matr{{rrrrr}    -1 &  1 &-2& -1& 1\\
%	2 &-1 & 0  & 2 &  1\\
%	-1&0 & 2 &1 &0}
=\left[ \begin{array}{rrrrr} -1 &  1 &-2& -1& 1\\
	2 &-1 & 0  & 2 &  1\\
	-1&0 & 2 &1 &0 \end{array}\right]
\eeq
(a)~If $\vct x (5)=[1,1,1]^T$ is  the desired state, then this vector
belongs to the max-plus span of the columns of $\Cc_5$ since
\beq
\Cc_5\mxsmp
\left[ \begin{array}{r} -1\\ 0\\ -1\\ -1\\ 0 \end{array}\right]
= \left[ \begin{array}{c} 1\\1\\1 \end{array}\right]
\eeq
Thus, $\hat{\vct u}=[ -1,0,-1,-1,0]^T$  is the greatest solution among all possible 5-step
control sequences that can reach the same state, which have values
$[a,b,-1,d,0]^T$ with $a\leq -1,\ b\leq 0,\ d\leq -1$.
\\
(b)~However, if the desired state is   $\vct x (5)=[-3,3,0]^T$ then this vector
does not belong to the column span of $\Cc_5$. Indeed,
(\ref{cntrloptsln2}) yields $\hat{\vct u}=[ -2,-4,-2,-2,-4]^T$ which is
only a greatest subsolution of (\ref{wcntrlmxseq})  since it can only reach $[-3,0,0]^T$
which is a lower state than desired.
} \end{Examples}

The above ideas  can also be applied to the observability problem.
A max-$\mgop$  system is \emph{observable} if
we can estimate the initial state by observing a sequence of output
values. By (\ref{sormxgsemtrti}), this can be done if the following system
of nonlinear equations can be solved:
\beq
\left[
\begin{array}{c} y(1) \\  \vdots \\ y(k) \end{array}
\right]
 =
\underbrace{ \left[
\begin{array}{c} \mtr{C}\mxgmp \mtr{A} \\  \vdots \\
\mtr{C}\mxgmp \mtr{A}^{(k)}  \end{array}
\right] }_{\dsty \Oo_k}
\mxgmp
\vct{x}(0)
\ \vee \
\left[
\begin{array}{c} y_{ns}(1) \\ \vdots \\ y_{ns}(k) \end{array}
\right]
\eeq
%where
%the null-state response values $y_{ns}(t)$ are given by (\ref{mxsoutr}).
Assuming that the first term of the right hand side containing the
initial state dominates the second term that contains the input
(e.g. by assuming inputs with sufficiently small values),
we can rewrite the above as
\beq
\Oo_k \mxgmp \vct{x}(0) = \vct{y}_k = [y(1),...,y(k)]^T
\eeq
This equation can be solved  exactly or approximately
by using the same methods as for the reachability equation.
Thus, if $\vset$ is a clog, the general solution is
\beq
\hat{\vct{x}}(0) = \conjtranmtr{\Oo_k} \mngmp \vct{y}_k
\eeq
and has the property that it is the largest solution with
$\Oo_k \mxgmp \hat{\vct{x}}(0) \leq \vct{y}_k$.

%###################################################
\section{Applications, Special Cases}

\subsection{Max-Sum systems}

One broad class of nonlinear dynamical systems
is described by (\ref{mxgse}) or (\ref{mngse}) by using the
max-plus clog $(\EREAL,\vee,\wedge,+)$ for scalar arithmetic and
the max-sum $\mxsmp$ and min-sum $\mnsmp$ matrix products (\ref{mxsmpr}),(\ref{mnsmpr}),
which are the basis of  minimax algebra \cite{Cuni79}.
%$\vset =\EREAL$ as the set of scalars, % $\vset =\EREAL =\REAL \cup \{ -\infty,+\infty\}$,
%the standard addition $+$ (extended over $\EREAL$) as the scalar `multiplication' $\mgop$,
%and the following max-sum $\mxsmp$ and min-sum $\mnsmp$ matrix products
%and the max-sum $\mxsmp$ and min-sum $\mnsmp$ matrix products defined in (\ref{mxsmpr}),(\ref{mnsmpr})
%as the general matrix `product' and its dual, respectively.
%\bea
%\mtr{C} = \mtr{A} \mxsmp \mtr{B}  & , &
% c_{ij} = {\dsty \bigvee _{k=1}^n} a_{ik}+b_{kj}
%\label{mxsmpr} \\
%\mtr{C} = \mtr{A} \mnsmp \mtr{B}  & , &
% c_{ij} = {\dsty \bigwedge _{k=1}^n} a_{ik}+'b_{kj}
%\label{mnsmpr}
%\eea
%where $+$ and $+'$ are identical for finite reals, but $a+(-\infty)=-\infty$
%and $a+'(+\infty)=+\infty$ for all $a\in \EREAL$.
%Over $\EREAL$ we use two operations $+$ and $+'$ which are identical for finite reals,
%but $a+(-\infty)=-\infty$ and $a+'(+\infty)=+\infty$ for all $a\in \EREAL$.
%Henceforth, we shall use the same symbol $+$ for both scalar operations.
%
Special cases of max-sum or min-sum dynamical systems
have been used for modeling, control and optimization in
(i)~discrete event dynamical systems (DES)
for  applications including  scheduling,  manufacturing and transportation,
(ii)~shortest path and related dynamic programming problems,
 and (iii)~operations research;
 see
\cite{BCOQ01,Butk10,CaLa99,CDQV85,CGQ04,DoKa95,GaKa99,HOW06,Kame93,KaDo94}
and the references therein.
 %(see \cite{CaHo90,CaLa99} for surveys of DES)

Next,  we examine state-space formulations and stability issues
for two classes of max-sum or min-sum dynamical systems   modeling recursive nonlinear filtering
and shortest path computation, which can be described by generalized versions of
the max-sum recursion (\ref{mxsrecurs}) or its dual.

%+++++++++++++++++++++++++++++++++++++++++++++++++++++++

\subsubsection{State-Space Models of Recursive Nonlinear Filters}

A very large class of discrete linear time-invariant
 systems used in control and signal processing \cite{Brog74,OpSc89}
% can be described by constant-coefficient linear difference equations.
is modeled via the following class of linear difference equations:
\beq
y(t) = \sum _{i=1}^n a_iy(t-i)+ \sum _{j=0}^m b_ju(t-j)
\label{lde}
\eeq
%A large class of linear discrete-time filters is described
%by linear difference equations.
Replacing sum with maximum and multiplication with addition
gives us the following nonlinear {\em max-sum difference equation}
\cite{Mara94a}
\beq
y(t) = \left( \bigvee _{i=1}^n a_i+y(t-i) \right)
\vee \left( \bigvee _{j=0}^m b_j+u(t-j) \right)
\label{mxsde}
\eeq
%capable of modeling a large class of
%morphological systems used in nonlinear filtering
%and image analysis. % \cite{MaSc90,Serr82}.
The signal values and all coefficients $a_i,b_j$ are from the max-plus clog.
If some $a_i=-\infty$, the term with $y(t-i)$ is not used in the equation.
%$n$ is the order of the equation, assuming $a_n>-\infty $.
Special (mainly non-recursive) cases of such  nonlinear difference equations
have found many applications in
morphological signal and image processing \cite{Heij94,MaSc90,Serr88,Ster86},
convex analysis \cite{Luce10,Rock70}, and optimization \cite{BeKa61,BeKa63a}. % \cite{BeKa61,BeKa63b}

The max-plus version of the general state equations (\ref{mxgse})
can model the dynamics of recursive  discrete-time filters
 described by the above max-sum difference equation.
Specifically, if $m=0$,
%setting $x_1(k)=y(k-n)$, $x_{2}(k)=y(k-n+1)$, ..., $x_n(k)=y(k-1)$
setting $x_i(t)=y(t-n+i-1)$, $i=1,...,n$,
and choosing matrices
%$A,B,C,D$ appropriately in terms of
%the coefficients $a_i$ and $b_j$
\begin{eqnarray}
\mtr{A} & = & \left[ \begin{array}{ccccc}
-\infty & 0 & -\infty & \ldots & -\infty \\
-\infty & -\infty & 0 & \ldots & -\infty \\
\vdots & \vdots & & & \vdots \\
-\infty & -\infty & -\infty & \ldots & 0 \\
a_n & a_{n-1} & a_{n-2} & \ldots & a_1
\end{array} \right] \; , \;
B = [b_0]
%\; , \;
%\mtr{B} = \left[ \begin{array}{c}
%-\infty \\ -\infty \\ \vdots \\ -\infty \\ b_0
%\end{array} \right]
\nonumber \\
\mtr{C} & = &  [a_n, \ldots, a_1]
\; , \quad
D = [b_0]
\label{ssmxsde}
\end{eqnarray}
models (\ref{mxsde}) as a max-sum special case of (\ref{mxgse}).

Consider now the following \emph{min-sum difference equation},
which describes a dual system to that of (\ref{mxsde}):
\beq
y(t) = \left( \bigwedge _{i=1}^n a_i+y(t-i) \right)
\wedge \left( \bigwedge _{j=0}^m b_j+u(t-j) \right)
\label{mnsde}
\eeq
Its dynamics can be modeled by the min-sum version of the general
 state equations (\ref{mngse}).
 For $m=0$, it admits a state space model as in (\ref{ssmxsde}),
 the only difference being that the null elements  in
the system matrices should be $+\infty$.

%As shown in \cite{Mara94a},
The system described by (\ref{mxsde}) or (\ref{ssmxsde}) is a dilation time-invariant (DTI)
system
iff  all its initial conditions are null $(-\infty)$ and  is initially at rest,
i.e. if $u(t)=-\infty$ for $t\leq t_0$ then $y(t)=-\infty$ for $t\leq t_0$.
Similar conditions apply for (\ref{mnsde}) to make it correspond to an
 erosion time-invariant (ETI) system.

\begin{figure}[!h]
\centerline{ \psfig{figure=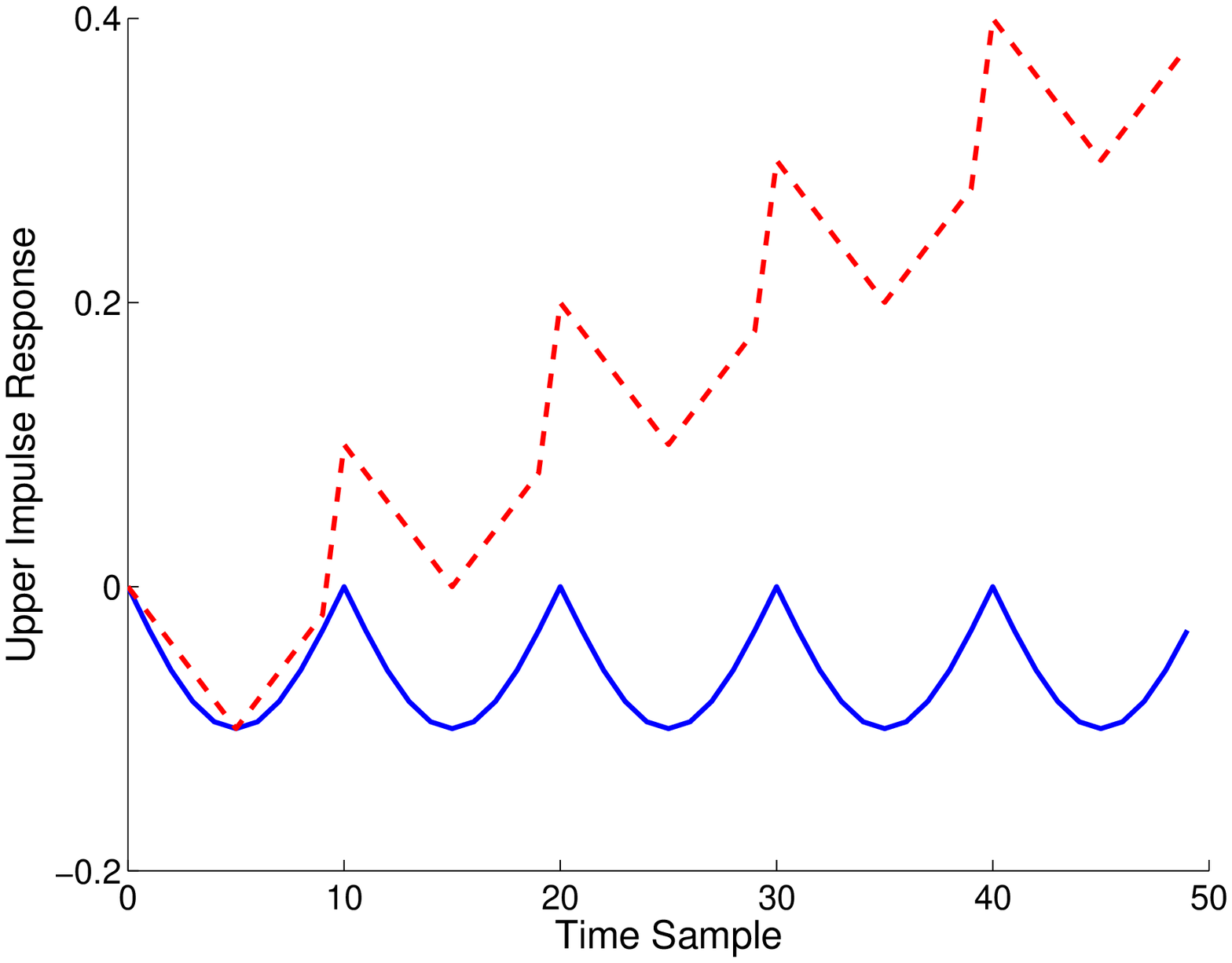,width=5.0cm}
 \psfig{figure=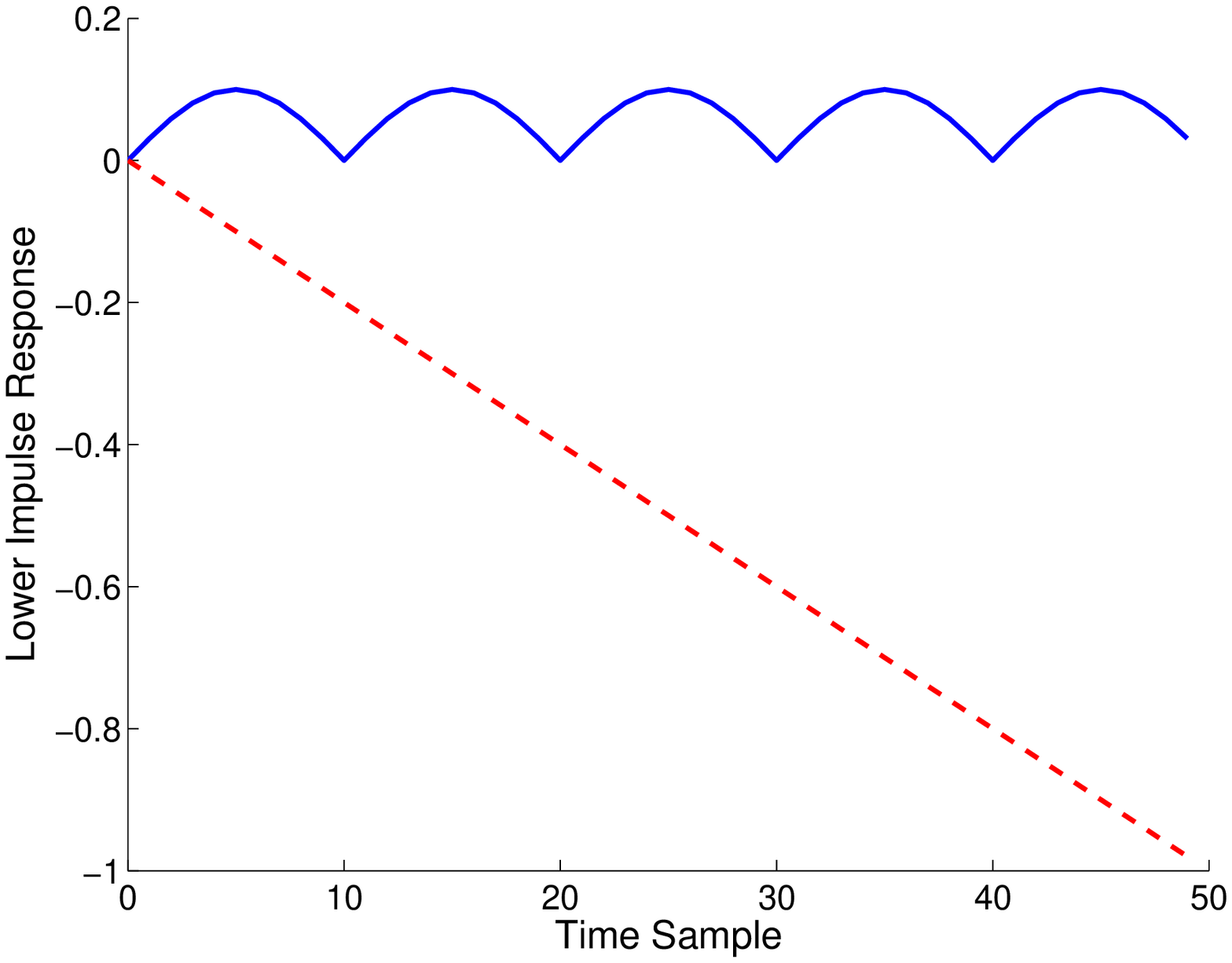,width=5.0cm} }
\centerline{(a)\hspace{5.0cm} (b)}
\centerline{ \psfig{figure=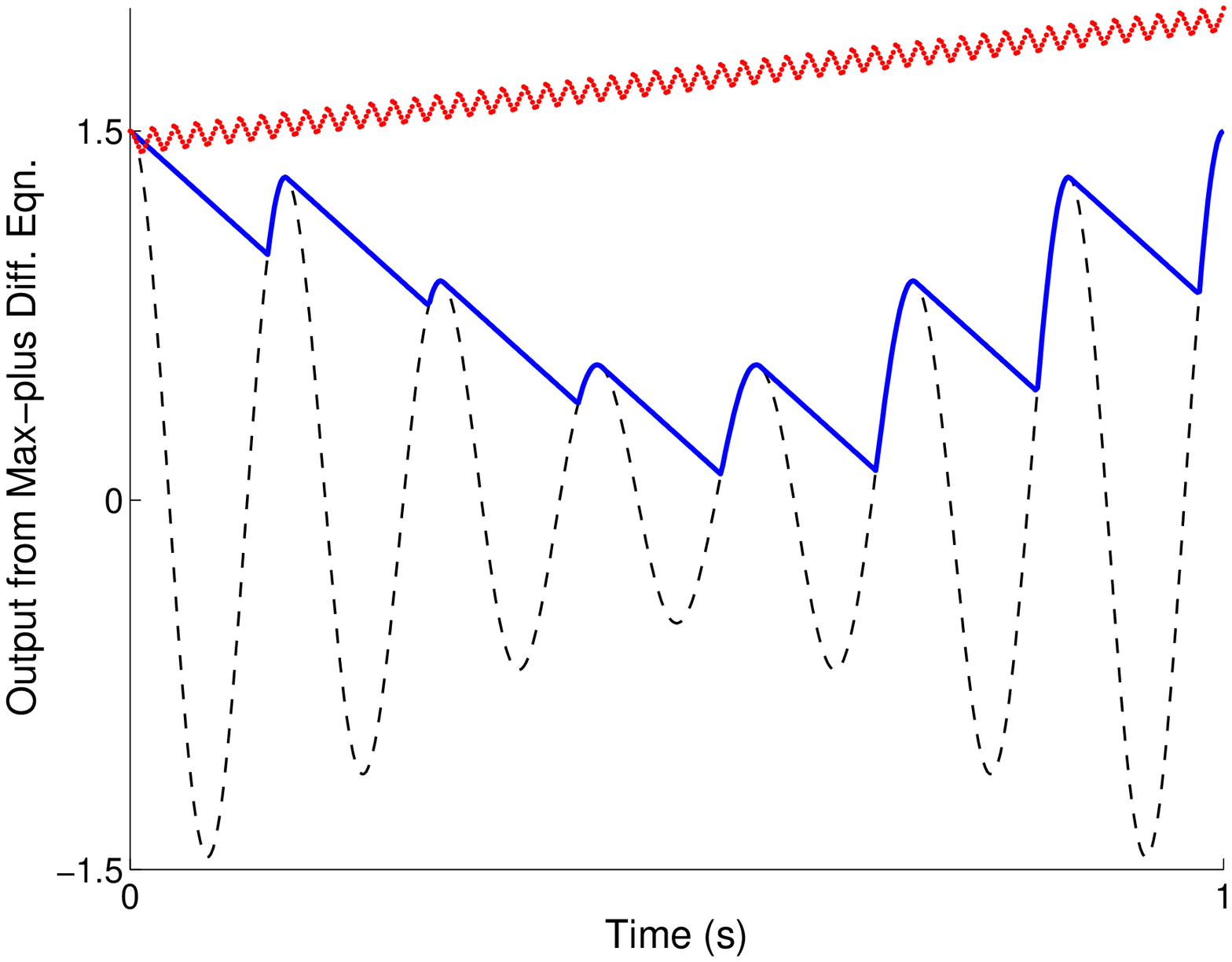,width=5.0cm}
 \psfig{figure=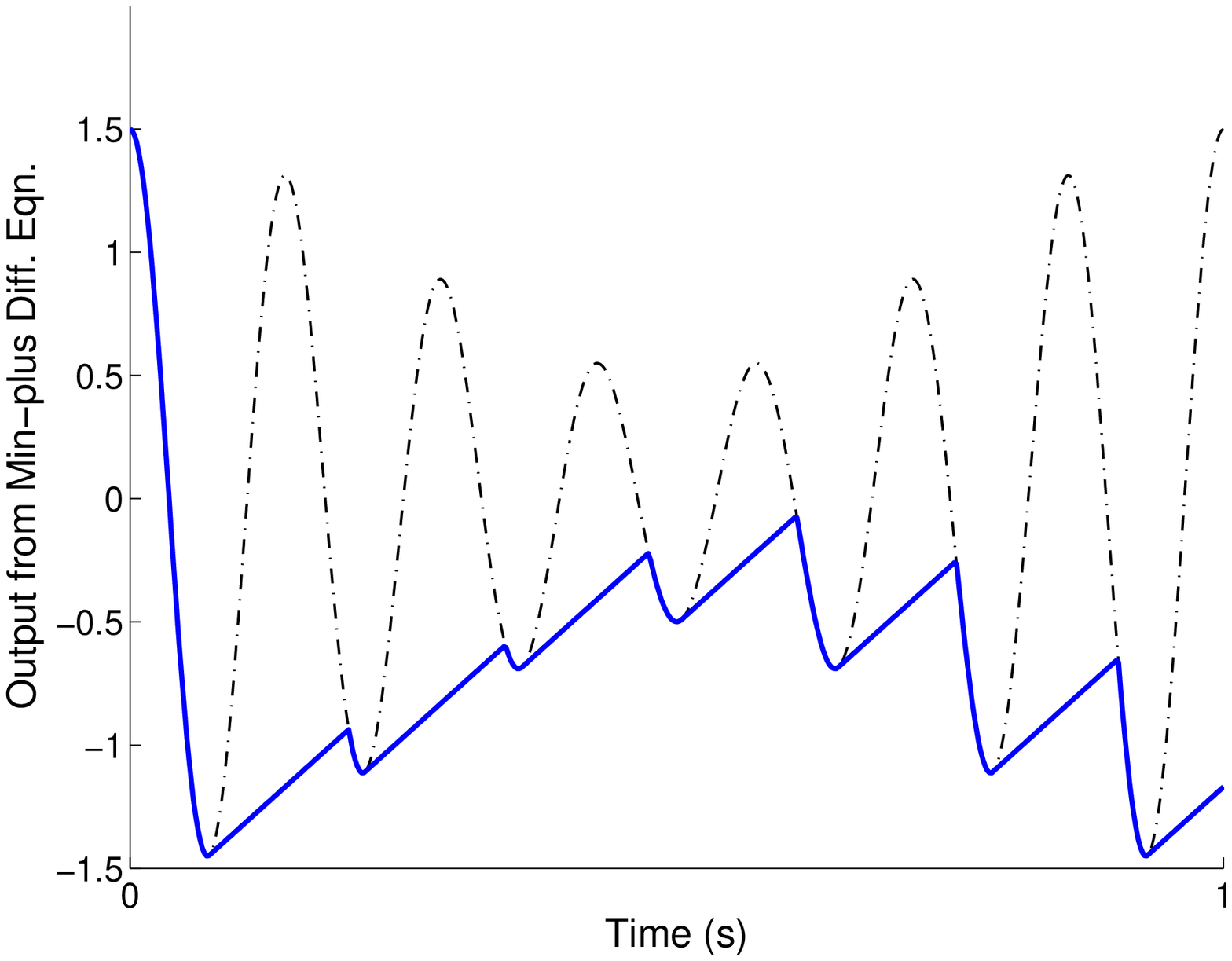,width=5.0cm} }
\centerline{(c)\hspace{5.0cm} (d)}
\caption{Responses of DTI or ETI systems described by the recursive max-sum  equation (\ref{mxsde})
or its min-sum version (\ref{mnsde}); in all cases $m=0$, $b_0=0$.
(a)~Impulse response (first 50 samples) of a $n=11^{th}$-order  DTI
 system for two coefficient sequences $\{ a_k\}$: in solid line $a_k=-\sin (\pi (k-1)/10)/10$ for $k=1,..,10$ and $a_{11}=0$,
 whereas  in dash line $a_k'=(|k-6|-5)/50$ for $k=1,..,10$ and $a_{11}'=0.1$.
(b)~Dual impulse response (first 50 samples) of a $11^{th}$-order recursive ETI
 system  for two coefficient sequences:
in solid line $a_k=\sin (\pi (k-1)/10)/10$ for $k=1,..,10$ and $a_{11}=0$,
 whereas in dash line $a_k'=(|k-6|-5)/50$ for $k=1,..,11$.
%(c)~Output signals from two DTI systems for an AM input signal (dashed line).
(c)~Output signals from two DTI systems whose input (dashed line) is an amplitude-modulated sine.
The first output (solid blue line) is from the stable system $y(t)=\max [y(t-1)+a_1,u(t)]$ with $a_1=-0.008$.
The second output (dotted red line) is from the unstable system that generated the unstable impulse response of (a).
(d)~Input signal as in (c) and output from the min-sum system $y(t)=\min [y(t-1)-a_1,u(t)]$.
}
\label{fg-mxsirout}
\end{figure}

\begin{Theorem}
 The max-plus principal eigenvalue of the matrix
$\mtr A$ in (\ref{ssmxsde}) is equal to $\prnceval{\mtr{A}}  = \bigvee _{k=1}^n a_k/k$.
\end{Theorem}
\emph{Proof}: \ The directed weighted graph of $\mtr A$ has $n$ nodes and $n$ elementary cycles
$(j,j+1,...,n,j)$ for $j=1,...,n$,  each with average weight $a_{n-j+1}/(n-j+1)$.
Hence, $\prnceval{\mtr{A}}  = \bigvee _{k=1}^n a_k/k$.
\hfill Q.E.D.
\\
%
%The max-sum system corresponding to the recursive nonlinear filter
%described by (\ref{mxsde}) has a principal eigenvalue equal to
%$\prnceval{\mtr{A}}  = \bigvee _{k=1}^n a_k/k$. Hence, for
%this system to be stable, all the coefficients $a_k$ must
%be non-positive and at least one of them must be zero.
%
Thus, the max-sum system corresponding to the recursive nonlinear filter
described by (\ref{mxsde}) is upper stable iff all the coefficients $a_k$
are non-positive and absolutely stable if additionally at least one of them is zero.
Such a numerical example is shown in Fig.~\ref{fg-mxsirout}(a), where
Theorem~\ref{th-stabil-peval} also applies and predicts a periodic impulse response.
Further, responses from stable and unstable DTI and ETI systems are shown in Fig.~\ref{fg-mxsirout}.
The stable outputs of Figs.~\ref{fg-mxsirout}(c,d) illustrate the applicability
of recursive DTI (ETI) for upper (lower) envelope detection, as explored in \cite{Mara94a}.

\subsubsection{Dynamic Programming}

The max-sum or min-sum recursive equations can also express various forms of dynamic programming,
either of maximizing some gain or minimizing some cost or distance \cite{Bert12,BCOQ01}.
For example consider (\ref{mxsrecurs}) and assume that $a_{ij}$ is the transition gain from state $i$ to state $j$
between two consecutive time instants
and that $x_i(t)$ represents the maximum possible gain to reach state $i$ in $t$ steps
starting from some initial state at $t=0$.
Then (\ref{mxsrecurs}) with a transposed transition matrix,
i.e. the max-sum system
\beq
\vct x(t)=\mtr A^T \mxsmp \vct x(t-1), \quad \vct x (0)=[0,-\infty,...,-\infty]^T,
\eeq
models a dynamic programming algorithm where, starting from state 1 with zero gain,
we move from state to state aiming at solving the above optimization problem by
sequentially maximizing the gain.
The optimum path can be found by backtracking.

Instead of max-sum, there is also a max-product example of dynamic programming
presented in Sec.~\ref{sc-viterbi-dp}.
Other abstract models of dynamic programming  have been studied in \cite{VePo87}.

\subsubsection{Distance maps and min-plus recursions}

\begin{figure}[!ht]
\centerline{
\psfig{figure=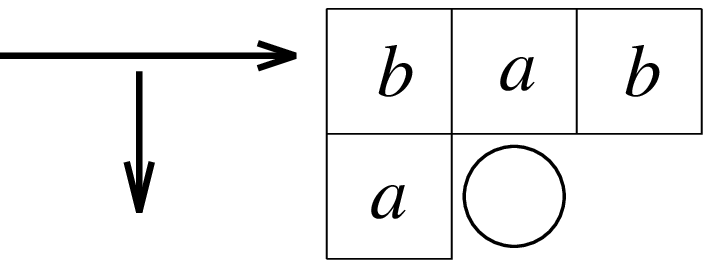,width=4.0cm}(a)
\hspace{2mm}
\psfig{figure=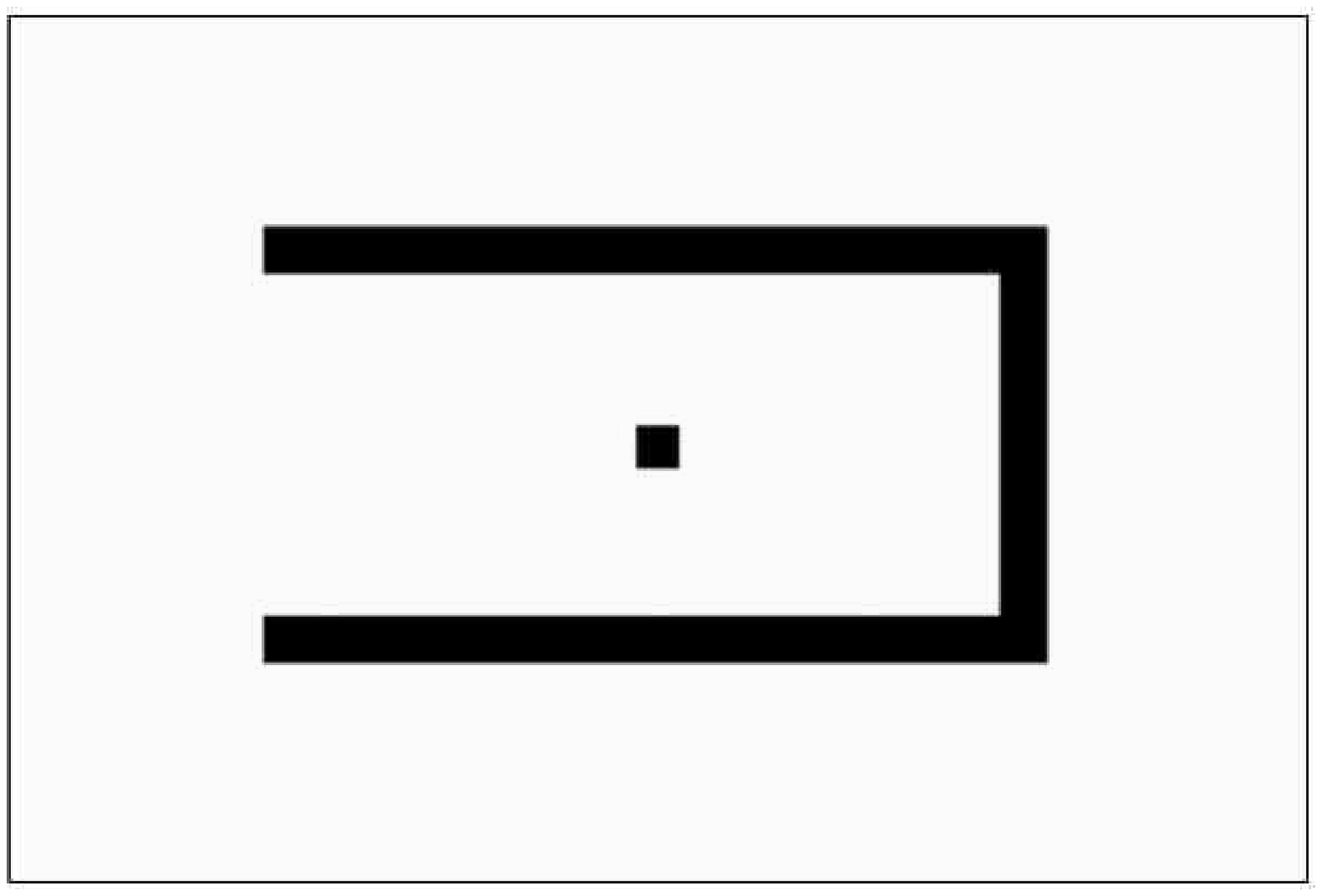,width=4.0cm}(b)
}
\centerline{
\psfig{figure=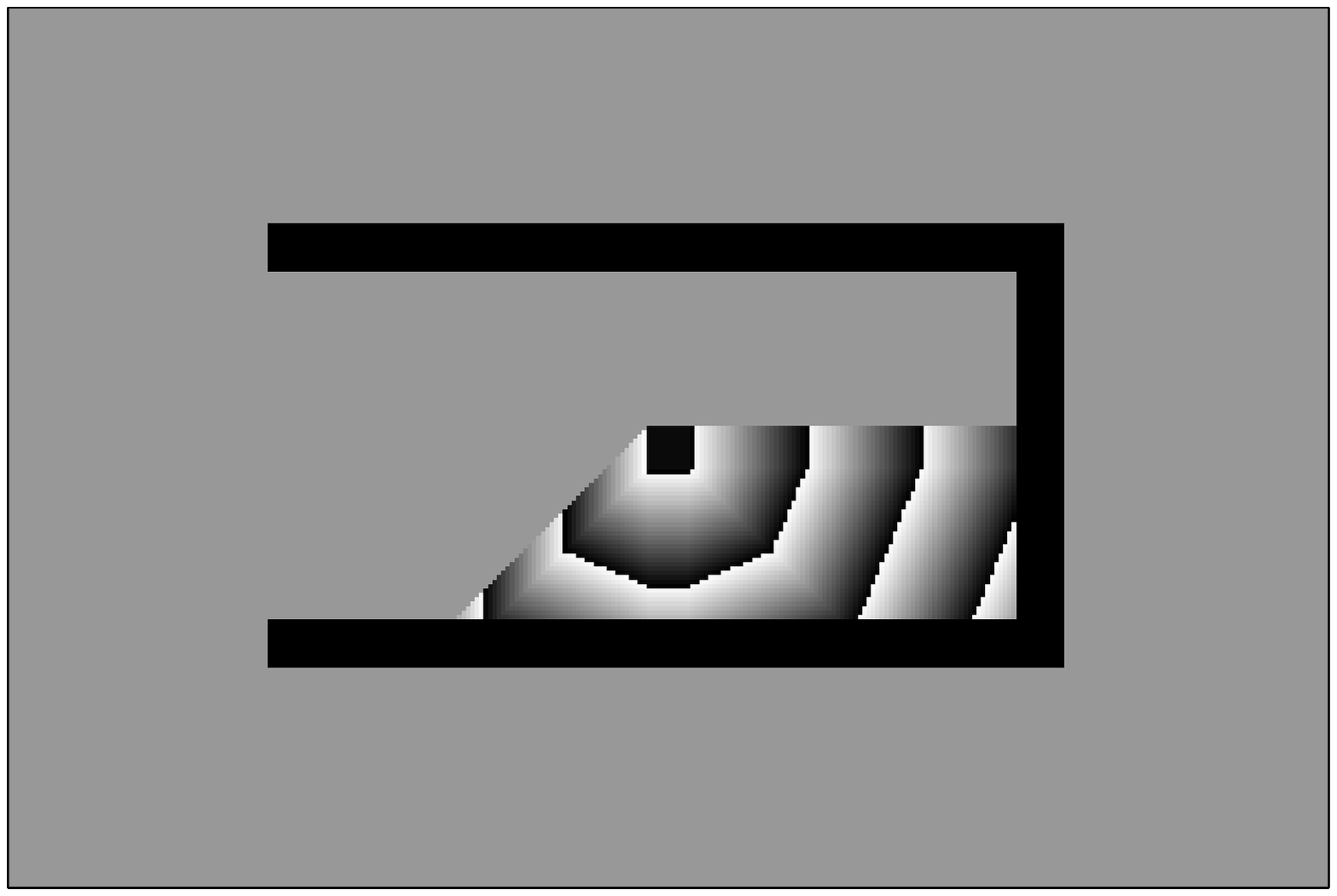,width=4.0cm}(c)
\psfig{figure=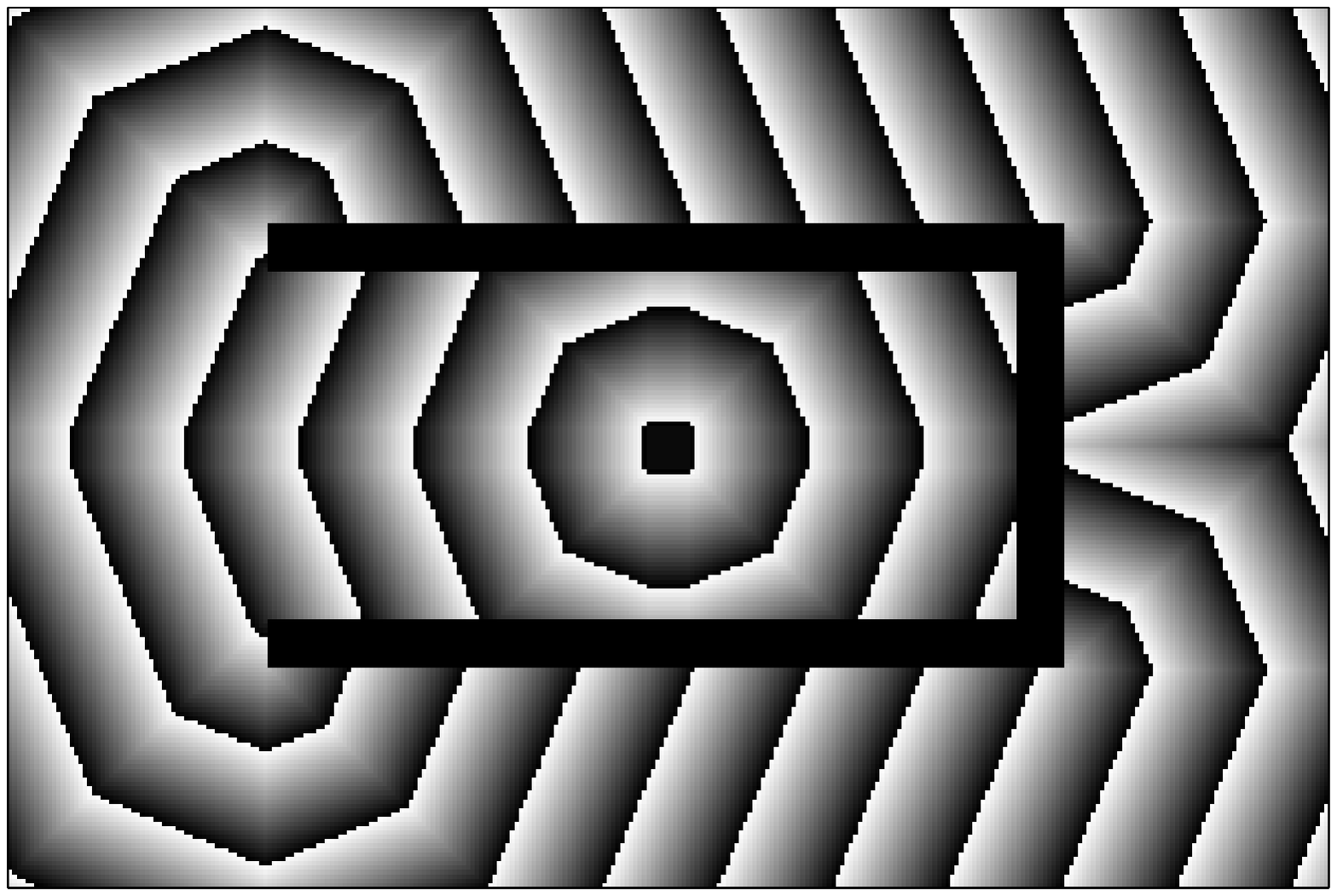,width=4.0cm}(d)
}
\caption{
(a)~Coefficient submask  for forward pass of sequential distance transform.
(b)~Source set $S$ and the obstacle wall set $W$.
(c)~First (forward) pass of
constrained distance transform with steps $(a,b)=(24,34)/25$.
(d)~Fourth (backward) pass and final result, shown with gray values modulo a constant.
}
\label{fg-cmfmask-cdt}
\end{figure}

%The max-plus or min-plus recursive equations can also express various forms of dynamic programming,
%either of maximizing some gain or minimizing some cost or distance.
The min-sum version of (\ref{mxsrecurs}) models shortest path problems.
Given a 2D rectangular field $f:\Vv \rightarrow \REAL$ on a grid $\Vv$ of $M\times N$ pixels,
its weighted distance transform is defined by
\beq
D_f(i,j)=\bigwedge_{(k,\ell)\in \Vv} d(i-k,j-\ell)+f(k,\ell)
\eeq
where $d(\cdot )$ is the Euclidean  distance.
For various cases of $f$, the above distance computation problem is at the heart of
several well-known optimization problems \cite{FeHu04a},\cite{Tsit95}.
If  $D_f$ is available, we can solve the shortest path problem from any point by following the gradient
 of the distance map.
 %
%If $f$ is two-valued and equals $\eimpls _S$
%where the latter function equals 0 on a set $S\sbs \Vv$ and $+\infty$ on $\Vv\setminus S$,
If $f$ equals $\eimpls _S$, which is the lower indicator function of a set $S\sbs \Vv$
with values 0 on $S$ and $+\infty$ on $\Vv\setminus S$,
then $D_f$ becomes  the \emph{distance transform} of the set $S$:
\beq
D_S(i,j)=\min_{(k,\ell)\in S}\| i-k,j-\ell\|
\eeq
which measures distances from $S$ out into its containing field.
Consider indexing rowwise the 2D rectangular grid $\Vv$ of $M\times N$ pixels $(i,j)$
  as a 1D sequence of points $t=N(i-1)+j$, $i=1,...,M, j=1,...,N$.
A good approximation to the Euclidean  distance function
$D_S(t)$ is to compute the chamfer distance \cite{Borg84} by propagating
a $3\times 3$ mask (8-pixel neighborhood) of local distance steps $(a,b)$.
%\begin{eqnarray}
%d_{n+1}[i,j] & = & \bigwedge_{(k,\ell) \in N} d_{n}[i-k,j-\ell] + w_{k\ell},
%\quad d_0=\eimpls _S
%\label{cdtparitr}
%\\
%d_S & = & \lim _{n\rightarrow \infty} d_n =d
%\label{cdtparlim}
%\end{eqnarray}
A serial implementation is an iterative  algorithm
where the 8-pixel neighborhood  is partitioned into two 4-pixel subneighborhoods,
and each new array of results  sequentially passes through  recursive
infimal convolutions $y_i(t)$, $i=1,2,3,...$, which for odd $i$ are a
forward pass  with the submask of Fig.~\ref{fg-cmfmask-cdt}(a)
scanning rowwise the 2D field from top to bottom
and for even $i$  are a backward pass with the reflected submask   in the reverse
 scanning order. The $i$-th forward pass is described by the min-sum difference equation
\beq
y_i(t)  = [ \bigwedge _{k=1}^{N+1} w_k+y_{i}(t-N+k-2) ] \wedge u_{i-1}(t)
\label{cdtserf}
\eeq
where $w_1=b,w_2=a,w_3=b,w_{N+1}=a$ and all other $w_k$ are $+\infty$, $u_0 = \eimpls _S$
and $u_i=y_{i}$ for $i\geq 1$.
%For each iteration $i$, this is a \emph{min-sum difference equation}, which describes
%This corresponds to  a dual system  of (\ref{mxsde}).
Its dynamics can be modeled by the min-sum version of the general
 state equations (\ref{mngse}).
 It admits a state space model as in (\ref{ssmxsde})
 with $n=N+1$ states  $x_k(t)=y(t-N+k-2)$, the
only differences being that the null elements ($-\infty$) in
the sparse system matrices should be replaced with $+\infty$, and all elements in the last row
of $\mtr A$ and in $\mtr C$ are $+\infty$ except at four positions ($k=1,2,3,N+1$) where
they are equal to the corresponding local distances.
The source set $S$ could be a small region from which we propagate distances; see Fig.~\ref{fg-cmfmask-cdt}(b,c).
If the field contains impenetrable obstacles (like `walls') $W$,
distance maps can be produced that  account for this impenetrability,
and then  shortest paths can be found that avoid collision with the walls,
which is useful in robotics \cite{VDVG86}. This can be done by imposing in each iteration
values $+\infty$ at all points of the wall $W$.
The algorithm (\ref{cdtserf}) generally converges to $D_S(t)=\lim y_i(t)$,
and the number of required  passes is two
 if there are no obstacles; see Fig.~\ref{fg-cmfmask-cdt}(d).
 For a 1D sequence $S$ of points,  we need only two passes as the following example  illustrates
 with recursions
 $y_1(t)=\min[y_1(t-1)+1,u_0(t)]$ and $y_2(t)=\min[y_2(t+1)+1,y_1(t)]$:
 \[
 \begin{array}{|c|c|c|c|c|c|c|c|c|c|c|c|c|} \hline % 12 pt sequence
 u_0 & \infty & 0 & \infty & \infty & \infty  & 0 & \infty & \infty & \infty & \infty & \infty & 0  \\ \hline
 y_1 & \infty & 0 & 1 & 2 & 3  & 0 & 1 & 2 & 3 & 4 & 5 & 0 \\ \hline
 y_2 & 1 & 0 & 1 & 2 &  1 & 0 & 1 & 2 & 3 & 2 & 1 & 0  \\
 %\begin{array}{|c|c|c|c|c|c|c|c|c|c|c|c|c|c|} \hline  % 13 pt sequence
% u_0 & \infty & 0 & \infty & \infty & \infty & \infty & 0 & \infty & \infty & \infty & \infty & \infty & 0  \\ \hline
% y_1 & \infty & 0 & 1 & 2 & 3 & 4 & 0 & 1 & 2 & 3 & 4 & 5 & 0 \\ \hline
% y_2 & 1 & 0 & 1 & 2 & 2 & 1 & 0 & 1 & 2 & 3 & 2 & 1 & 0  \\
 \hline \end{array}
 \]
 Note that both recursive equations are stable min-plus systems.

\subsection{Max-Product systems}

Another class of nonlinear dynamical systems is obtained
by using the nonnegative numbers $\vset =[0,+\infty]$ as scalars,
the standard product ($\times$) as scalar `multiplication',
and the following max-product $\mxpmp$ and its dual $\mnpmp$
as generalized matrix `multiplications' in (\ref{mxgse}) and (\ref{mngse}):
\bea
 \mtr{C} = \mtr{A} \mxpmp \mtr{B}=[c_{ij}]  & , &
    c_{ij} = {\dsty  \bigvee _{k=1}^n} a_{ik}\times b_{kj}
\label{mxpmpr}
\\
\mtr{C} = \mtr{A} \mnpmp \mtr{B}=[c_{ij}]  & , &
 c_{ij} = {\dsty \bigwedge _{k=1}^n} a_{ik}\times' b_{kj}
\label{mnpmpr}
\eea
%For the dual matrix `product' $\mnpmp$ we need to replace $\vee$ with $\wedge$
%and $\times$ with $\times'$, where $\times$ and $\times'$ coincide over $(0,+\infty)$,
The scalar multiplications $\times$ and $\times'$ coincide over $(0,+\infty)$,
but $a\times 0=0$ and $a\times '(+\infty) =+\infty$ for all $a\in [0,+\infty]$.
Henceforth, we shall use the same symbol $\times$ for both scalar operations.
Here the scalar arithmetic is based on  the max-times clog $([0,\infty],\vee,\wedge,\times)$.
This max-product formalism can model dynamical
systems whose inputs, states, and outputs  are constrained to
be nonnegative.
Note that there is an isomorphism between the max-sum and the max-product
 systems because, if we have the following max-product
state equations
\beq
\vct{x}(t) =  \mtr{A}\mxpmp \vct{x}(t-1) \vee \mtr{B}\mxpmp \vct{u}(t)
\eeq
and take  logarithms of both sides element-wise, we obtain the max-sum equations
\beq
\log \vct{x}(t) = \log \mtr{A} \mxsmp \log \vct{x}(t-1)
\vee \log \mtr{B} \mxsmp \log \vct{u}(t)
\eeq
Such systems have found applications in speech recognition and other natural language processing tasks
using finite-state automata \cite{MPR02,HoNa13},
in computer vision \cite{FeHu04a},
the max-product algorithm in belief propagation \cite{Pear88} and related probabilistic graphical models
 used in machine learning \cite{Bish06}.
%They also have a great potential for image processing due to the
%nonnegativity of the signals involved.

\subsubsection{Viterbi Algorithm and HMMs}
\label{sc-viterbi-dp}

Given a time sequence of observations (feature vectors) $\mtr O=(\vct o_t)_{t=0}^T$,
a fundamental problem in their statistical modeling using  hidden Markov models (HMMs)  \cite{RaJu93}
with $n$ discrete states $\{ 1,...,n\}$
 is to find the best sequence of states $\hat{\vct s}=(s_0,s_1,...,s_T)$ that maximizes the probability
 $\mathrm{Pr}(\mtr O,\vct s|\boldsymbol{\theta})$,
 where $\boldsymbol{\theta}=([\pi_i],[a_{ij}],[p_i])$ are the HMM parameters:
 $\pi_i$ are the initial state probabilities at $t=0$,
$a_{ij}=\mathrm{Pr}(s_t=j|s_{t-1}=i)$ are state transition probabilities, and $p_i(t)$ are the
state-conditional observation probabilities $p(\vct o_t|s_t=i)$ often modeled by Gaussian Mixture models (GMMs).
 Consider the highest probability of a single partial state sequence ending at state $i$ at time $t$
 and accounting for the first $t+1$ observations:
\beq
x_i(t)=\max_{s_0,...,s_{t-1}}\mathrm{Pr}[s_0,...,s_{t-1},s_t=i,\vct o_0,...,\vct o_t|\boldsymbol{\theta}]
\eeq
%where $\boldsymbol{\theta}=([\pi_i],[a_{ij}],[p_i])$ are the HMM parameters: $\pi_i$ the initial state probability,
%$a_{ij}=\mathrm{Pr}(s_t=j|s_{t-1}=i)$ are state transition probabilities, and $p_i$ are the
%state-conditional continuous densities $p_i(\vct o_t)$ represented by Gaussian Mixture models (GMMs).
One solution is to use the Viterbi algorithm to find the max global score
%$\hat{P}=\mathrm{Pr}(\mtr O,\vct s|\boldsymbol{\theta})=\max_i x_i(T)$
\beq
\hat{P}=\mathrm{Pr}(\mtr O,\hat{\vct s}|\boldsymbol{\theta})=\max_i x_i(T)
\label{viterbiscore}
\eeq
and then find the optimal state sequence via backtracking.
This is essentially dynamic programming and amounts to evolving
the following system, for $t=1,...,T$,
\beq
\begin{array}{rcl}
x_i(t) & = & \left( \bigvee _{j=1}^n a_{ji} x_j(t-1) \right)\cdot  p_i(t) \\
y(t) & = & \bigvee _{i=1}^n x_i(t)
\end{array}
\label{mpseviterbi}
\eeq
with $x_i(0)=\pi_i p_i(0)$.
Then, this is a max-product system with matrices $\mtr A(t)=[a_{ji}]p_i(t)$, $\mtr C=[1,1,...,1]$ and zero input.
The Viterbi score is given by the final output $\hat{P}=y(T)$.

\subsubsection{Attention Control and Multimodal Saliencies}

Assume a  video sequence of audio-visual (AV) events each to be scored with some degree of saliency in $[0,1]$
where `saliency' is some bottom-up low-level sensory form of  attention by a human watching this video.
The states $x_1,x_2,x_3,x_4$ represent time-evolving mono- or multi-modal saliencies,
where 1=audio, 2=visual, 3=audiovisual, and 4=non-salient.
Peaks in these saliency trajectories signify important events, which can be automatically detected
and produce video summaries that agree well with human attention \cite{Eva+13}.
The following state equations are a possible
max-product dynamical model
we have proposed for the evolution of these saliency states \cite{MaKo15}:
%
%\beq
%\left[ \begin{array}{c} x_1(t) \\ x_2(t) \\ x_3(t) \\ x_4 (t) \end{array} \right]
%=
%\left[ \begin{array}{cccc}
%a_{11}p_1(t) & a_{12}p_1(t) & a_{13}p_1(t) & a_{14}p_1(t) \\
%a_{21}p_2(t) & a_{22}p_2(t) & a_{23}p_2(t) & a_{24}p_2(t) \\
% a_{31}p_3(t) & a_{32}p_3(t) & a_{33}p_3(t) & a_{34}p_3(t) \\
%a_{41}p_4(t) & a_{42}p_4(t) & a_{43}p_4(t) & a_{44}p_4(t)  \end{array} \right]
%\mxpmp \left[ \begin{array}{c} x_1(t-1) \\ x_2(t-1) \\ x_3(t-1) \\ x_4 (t-1) \end{array} \right]
%\bigvee
%\mtr{B} \mxpmp
%\left[ \begin{array}{c} u_1(t) \\ u_2(t) \\ u_3(t) \\ u_4(t) \end{array} \right]
%\label{mpseviterbi}
%\eeq
%
\beq
x_i(t) = \left( \bigvee _{j=1}^4 a_{ji} x_j(t-1) \right)\mgop p_i(t) \vee
\left( \bigvee _{j=1}^4 b_{ij} u_j(t) \right)
\label{mpsevitscal}
\eeq
for state $i=1,2,3,4$.
The constants $a_{ij}$ represent state transitions probabilities and
$p_i(t)=p(\vct o_t|s_t=i)$ denote the % state-conditional
probabilities of observed  low-level feature vectors $\vct o_t$
while being at the $i$-th saliency state.
We assume that the parameters $a_{ij},\ b_{ij}$ and $p_i(t)$ are given.

Given a time sequence of such observations $(\vct o_t)$ one can fit HMMs to  these data
using maximum likelihood. Then,
the first term in the RHS of (\ref{mpsevitscal}) models the evolution of the Viterbi dynamic programming
algorithm (\ref{mpseviterbi}) used in  HMMs.
%for optimal  state estimation,
%if we initialize at $t=0$ the four states by setting $x_i(0)=\pi_i p_i(0)$.
%where $\pi_i$ denotes the probability of the system being at the $i$th state  at $t=0$.
For example, if the inputs $u_i(t)$ are all null, then the single output
$
y(t)=\bigvee_{i} x_i(t)
$
 computes the Viterbi  score (\ref{viterbiscore}).
 %, which is the probability for having observed the data $(\vct o_0,...,\vct o_t)$
% and the HMM having passed through the optimum state sequence (that maximizes this probability).
One main difference of our system (\ref{mpsevitscal}) with the Viterbi algorithm (\ref{mpseviterbi}) is that we have
 the probability-like signals $u_i(t)$ which can act as control inputs coming possibly from previous human attention states
  or higher-level events; e.g. detected human faces, voice activity, or text semantics.
Another difference is that the  outputs
 of the dynamical system can be various min-max combinations of the saliency states
of various modalities; e.g. the single output
\beq
y(t) = c_1x_1(t) \vee c_2x_2(t) \vee d_1u_1(t) \vee d_2u_2(t)
\label{mpoutmxstainp}
\eeq
forms a weighted max-product fusion of the audio and visual saliencies as well as the two corresponding inputs.
In such modality and input combinations, the max rule can be replaced by min too.
A third difference is that the data-controlled probabilities $p_i(t)$ can enter not only via multiplication
but also via any commutative binary operation $\mgop$ that distributes over maximum.
If $\mgop=\max$, then the $p_i(t)$ can be viewed as control inputs.
Finally,
our CWL theoretical formulation allows us to also compute
analytically the responses of such max-product dynamical systems; see Sec.~\ref{sc-soresp}.

\begin{figure*}[!h]
\centerline{\psfig{figure=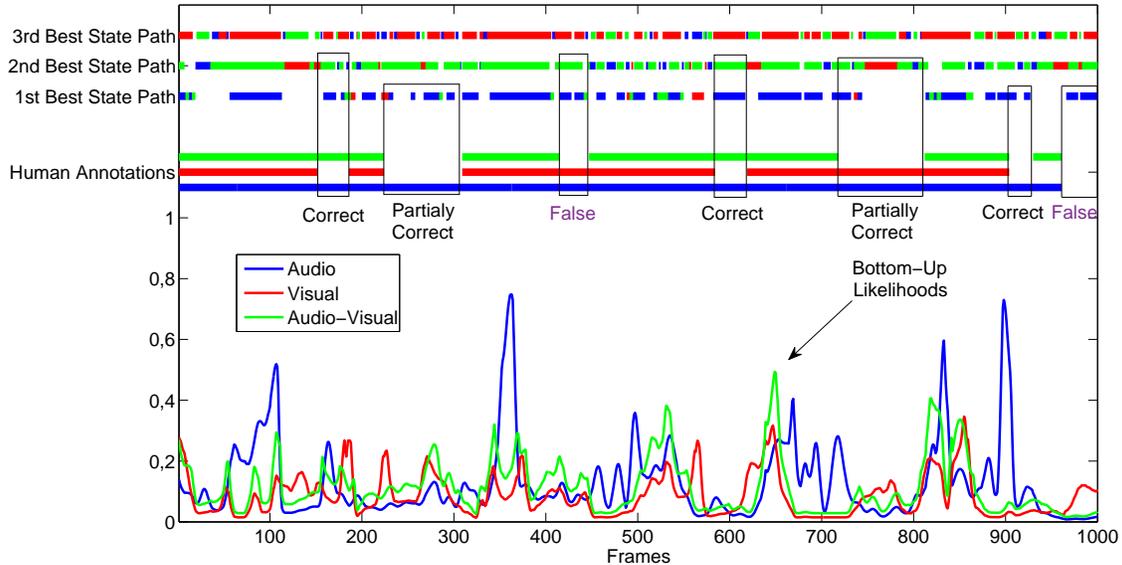,width=0.90\textwidth}}
\caption{Evolution of audio (blue), visual (red) and audio-visual (green) bottom-up
 likelihoods computed from observed features.  We also see the human annotations and the 3-Best state paths
using the max-product dynamical system (\protect{\ref{mpsevitscal}}) with $\mgop$ being product operation
and two  control inputs $u_1(t)$ and $u_2(t)$ providing binary information from voice and face detection respectively.
(This figure is best viewed in color.)}
\label{fg-avsal}
\end{figure*}

In our experiments \cite{MaKo15} we estimated the state transition probabilities $a_{ij}$ using the EM algorithm on
some training data from movie videos. For estimating the observation data probabilities $p_i(t)$
we fitted GMMs to audio and visual feature vectors extracted from the video data at each frame $t$.
Figure~\ref{fg-avsal} shows the results (on testing data from the same movie videos) of various approaches
we have initiated to track the joint audio-visual (AV) saliency state
and compare it (i)~with
human annotations, i.e. binary AV saliency manually annotated by a human who observed these movie videos,
and (ii)~with an AV saliency automatically computed in \cite{Eva+13} by fusing saliencies
of the audio and visual streams measured from monomodal cues.
Our ongoing research goal here is to develop a computational model that can track human attention
in the form of audio-visual saliency states
based on multimodal sensory inputs.
  As shown in Fig.~\ref{fg-avsal} and explained numerically in \cite{MaKo15},
our results using the max-product dynamical system are encouraging;
they can track audio-visual saliencies with smaller error than bottom-up feature-based local measurements
and can improve with higher-level control input.

% --- Max-Fuzzy norms ------------------------
\subsection{Max-$\inorm$norm systems and Fuzzy Markov Chains}
\label{sc-fuzzy}

There are many types of nonlinear dynamical systems where
the  elements of the state, input and output vectors  represent probabilities
or memberships. Examples include
probabilistic or fuzzy  control systems \cite{AvSa02,KlYu95,MST00},
fuzzy image convolutions \cite{BlMa95,Mara05a},
as well as certain types of neural nets with max-min combinations of inputs \cite{KlYu95,YaMa95}.
The dynamics of large classes of such systems can be described
by  the general model if we restrict the set of scalars to be
$\vset =[0,1]$ and use as scalar `multiplication' $a\mgop b=\inorm (a,b)$
a  \emph{fuzzy intersection norm} \cite{KlYu95}, i.e.,
a binary operation $\inorm : [0,1]^2\rightarrow [0,1]$ that
is i)~commutative, ii)~associative,  iii)~increasing and iv)~satisfies
the boundary condition $\inorm (a,1)=a$ for all $a\in [0,1]$.
This is also known as `triangular norm' (t-norm) in statistics.
We also require that $\inorm$ is continuous, which makes it a scalar dilation \cite{MST00}.
As dual `scalar multiplication', we use a continuous binary operation $\dinorm (a,b)=a\dmgop b$ % on $[0,1]$
that satisfies (i)-(iii) and the dual boundary condition $\dinorm (a,0)=a$.
This is a \emph{fuzzy union norm} \cite{KlYu95}, also known as `t-conorm', and is a scalar erosion on $[0,1]$.
Choosing in the general lattice dynamical model the above set of scalars
and `multiplications' among them creates the case of
% \emph{max-$\inorm$norm and min-$\dinorm$norm}
max-$\inorm$norm and min-$\dinorm$norm systems, obtained
by replacing in (\ref{mxgse}) and (\ref{mngse}) the general max-$\mgop$ matrix multiplication
and its dual with the following % max-$\inorm$norm and min-$\dinorm$norm
versions:
\bea
\mtr{C} %= \mtr{A} \mxgmp \mtr{B}
= \mtr{A} \mxfmp \mtr{B}=[c_{ij}]  & , &
 c_{ij} = {\dsty \bigvee _{k=1}^n} \inorm (a_{ik},b_{kj})
\label{mxfmpr} \\
\mtr{C} %= \mtr{A} \mngmp \mtr{B}
= \mtr{A} \mnfmp \mtr{B}=[c_{ij}]  & , &
 c_{ij} = {\dsty \bigwedge _{k=1}^n} \dinorm (a_{ik},b_{kj})
\label{mnfmpr}
\eea
Usually we select $\dinorm(a,b)=a\dmgop b=\cinorm(a,b)$ where
$\cinorm$ is the conjugate norm obtained via fuzzy complementation:
\beq
\cinorm(a,b)=1-\inorm(1-a,1-b)
\eeq
then $(\inorm,\cinorm)$ form a negation duality, but not an adjunction.
The most well-studied choice for the $\inorm$ norm and its dual norm $\cinorm$
are the min and max operations, respectively.
Another known case is for $\inorm$ to equal the product operation.
Table~\ref{tb-tncnadj} shows these cases  and their adjoints so that
$(\inorm,\asbdl)$ and $(\asber,\cinorm)$ are scalar adjunctions.
There are also numerous other choices.
\begin{table}[h!]
\caption{T-norms, Conorms and their Adjoints}
\centerline{
\begin{tabular}{|c|c||c|c|} \hline
t-norm & adjoint t-norm & t-conorm & adjoint t-conorm \\ \hline
$\inorm (a,v)$    & $\asbdl (a,w)$ & $\unorm=\glconj{\inorm}(a,w)$ & $\asber(a,v)$ \\ \hline
$\min (a,v)$ & $\max([w\geq a],w)^{\dagger}$
%\footnote{$[P]$ denotes the Iverson bracket, with value 1 (0) if predicate $P$ is true (false).}
 & $\max(a,w)$ & $\min([v>a],v)$ \\ \hline
$a\cdot v$ & $\min(w/a,1)$ & $a+w-a\cdot w$ & $\max(\frac{v-a}{1-a},0)$ \\ \hline
\end{tabular}
} $\;$\\
{\footnotesize $^{\dagger}\ [P]$ is the Iverson bracket with value 1 (0) if  $P$ is true (false).}
\label{tb-tncnadj}
\end{table}

An application of the above ideas to state-space description and control of
fuzzy dynamical systems is presented in \cite{MST00}.
Further, dynamical systems with states $\vct x(t) \in [0,1]^n$
and transition rule based on the max-min matrix `multiplication' ($\mgop=\min$) % ($T=\min$)
\beq
\vct x(t+1) = \mtr A \mxfmp \vct  x(t)=\mtr A^{(t)} \mxfmp \vct x(0), \quad \mtr A = \mtr P^T
\eeq
where $\mtr P=[p_{ij}]\in [0,1]^{n\times n}$ is the matrix of state transition probabilities
or fuzzy relations among states $(i,j)$,
have been called \emph{fuzzy Markov chains (FMCs)} in \cite{AvSa02}
and studied for decision-making.  An advantage they have over classical Markov chains
(whose transition rule is based on the sum-product matrix multiplication)
is that the powers of the transition matrix always reach a stationary solution $\vct x(\infty)$
in a finite number of steps.
Namely, the max-min powers of any matrix $\mtr A$ either converge in a finite time $\tau$,
i.e. $\mtr A^{(\tau +1)}=\mtr A^{(\tau )}$,
or oscillate  with a finite period  $\nu$ after some finite power $\tau$. % $\nu$.
In the aperiodic case ($\nu=1$), if the limiting matrix $\mtr A^{(\tau )}$ has identical columns,
then the stationary solution
$\vct x(\infty)$ is independent of the initial state $\vct x(0)$
and the FMC is called \emph{ergodic}.

We can extend these results for more general FMCs by using alternative $\inorm$-norms, e.g. the product.
Specifically, for both cases of Table~\ref{tb-tncnadj}
(i.e. when $\inorm$ is the minimum or product operation on $[0,1]$)
%                                                                                                                                                                                                                                                                                                                                                                                                                                                                                                                                                                                                                                                                                                                                                                                                                                                                                                                                                                                                                                                                                                                                                                                                                                                                                                                                                                        on $[0,1]$)
 Theorem~\ref{th-metmtr} applies and in particular (\ref{metmtrfin}) always holds.
From this we can deduce the finite convergence properties of generalized FMCs.
Further, if $a_{ii}=1$ for all $i$, then it follows
that $\mtr A^{(t)}\leq \mtr  A^{(t+1)}$ for all $t\geq 1$;
hence from  (\ref{metmtrfin}) we can prove an aperiodic finite convergence since
\beq
\metmtr{\mtr A} =\mtr A^{(n)}=\mtr A^{(t)} \quad \forall t > n
\eeq
Thus, $\mtr A \mxfmp \metmtr{\mtr A}=\metmtr{\mtr A}$.
This implies that all columns of the metric matrix  $\metmtr{\mtr A}$
are solutions of
\beq
\mtr A \mxfmp \vct  x = \vct x
\eeq
Such vectors are max-$\inorm$ eigenvectors of $\mtr A$ whose
principal eigenvalue is $\lambda (\mtr{A})=1$ and provide stationary solutions of the FMC.
As a numerical example, consider the transition matrix $\mtr A$ and its powers of a max-min FMC:
\beq
\mtr A=\left[ \begin{array}{ccc} 1 & 0.4 & 0 \\ 0.3 & 1 & 0.5 \\ 0.7 & 0.2 & 1 \end{array} \right]\leq \mtr A^{(2)}=\mtr A^{(3)}
=\metmtr{\mtr A}=\left[ \begin{array}{ccc} 1 & 0.4 & 0.4 \\ 0.5 & 1 & 0.5 \\ 0.7 & 0.4 & 1 \end{array} \right]
\eeq
%\begin{eqnarray}
%\mtr A & = & \left[ \begin{array}{ccc} 1 & 0.4 & 0 \\ 0.3 & 1 & 0.5 \\ 0.7 & 0.2 & 1 \end{array} \right]
% \leq \mtr A^{(2)}=\mtr A^{(3)}=\metmtr{\mtr A} \nonumber \\
%\metmtr{\mtr A} & = & \left[ \begin{array}{ccc} 1 & 0.4 & 0.4 \\ 0.5 & 1 & 0.5 \\ 0.7 & 0.4 & 1 \end{array} \right]
%\end{eqnarray}
The columns of  $\metmtr{\mtr A}$ provide stationary solutions of this FMC;
e.g. % $\mtr A \mxfmp [1,0.5,0.7]^T=[1,0.5,0.7]^T$.
\beq
\mtr A \mxfmp [1,0.5,0.7]^T=[1,0.5,0.7]^T .
\eeq

%###################################################
\section{Conclusions}

In this work we have a developed a unified theory of nonlinear dynamical systems of
the max-$\mgop$ type and their dual min-$\dmgop$ type over nonlinear vector and signal spaces
 which we call complete weighted lattices (CWLs).
 Special cases include max-sum or min-sum systems encountered in discrete event systems and shortest path problems,
max-product systems in statistical inference like the Viterbi algorithm, and the max-fuzzy-norms systems encountered in
certain types of neural nets and fuzzy control.
We have studied several control-theoretic and signal processing aspects of such systems, both by using CWLs for
shorter proofs of known cases and by extending the theory to more general cases.
Further we have also outlined several application areas that are either new or not often encountered in the literature,
which has emphasized so far the max-plus case and its application to discrete events systems;
examples include state-space representation and stability analysis of geometric filtering,
distance maps, fuzzy Markov chains, a generalized Viterbi algorithm for HMMs with control inputs
and its application to tracking salient events in multimodal videos.

Overall, the unified formulation of the above systems and the corresponding CWL framework
provide several advantages over minimax algebra which include:
capability of handling both finite- and infinite-dimensional cases;
 co-existence over the same space of the max-$\mgop$  and the dual min-$\dmgop$ systems;
lattice monotone operators that can represent both matrix-vector multiplications in state-space
as well as sup/inf input-output signal convolutions;
lattice adjunctions (pairs of dual operators)
that yield optimal solutions to max-$\mgop$ and  min-$\dmgop$ equations via lattice projections.
\\[3mm]

\section*{Acknowledgements}
The author wishes to thank the anonymous Reviewers for their constructive comments.
He also wishes to thank Petros Koutras at NTUA CVSP lab for producing Figure~\ref{fg-avsal}
and Anastasios Tsiamis at the University of Pennsylvania for Example~\ref{ex-reachabil}.
% --- For ARCHIV ----
\\
This research was
supported by the project ``COGNIMUSE"  under the ``ARISTEIA"
Action of the Operational Program “Education and Lifelong Learning” and was
co-funded by the European Social Fund and Greek National Resources.
It was also partially supported   by the European
Union under the projects  MOBOT with grant FP7-600796
and BabyRobot with grant H2020-687831.
\\[3mm]

%################# BIB ##################################
\bibliographystyle{plain}
%
%\bibliography{maragos_dswl_bibliography}  % name your BibTeX data base
%

\footnotesize{
\bibliographystyle{plain}

} % end of footnotesize

\end{document}